\documentclass[11pt,a4paper]{article}

\usepackage[T1]{fontenc}
\usepackage[utf8]{inputenc}
\usepackage{lmodern}
\usepackage{amsmath,amssymb,mathtools,bm}
\usepackage{microtype}
\usepackage[a4paper,margin=2.45cm]{geometry}
\usepackage{booktabs,array,longtable}
\usepackage{enumitem}
\usepackage{xcolor}
\definecolor{mygreen}{RGB}{220, 245, 220}
\usepackage{hyperref}
\usepackage[nameinlink,noabbrev]{cleveref}
\usepackage{graphicx}

\hypersetup{
 colorlinks=true,
 linkcolor=blue!50!black,
 citecolor=blue!50!black,
 urlcolor=blue!50!black,
 pdftitle={Jacobi Modular Flavour at Level Three},
 pdfauthor={Ferruccio Feruglio}
}
\setlist[itemize]{topsep=4pt,itemsep=2pt,parsep=1pt}
\setlist[enumerate]{topsep=4pt,itemsep=3pt,parsep=1pt}

\newcommand{\ii}{\mathrm{i}}

\newcommand{\one}{\mathbbm{1}}
\IfFileExists{bbm.sty}{\usepackage{bbm}}{\renewcommand{\one}{\mathbf{1}}}
\newcommand{\C}{\mathbb C}
\newcommand{\R}{\mathbb R}
\newcommand{\Z}{\mathbb Z}
\newcommand{\F}{\mathbb F}
\newcommand{\Hone}{\mathbb H}
\newcommand{\Htwo}{\mathbb H_2}
\newcommand{\Sp}{\mathrm{Sp}}
\newcommand{\SL}{\mathrm{SL}}

\newcommand{\Heis}{\mathrm H}
\DeclareRobustCommand{\Img}{\operatorname{Im}}

\newcommand{\Dtw}{\Delta(27)}

\newcommand{\Jthree}{J_3}
\newcommand{\Omec}{\Omega(1)}

\newcommand{\GJ}{\Gamma^{J}}

\newcommand{\eps}{\varepsilon}

\newcommand{\three}[1]{\mathbf 3^{(#1)}}
\newcommand{\six}[1]{\mathbf 6^{(#1)}}

\title{\bfseries Eclectic flavour symmetries without flavons\\[3pt]}

\renewcommand{\thefootnote}{\fnsymbol{footnote}}
\author{
Ferruccio Feruglio$^{1}$\thanks{E-mail: \texttt{feruglio@pd.infn.it}}
\quad
Antonio Marrone$^{2,3}$\thanks{E-mail: \texttt{antonio.marrone@ba.infn.it}}
\\[10pt]
\normalsize $^{1}$\textit{INFN, Sezione di Padova, Italia}\\
\normalsize $^{2}$\textit{Dipartimento Interateneo di Fisica, Bari, Italia}\\
\normalsize $^{3}$\textit{INFN, Sezione di Bari, Italia}
}
\date{}

\begin{document}
\maketitle
\renewcommand{\thefootnote}{\arabic{footnote}}

\begin{abstract}
We develop a new framework in which eclectic groups are realised as flavour symmetries without flavons.
Focusing on the group $\Omega(1)$, we show that it can be realised as a finite image of the
integral Jacobi group $\Heis(\Z)\rtimes\SL(2,\Z)$, acting on two moduli $(\tau,z)$.
The Heisenberg subgroup $\Heis(\Z)$ leaves the modulus $\tau$ unchanged,
and its finite image $\Delta(27)$ plays the role of the traditional flavour symmetry.
We provide the ingredients needed to construct supersymmetric Jacobi-invariant theories, both in globally supersymmetric theories and in supergravity. We develop a realistic model of fermion masses within this framework and
formulate a consistent CP symmetry, which can be imposed at the
Lagrangian level and broken spontaneously by the moduli. We discuss how the conventional eclectic-flavour description is recovered in an appropriate limit.
We also extend this geometric interpretation to the $T^2/\mathbb Z_4$ and $T^2/\mathbb Z_2$ eclectic building blocks,
and clarify why the $T^2/\mathbb Z_6$ case is structurally different.
\end{abstract}

\tableofcontents
\section{Motivation and scope}
The origin of flavour remains one of the major open questions of particle
physics.  Fermion masses and mixing parameters are now measured with
remarkable precision, yet their hierarchical pattern and the very
different structures observed in the quark and lepton sectors are not
explained by the Standard Model.  Symmetry-based approaches have long
provided one of the main strategies to address this problem, but no
single framework has emerged as a baseline theory of flavour
\cite{Feruglio:2019ybq,Ding:2024ozt}.  In the neutrino sector, in
particular, much of the mixing pattern is already accurately known,
while a comparatively small number of qualitatively important
quantities---such as the absolute neutrino-mass scale, the Majorana
nature of neutrinos and the pattern of leptonic CP violation---remain
undetermined.  This situation makes flavour both a precision problem
and a structural one: the challenge is not only to reproduce the data,
but to understand why they take the form they do.

The modular
approach to flavor~\cite{
Feruglio:2017spp}, reviewed in Refs.~\cite{Ding:2023htn,Nilles:2023shk,Kobayashi:2023zzc,Ratz:2024imd}
has provided an economical organising principle for fermion masses. In the standard construction,
a modulus $\tau$ transforms under the infinite modular group, whereas matter multiplets and modular forms
transform in finite-dimensional representations with finite image. A chiral supermultiplet typically obeys
\begin{equation}
 \tau\longmapsto\frac{a\tau+b}{c\tau+d}\,,\qquad
 \Phi_I\longmapsto(c\tau+d)^{-k_I}\rho_I(\gamma)\Phi_I\,,
 \label{eq:standard-modular}
\end{equation}
where $\rho_I(\gamma)$ is a unitary representation of a finite image of $\SL(2,\Z)$.
This differs from the more familiar realisation in terms of a traditional flavour symmetry, acting as
\begin{equation}
 \tau\longmapsto\tau\,,\qquad
 \Phi_I\longmapsto U_I(g)\Phi_I\,,
 \label{eq:traditional}
\end{equation}
where $U_I(g)$ is a unitary representation of the flavour group and the modulus $\tau$ is inert.
The difference becomes particularly relevant when the symmetry must be broken in order to reproduce the
observed fermion spectrum. In conventional bottom-up realisations, the breaking of a traditional flavour
symmetry is usually implemented through flavon vacuum expectation values, whereas purely modular
constructions can avoid an independent flavon sector. From this viewpoint, the often elaborate flavon
sector makes the traditional flavour-symmetry setup less economical.

At first sight, the two types of symmetry may appear difficult to reconcile. In top-down constructions,
however, they coexist and are related by a consistency condition of the form
\begin{equation}
 \rho_I(\gamma) U_I(g) \rho_I(\gamma)^{-1}=U_I(v_\gamma(g))\,,
 \label{eq:eclectic}
\end{equation}
where $v_\gamma(g)$ defines a non-trivial automorphism of the discrete traditional flavour group.
Traditional and modular transformations then generate an eclectic
group~\cite{
Baur:2019kwi,Baur:2019iai,
Nilles:2020nnc,Nilles:2020kgo,
Baur:2021TopDown,Baur:2024qzo},
of which the traditional flavour group is a normal subgroup. 
A closely related structure arises in magnetized toroidal and D-brane
compactifications~\cite{
Cremades:2003qj,Cremades:2004wa,
Abe:2009vi,Marchesano:2013Dbrane,
Kobayashi:2018rad,Ohki:2020bpo,
Kikuchi:2020frp,Kikuchi:2021ogn,
Almumin:2021Metaplectic}.
In these constructions, chiral zero-mode wave functions are naturally
described by theta functions and transform non-trivially under
translations and modular transformations of the compact torus.
Magnetic translations can generate finite Heisenberg-type flavour
groups, while modular transformations act on
the zero-mode multiplets through finite, and in general metaplectic,
representations.  Thus conventional discrete flavour transformations
and modular transformations need not commute.

A similar interplay is realised in heterotic orbifold
compactifications.
For example, in heterotic orbifold
compactifications containing a $T^2/\Z_3$ sector, the twisted states localised at the three fixed points
exhibit the traditional flavour symmetry $\Delta(54)$, while the target-space modular symmetry gives rise
to the finite modular group $T'\simeq\SL(2,3)$. Their multiplicative closure is the eclectic group
$\Omec$~\cite{Nilles:2020tdp,Nilles:2020gvu} of order $648$~\footnote{
See also Refs.~\cite{Kobayashi:2006StringyOrigin,
Baur:2019kwi,Baur:2019iai,KnappPerez:2025Demystifying}
for the string-theoretic origin of non-Abelian flavour symmetries
and their relation to string selection rules.}.
The two subgroups are not independent: $\Delta(54)$ and $T'$ share a non-trivial order-two element,
conventionally identified with $S^2$. Consequently, $\Omec$ can be organised as
\begin{equation}
 \Omec\simeq \Delta(27)\rtimes T'\,,
 \label{eq:omega1-J3}
\end{equation}
a form that will be directly relevant to our construction below.\footnote{$\Omec$ is further enhanced to
$\Omega(2)$ when the additional R-symmetry of the six-dimensional compactification is included.}

In the top-down orbifold constructions in which eclectic flavour is realised, the modular symmetry is
therefore accompanied by a larger non-commuting flavour structure. This property can substantially
constrain the otherwise less controlled K\"ahler potential of the low-energy theory~\cite{Chen:2019ewa}
and has been exploited in a series of explicit constructions~\cite{Chen:2021prl,Ding:2023ynd,Li:2023dvm,Li:2024pff}.
In the presently known realistic implementations, however, the breaking of the traditional factor still
relies on flavon multiplets, which one would like to avoid. The role played by the flavon is particularly
transparent in the Yukawa sector. A generic mass matrix in the $T^2/\Z_3$ eclectic construction has the
form~\cite{Baur:2022hma}
\begin{equation}
 M_{\rm ecl}(\tau,\varphi)=c
 \begin{pmatrix}
 \widehat Y_2\,\varphi_1 & -\dfrac{\widehat Y_1}{\sqrt2}\,\varphi_3 & -\dfrac{\widehat Y_1}{\sqrt2}\,\varphi_2\\[2mm]
 -\dfrac{\widehat Y_1}{\sqrt2}\,\varphi_3 & \widehat Y_2\,\varphi_2 & -\dfrac{\widehat Y_1}{\sqrt2}\,\varphi_1\\[2mm]
 -\dfrac{\widehat Y_1}{\sqrt2}\,\varphi_2 & -\dfrac{\widehat Y_1}{\sqrt2}\,\varphi_1 & \widehat Y_2\,\varphi_3
 \end{pmatrix},
 \label{eq:eclectic-texture}
\end{equation}
where $\widehat{\bm Y}_{\bm 2}(\tau)=(\widehat Y_1,\widehat Y_2)^T$ is a modular doublet of $T'$ and
$\varphi=(\varphi_1,\varphi_2,\varphi_3)^T$ is a triplet flavon. Group theoretically, the corresponding
coupling structure is associated with the product
$\widehat{\bm Y}_{\bm 2}(\tau)\otimes\varphi_{\bm 3}$.

The central idea pursued here is that the second factor need not originate from an independent scalar
multiplet: it may instead be encoded in the modular geometry itself. More generally, the entire coupling
structure represented by $M_{\rm ecl}$ may be replaced by a multiplet of functions of two moduli, $\tau$
and $z$. Rather than
starting from two finite symmetry factors and their mutual
automorphisms, we seek a common infinite geometric parent acting on the moduli $(\tau,z)$, from which
both the traditional and modular components arise as finite remnants. This opens the possibility of retaining the successful group-theoretical structure while
eliminating both flavons and a separate vacuum-alignment sector. The present work therefore addresses the
following question: can an eclectic-like flavour construction be realised entirely through transformations
of moduli, without introducing flavons?

We show that the answer is affirmative and we provide an explicit realization for $\Omega(1)$. More precisely, $\Omega(1)$ can be realised as the
finite level-three image of the integral Jacobi group
\begin{equation}
 \GJ=\Heis(\Z)\rtimes\SL(2,\Z)\,.
\end{equation}
The relevant finite group is
\begin{equation}
 J_3\simeq \mathrm H_3(\F_3)\rtimes\SL(2,\F_3)
 \simeq \Delta(27)\rtimes T'
 \simeq \Omega(1)\,.
 \label{eq:J3-Omega1}
\end{equation}
Thus the subgroup that plays the role of traditional $\Delta(27)$ flavour symmetry is promoted to the
finite image of genuine Jacobi transformations. Its Heisenberg translations act on the additional elliptic
modulus $z$ while leaving $\tau$ invariant, whereas the $\SL(2,\Z)$ factor acts on both $\tau$ and $z$.
The traditional and modular transformations are therefore unified as different components of a single
infinite geometric symmetry.

We provide the ingredients needed to build supersymmetric Jacobi-invariant theories, both in globally supersymmetric theories and in supergravity. We outline a low-weight catalogue of holomorphic level-three Jacobi forms,
identify the corresponding automorphy factors, and define the general Jacobi transformation law of chiral
supermultiplets. We construct minimal and non-minimal K\"ahler potentials and derive the general invariance
conditions on the superpotential. Finally, we show how Jacobi multiplets can replace the combination of
modular forms and flavons appearing in conventional eclectic models. In particular, a Jacobi sextet can
realise the six-dimensional coupling structure associated with a modular doublet times a flavour triplet,
without introducing a flavon multiplet or a separate vacuum-alignment sector. We then use this mechanism
to construct a realistic model of fermion masses.

Although eclectic flavour symmetry can be formulated in a bottom-up language, its strongest motivation comes from string theory. Our proposal is deliberately bottom-up: the weights, Jacobi indices and $J_3$ representations of matter multiplets and Jacobi forms are treated as input data rather than being derived from a compactification. Nevertheless, closely related geometric ingredients are known to occur in string theory. In particular, Jacobi forms arise naturally in heterotic compactifications with Wilson-line moduli, while enlarged modular symmetries involving $\Sp(4,\Z)$ have appeared both in heterotic compactifications and in their flavour interpretation
\cite{Mayr:1995Stieberger,Cardoso:1996JacobiWilson,Nazaroglu:2013Jacobi,
Baur:2020Siegel,Nilles:2021Sp4Orbifolds,Ishiguro:2021Symplectic}.
The appearance of Jacobi modular symmetry as a remnant of a larger modular structure is also familiar in string compactifications. In heterotic/F-theory duality, for instance, Jacobi forms arise as Fourier--Jacobi coefficients of orthogonal modular forms \cite{Sakai2026FtheoryK3}. 
These observations provide useful motivation for the Jacobi framework considered here, without requiring a specific string interpretation of the elliptic modulus $z$.

\section{Jacobi flavour symmetry}
\label{PR}
To avoid an independent flavon sector, we seek a geometric realisation in which
the two non-commuting finite actions in Eq.~\eqref{eq:eclectic} both descend from
transformations of moduli. The simplest construction with two independent
$\SL(2,\Z)$ factors acting separately on two moduli does not produce the desired
result: the two actions commute, and therefore $\rho_I(\gamma)$ and $U_I(g)$
would have to commute as well. In this case Eq.~\eqref{eq:eclectic} would be
satisfied only by the trivial automorphism $v_\gamma(g)=g$. The minimal setting
in which the two transformations are intrinsically intertwined is provided by
the Jacobi group, which combines an ordinary modular transformation of one
complex modulus with elliptic translations of a second one. In the bottom-up
approach adopted here, the Jacobi group is taken as the fundamental flavour
symmetry.

\subsection{The Jacobi group}
\label{TJG}
We start by considering two moduli $(\tau,z)$, $\Img\tau>0$, living in the Jacobi upper half-plane $\Hone\times\C$.
We extend the standard action of the modular group on $\tau$, by letting $z$ transform with modular weight $-1$:
\begin{equation}
 \tau\longmapsto\frac{a\tau+b}{c\tau+d},
 \qquad
 z\longmapsto\frac{z}{c\tau+d}\,,
 \label{eq:mod-action-Jacobi}
\end{equation}
where
\begin{equation}
 \gamma=\begin{pmatrix}a&b\\c&d\end{pmatrix}\in\SL(2,\Z),
 \qquad ad-bc=1\,.
\end{equation}
This is the {\em modular} part of our set of transformations. As for the geometric counterpart of the traditional flavour transformations, we ask that the modulus $\tau$ remains unchanged, while $z$ is mapped to an equivalent point on the complex torus
\begin{align}
 E_\tau=\mathbb C/(\mathbb Z\tau+\mathbb Z)\,.
\end{align}
This request is satisfied by the translations of the $z$ modulus:
\begin{equation}
 \tau\longmapsto\tau,\qquad
 z\longmapsto z+\lambda\tau+\mu\,,~~~~~~~~~~~~~~~\lambda,\mu\in\Z\,.
 \label{eq:Heis-action-Jacobi}
\end{equation}
Thus $\tau$ determines the complex structure of the torus, while $z$ specifies a point on it.
The transformations of eq. (\ref{eq:Heis-action-Jacobi}) are generated by $X:(\tau,z)\to(\tau,z+\tau)$ and $Y:(\tau,z)\to(\tau,z+1)$. 
 The translations \(X\) and \(Y\) commute on the moduli space, but their lifts to fields need not do so.
There is a useful physical analogy with magnetic translations.
For a charged particle on a torus in the presence of a magnetic flux,
ordinary translations commute as transformations of the coordinates,
but their action on wave functions must be accompanied by gauge
transformations.  The corresponding magnetic-translation operators $T_X$ and $T_Y$
therefore commute only up to a phase,
\begin{equation}
T_X\,T_Y
 =
 e^{2\pi i\Phi/\Phi_0}\,
T_Y\,T_X,
\end{equation}
where $\Phi$ is the magnetic flux enclosed by the two translations.
This phase acts trivially on the coordinates but non-trivially on the
wave functions and provides the central element of the magnetic
translation group~\footnote{Closely related structures occur in magnetized toroidal
compactifications, where degenerate zero-mode wave functions transform
under non-commuting magnetic translations and finite Heisenberg-type
flavour groups, including $\Delta(27)$, can emerge
\cite{Abe:2009vi,Kobayashi:2018rad,Ohki:2020bpo}.}. 

In our case, as transformations of the elliptic coordinate $z$, $X$ and $Y$ are
ordinary lattice translations and therefore commute on the base torus
$E_\tau$. Matter multiplets and Jacobi forms, however, transform with non-trivial automorphy factors. Their lifts under \(X\) and \(Y\) need not commute and, in the Jacobi realization relevant here, close only up to a phase represented by a central generator \(Z\). The latter acts trivially on $(\tau,z)$ but
non-trivially on the multiplets.  The natural symmetry of the full
theory is thus the central extension of the lattice translations,
namely the integral Heisenberg group $\Heis(\mathbb Z)$,
whose generators  $X,Y,Z$ satisfy~\footnote{It is useful to comment briefly on the normalization adopted in
Eq.~\eqref{eq:integral-Heis}.  The central generator $Z$ is chosen as a primitive central generator, while the commutator of $X$ and $Y$ is $Z^2$,
\begin{equation}
   XY=Z^2YX .
\end{equation}
This distinction has no effect at level three: $Z$ and $Z^2$ generate the same group, so that the finite
Heisenberg group is simply
\begin{equation}
   H_3(\mathbb F_3)\simeq\Delta(27).
\end{equation}
It becomes relevant, however, at level two.  In that case keeping the
primitive generator $Z$ leads to a larger finite Heisenberg group than
one would obtain by retaining only the commutator $Z^2$.  As we shall
see below, this is precisely the structure needed for the
$T^2/\mathbb Z_4$ eclectic symmetry.}
\begin{equation}
\Heis(\Z)
=
\left\langle X,Y,Z \,\middle|\,
XZ=ZX,\;
YZ=ZY,\;
XY=Z^{2}YX
\right\rangle\,.
\label{eq:integral-Heis}
\end{equation}
The element $Z$ is the center of the group. Its action on the moduli $(\tau,z)$ is trivial, but it will act non-trivially on matter multiplets and modular forms. 

A generic element of $\Heis(\Z)$ will be denoted by $h(\lambda,\mu,\kappa)$, with $\lambda,\mu,\kappa\in\Z$. The parameter $\kappa$ is associated to the central element $Z$. The modular transformations of Eq.~\eqref{eq:mod-action-Jacobi}, together with the centrally extended lattice translations forming the integral Heisenberg group \(\Heis(\mathbb Z)\), generate the integral Jacobi group $\GJ$
\begin{equation}
 \GJ=\Heis(\Z)\rtimes\SL(2,\Z)\,.
 \label{eq:Jacobi-infinite}
\end{equation}
The Jacobi group has the structure of a semidirect product since
conjugation by $\gamma$ defines an
automorphism $v_\gamma$ of the Heisenberg subgroup,
\begin{align}
 \gamma\,h(\lambda,\mu,\kappa)\,\gamma^{-1}
 &=
 v_\gamma\!\left(h(\lambda,\mu,\kappa)\right)
 =
 h(\lambda',\mu',\kappa')\,,
 \label{eq:Heis-automorphism}
\end{align}
where
\begin{align}
 \begin{pmatrix}
 \lambda'\\[1mm]
 \mu'
 \end{pmatrix}
 &=
 \gamma^{-T}
 \begin{pmatrix}
 \lambda\\[1mm]
 \mu
 \end{pmatrix}
 =
 \begin{pmatrix}
 d&-c\\
 -b&a
 \end{pmatrix}
 \begin{pmatrix}
 \lambda\\[1mm]
 \mu
 \end{pmatrix},
 \label{eq:Heis-SL-action}
\end{align}
up to the corresponding shift of the central coordinate. On a matter
multiplet $\Phi_I$, the same semidirect-product structure implies
\begin{align}
 \rho_I(\gamma)\,
 U_I(h)\,
 \rho_I(\gamma)^{-1}
 =
 U_I\!\left(v_\gamma(h)\right)\,.
 \label{eq:Jacobi-eclectic-relation}
\end{align}
In general, the transformation law of a matter multiplet $\Phi_I$ also involves an automorphy factor $J_I(g,x)$, which is discussed in Section \ref{sec:automorphy-bottomup}, which satisfies the cocycle condition
\begin{align}
J_I(g_1g_2,x)=J_I(g_1,g_2x)J_I(g_2,x)\,,
\label{cocycle}
\end{align}
where $x=(\tau,z)$.
From the cocycle condition we get
\begin{align}
J_I(\gamma,h\gamma^{-1} x)\,J_I(h,\gamma^{-1} x)\,J_I(\gamma^{-1} ,x)= J_I(v_\gamma(h),x)\,,
 \label{eq:Jacobi-eclectic-relation1}
\end{align}
which, in combination with eq.~\eqref{eq:Jacobi-eclectic-relation}, provides precisely the generalization of the consistency condition of eq.~\eqref{eq:eclectic}: the chain of transformations $\gamma\, h\, \gamma^{-1}$ on a chiral multiplet returns
a transformation $v_\gamma(h)$ of the Heisenberg group. In the conventional eclectic limit, in which the Heisenberg factor acts trivially on the moduli and carries no non-trivial automorphy factor, one recovers the familiar relation of eq.~\eqref{eq:eclectic}.
Thus, in the Jacobi construction, the eclectic consistency relation is not imposed as an additional requirement: it follows directly from the semidirect-product structure of the underlying geometric symmetry. The action of the integral Heisenberg subgroup $\Heis(\Z)$  leaves unchanged the modulus $\tau$ and represents the geometric counterpart of the traditional flavour transformations. 

A generic Jacobi transformation $g=(\gamma;\lambda,\mu,\kappa)$ therefore acts as
\begin{equation}
 (\tau,z)\longmapsto
 \left(
 \frac{a\tau+b}{c\tau+d},
 \frac{z+\lambda\tau+\mu}{c\tau+d}
 \right).
 \label{eq:generic-Jacobi-action-z}
\end{equation}
As $\tau$ varies, the torus $E_\tau$
varies with it, defining a family of elliptic curves over the modular parameter space, usually referred to as the universal elliptic curve. The Jacobi group acts simultaneously on the base coordinate $\tau$ and on the point $z$ in the corresponding fibre.

A convenient choice of fundamental domain ${\cal F}^{J}$ is obtained by quotienting the Jacobi upper half-plane by the discrete Jacobi group. One may first reduce the modular coordinate to the standard fundamental domain and then, at fixed $\tau$, reduce the elliptic coordinate modulo the lattice generated by $1$ and $\tau$. A convenient choice is therefore~\footnote{Equivalently, in terms of the convention $w=z/3$ adopted below,
\begin{align}
w&=\frac{u+v\tau}{3},\qquad
-\frac12\leq u<\frac12,\qquad
-\frac12\leq v<\frac12\,.
\label{eq:wFundamentalDomain}
\end{align}}
\begin{align}
{\cal F}^{J}
&=\{(\tau,z)\in\mathbb H\times\mathbb C:\ 
|\Re\tau|\leq \tfrac12,\quad |\tau|\geq1,
\nonumber\\
&\hspace{24mm} z=u+v\tau,\quad
-\tfrac12\leq u<\tfrac12,\quad
-\tfrac12\leq v<\tfrac12\}\,.
\label{eq:JacobiFundamentalDomain}
\end{align}
Thus $\tau$ parametrizes the usual modular fundamental region, while $z$ is restricted to a fundamental parallelogram of the elliptic fibre $E_\tau$. As usual, points on the boundary are subject to further identifications, and additional stabilizers occur at the elliptic fixed points of the modular action. The central Heisenberg generator acts trivially on $(\tau,z)$ and therefore does not modify this description.

\subsection{Level three and the finite Jacobi group}
\label{sec:level-three}
As in ordinary modular flavour constructions, the moduli transform under an infinite group, whereas matter multiplets and modular forms may transform through a finite image.  We shall be interested in level three.  Reducing the modular factor modulo three gives
\begin{equation}
 \SL(2,\Z)\longrightarrow\SL(2,\F_3)\simeq T'\,.
 \label{eq:SL2-reduction}
\end{equation}
Likewise, reducing $\lambda,\mu,\kappa$ modulo three gives the finite Heisenberg group
\begin{equation}
 H_3(\F_3)
 =\{h(\lambda,\mu,\kappa);\;\lambda,\mu,\kappa\in\F_3\}\,,
 \label{eq:finite-Heisenberg}
\end{equation}
whose order is $|H_3(\F_3)|=27$. With the convention of Eq.~\eqref{eq:integral-Heis}, $H_3(\F_3)$ is isomorphic to
$\Delta(27)$.
Since the modular action in Eq.~\eqref{eq:Heis-SL-action} preserves the Heisenberg commutator, the semidirect-product structure survives the reduction.  The finite image of the Jacobi group is therefore

\begin{equation}
\displaystyle
 J_3=H_3(\F_3)\rtimes\SL(2,\F_3)
 \simeq\Delta(27)\rtimes T'\,,
 \label{eq:J3-definition}
\end{equation}
of order $648$.  The full symmetry acting on the moduli remains $\Gamma^J$; the reduction modulo three only specifies the finite matrices acting on multiplets.

For the explicit forms it is convenient to introduce the rescaled elliptic coordinate $w=z/3$.
Then
\begin{equation}
 (\tau,w)\longmapsto
 \left(
 \frac{a\tau+b}{c\tau+d},
 \frac{w+(\lambda\tau+\mu)/3}{c\tau+d}
 \right).
 \label{eq:generic-Jacobi-action-w}
\end{equation}
This normalization is particularly useful because the Jacobi forms entering the flavour construction have conventional integral index.

\subsection{Automorphy factors and matter fields}
\label{sec:automorphy-bottomup}
A Jacobi multiplet is characterized by a modular weight $k$, a classical
Jacobi index $m$, and a finite representation $\rho$ of the level-three
Jacobi group $J_3$. Working in the rescaled elliptic coordinate $w=z/3$,
we define the associated level-three automorphy factor as
\begin{align}
 J_{k,m}(g;\tau,w)
 ={}&(c\tau+d)^k
 \exp\Bigg\{2\pi i m\Bigg[
 \frac{c\left(w+(\lambda\tau+\mu)/3\right)^2}{c\tau+d}
 -\frac{\lambda^2\tau+6\lambda w+\kappa}{9}
 \Bigg]\Bigg\}\,,
 \label{eq:Jacobi-automorphy-BB}
\end{align}
where $g=\big(\gamma,(\lambda,\mu),\kappa\big)\in \GJ$
and $\kappa$ --- not to be confused with the weight $k$ --- parametrizes the central element $Z$.
By construction this factor satisfies the cocycle
condition
\begin{equation}
 J_{k,m}(g_1g_2;\tau,w)
 =J_{k,m}(g_1;g_2\cdot(\tau,w))\,
  J_{k,m}(g_2;\tau,w)\,,
 \label{eq:cocycle-BB}
\end{equation}
which guarantees consistency under group
composition of the transformation properties of Jacobi forms and matter multiplets.
A vector-valued Jacobi form $Y_{k,m}$ of weight $k$, index $m$ and
representation $\rho$ then transforms as
\begin{equation}
\displaystyle
 Y_{k,m}(g\cdot(\tau,w))
 =J_{k,m}(g;\tau,w)\,
 \rho(\widetilde g)\,Y_{k,m}(\tau,w)\,,
 \label{eq:Jacobi-transformation-BB}
\end{equation}
where $\widetilde g$ denotes the image of $g$ under the projection onto
the finite level-three quotient $J_3$.
It is instructive to see what this implies for the central generator
$Z=h(0,0,1)$ of the Heisenberg subgroup, which acts trivially on the
moduli, $Z\cdot(\tau,w)=(\tau,w)$, but is not represented trivially in
the automorphy factor.
Consistency of \eqref{eq:Jacobi-transformation-BB} under $g=Z$ then
forces
\begin{equation}
\rho(Z)=e^{2\pi i m/9}\,\one\,,
 \label{eq:m-central-character-BB}
\end{equation}
i.e.\ a non-vanishing Jacobi multiplet exists only if $\rho(Z)$ is the
scalar fixed by the index $m$.
For the representations relevant here, $\rho$ is assumed to factor
through $J_3$ in such a way that the central kernel of the level-three
reduction acts trivially; in particular $Z^3$ lies in this kernel, so
$\rho(Z)^3=\one$. Combined with \eqref{eq:m-central-character-BB} this
gives $ e^{2\pi i m/3}=1$ and
\begin{equation}
 m\in3\mathbb{Z}\,.
 \label{eq:m-multiple-three-BB}
\end{equation}
Writing $m=3n$, eq.~\eqref{eq:m-central-character-BB} becomes
$\rho(Z)=e^{2\pi i n/3}$, so the value of $\rho(Z)$ only depends on
$m \bmod 9$: $m\equiv0\pmod9$ is centre-neutral ($\rho(Z)=1$), while
$m\equiv3\pmod9$ and $m\equiv6\pmod9$ correspond to the
$\omega$- and $\omega^2$-sectors, respectively, with
$\omega=e^{2\pi i/3}$.

Matter multiplets are treated differently. A chiral matter field
$\Phi_I$ is assigned its own weight $k_I$, index $m_I$, and unitary
representation $\rho_I$ of $J_3$, transforming with the \emph{inverse}
automorphy factor,
\begin{equation}
\displaystyle
 \Phi_I\longmapsto
 J_{k_I,m_I}(g;\tau,w)^{-1}
 \rho_I(\widetilde g)\,\Phi_I\,.
 \label{eq:matter-Jacobi-BB}
\end{equation}
Crucially, for matter fields the index $m_I$ and the finite central
character carried by $\rho_I$ are \emph{independent} data: there is no
constraint analogous to \eqref{eq:m-central-character-BB}. Under the
central element $Z$, the total transformation is simply
\begin{equation}
 \Phi_I\ \stackrel{Z}{\longmapsto}\
 e^{2\pi i m_I/9}\,\rho_I(Z)\,\Phi_I\,.
 \label{eq:matter-central-BB}
\end{equation}
The distinction between \eqref{eq:m-central-character-BB} and
\eqref{eq:matter-central-BB} is conceptually important: a Jacobi form
is a fixed function of the moduli $(\tau,w)$ and must obey the
consistency condition \eqref{eq:m-central-character-BB} in order to be
non-zero, whereas a matter field is an independent dynamical degree of
freedom, free to carry any combination of index and central character.
\section{Jacobi-invariant supersymmetric effective theory}
\label{sec:Jacobi-EFT}
We now formulate the supersymmetric effective theory directly on the Jacobi moduli space, without introducing any additional modulus.  Moduli and matter fields are described by chiral supermultiplets.  We first discuss the K\"ahler potential and then the superpotential selection rules.

\subsection{K\"ahler potential}
\label{sec:Kahler}
Write
\begin{equation}
 \tau=x+i y,\qquad w=u+i v,\qquad y>0\,.
\end{equation}
The Siegel--Jacobi upper half-plane admits a two-parameter family of invariant
K\"ahler metrics~\cite{Yang:2007Jacobi,Yang:2008Cayley}. A convenient potential is
\begin{equation}
\displaystyle
 K_{\rm mod}
 =-h\log y+4\pi\nu\,\frac{v^2}{y}\,,
 \qquad h>0,\quad\nu>0.
 \label{eq:Kmod-Jacobi}
\end{equation}
Under a generic Jacobi transformation
$g=(\gamma;\lambda,\mu,\kappa)$, the K\"ahler potential transforms as
\begin{align}
 K_{\rm mod}\big(g\cdot(\tau,w)\big)
 &=
 K_{\rm mod}(\tau,w)
 +f_g(\tau,w)+\overline{f_g(\tau,w)}\,,
 \label{eq:Kmod-Kahler-transformation}
\end{align}
where
\begin{align}
 f_g(\tau,w)
 ={}&h\log(c\tau+d)
 +2\pi i\nu
 \Bigg[
 \frac{c\left(w+(\lambda\tau+\mu)/3\right)^2}{c\tau+d}
 -\frac{\lambda^2\tau}{9}
 -\frac{2\lambda w}{3}
 -\frac{\kappa}{9}
 \Bigg] .
 \label{eq:Kmod-holomorphic-shift}
\end{align}
Thus
\begin{equation}
 f_g(\tau,w)=
 \log J_{h,\nu}(g;\tau,w)\,,
 \label{eq:Kmod-automorphy}
\end{equation}
up to the choice of branch of the logarithm. The term proportional to the
central coordinate $\kappa$ is purely imaginary and therefore drops out of
$f_g+\bar f_g$. Hence it contributes only an irrelevant constant phase to the
holomorphic automorphy factor. The K\"ahler metric derived from
Eq.~\eqref{eq:Kmod-Jacobi} is consequently invariant under the full Jacobi
group.

The corresponding line element can be written as
\begin{equation}
 ds^2
 =\frac{h}{4y^2}|d\tau|^2
 +\frac{2\pi\nu}{y}
 \left|dw-\frac{v}{y}d\tau\right|^2\,.
 \label{eq:Jacobi-Kahler-metric-BB}
\end{equation}
Thus both $\tau$ and the elliptic modulus $w$ have genuine kinetic terms.  
For a Jacobi multiplet of weight $k$ and index $m$ it is useful to introduce the Hermitian factor
\begin{equation}
\displaystyle
 {\cal H}^{J}_{k,m}(\tau,w)
 =y^k\exp\left[-\frac{4\pi m v^2}{y}\right].
 \label{eq:H-Jacobi}
\end{equation}
It transforms as
\begin{equation}
 {\cal H}^{J}_{k,m}(g\cdot(\tau,w))
 =\left|J_{k,m}(g;\tau,w)\right|^{-2}
 {\cal H}^{J}_{k,m}(\tau,w)\,.
 \label{eq:H-Jacobi-transform}
\end{equation}
A minimal invariant matter K\"ahler potential is therefore
\begin{equation}
\displaystyle
 K_{\rm matter}^{J}
 =\sum_I
 \left({\cal H}^{J}_{k_I,m_I}\right)^{-1}
 \Phi_I^\dagger\Phi_I
 =\sum_I
 y^{-k_I}
 e^{4\pi m_Iv^2/y}
 \Phi_I^\dagger\Phi_I.
 \label{eq:Kmatter-Jacobi-BB}
\end{equation}
Equations~\eqref{eq:Kmod-Jacobi} and \eqref{eq:Kmatter-Jacobi-BB} are minimal symmetry-compatible choices, not the most general K\"ahler potential. In particular the flavour universality of $K_{\rm matter}^{J}$ is not maintained if we allow more general terms.

\subsection{Jacobi-covariant kinetic tensors}
\label{sec:Jacobi-covariants}
Non-universal matter metrics can be constructed without introducing new flavour spurions.  If $Y_{k,m}(\tau,w)$ is a vector-valued Jacobi form, the Hermitian matrix
\begin{equation}
\displaystyle
 {\cal Q}_Y(\tau,w)
 ={\cal H}^{J}_{k,m}(\tau,w)
 Y_{k,m}(\tau,w)Y_{k,m}(\tau,w)^\dagger
 \label{eq:QY-BB}
\end{equation}
transforms only by finite-representation conjugation,
\begin{equation}
 {\cal Q}_Y\longmapsto
 \rho(\widetilde g){\cal Q}_Y\rho(\widetilde g)^\dagger\,.
 \label{eq:QY-transform}
\end{equation}
It can therefore enter local Jacobi-invariant K\"ahler metrics whenever the representation structure allows it.  Schematically,
\begin{equation}
 K_I=
 \left({\cal H}^{J}_{k_I,m_I}\right)^{-1}
 \Phi_I^\dagger
 \left[\one+\sum_a c_{I,a}{\cal Q}_{Y_a}+\cdots\right]
 \Phi_I\,.
 \label{eq:nonminimal-K-BB}
\end{equation}
The coefficients $c_{I,a}$ are not fixed by symmetry.  The important point is that the flavour directions are fixed functions of the same moduli that control the holomorphic couplings; no independent flavon alignment is required.
In the phenomenological analysis below we do not interpret
Eq.~\eqref{eq:nonminimal-K-BB} as a perturbative expansion in small
$c_{I,a}{\cal Q}_{Y_a}$.  Rather, the symmetry-selected rank--one directions
are used as a finite ansatz for the matter metric~\footnote{We parametrize the corrections by
normalized projectors.  Large fitted projector coefficients therefore
measure strong non-universality of the kinetic terms, not a controlled
small correction to the universal metric.  At a fixed vacuum repeated
powers of a single projector introduce no new flavour direction,
$P_A^n=P_A$; additional independent covariants would correspond to enlarging
the phenomenological ansatz.}.

\subsection{Superpotential and selection rules}
\label{sec:Susy}
Consider a superpotential term
\begin{equation}
 W\supset
 Y_{I_1\ldots I_n}(\tau,w)
 \Phi_{I_1}\cdots\Phi_{I_n}\,.
\end{equation}
It is useful to characterize the transformation of the superpotential itself
by an automorphy weight $k_W$, an index $m_W$, and a possible unitary
character $\chi_W$,
\begin{equation}
 W\big(g\cdot(\tau,w)\big)
 =
 J_{k_W,m_W}(g;\tau,w)\,
 \chi_W(\widetilde g)\,
 W(\tau,w)\,.
 \label{eq:W-Jacobi-transformation}
\end{equation}
Since the matter fields transform with the inverse Jacobi automorphy factors,
a coupling $Y_{I_1\ldots I_n}$ of weight $k_Y$, index $m_Y$ and finite
representation $\rho_Y$ can contribute only if
\begin{align}
 k_Y-\sum_a k_{I_a}&=k_W\,,
 \nonumber\\
 m_Y-\sum_a m_{I_a}&=m_W\,,
 \label{eq:superpotential-selection-BB}
\end{align}
and the finite representations admit the character carried by the
superpotential,
\begin{equation}
 \rho_Y\otimes\rho_{I_1}\otimes\cdots\otimes\rho_{I_n}
 \supset \chi_W\,.
 \label{eq:superpotential-rep-selection-BB}
\end{equation}
These equations provide a unified form of the Jacobi selection rules.
In rigid supersymmetry the superpotential is invariant, so that
\begin{equation}
 k_W=m_W=0,\qquad \chi_W=\mathbf 1\,.
\end{equation}

In supergravity, instead, the invariant quantity is
${\cal G}=K+\log|W|^2$.  The matter K\"ahler potential, including the
Jacobi-covariant non-universal corrections of
Eq.~\eqref{eq:nonminimal-K-BB}, is strictly invariant under $\Gamma^J$.
Hence the K\"ahler transformation of the full potential is entirely due to
$K_{\rm mod}$,
\begin{equation}
 K\longmapsto K+f_g+\bar f_g\,,
\end{equation}
and invariance of ${\cal G}$ requires
\begin{equation}
 W\longmapsto e^{-f_g}W\,.
\end{equation}
Using Eq.~\eqref{eq:Kmod-automorphy}, one obtains
\begin{equation}
 k_W=-h,\qquad m_W=-\nu\,,
 \label{eq:SUGRA-W-BB}
\end{equation}
up to an allowed unitary character $\chi_W$.  Thus the same selection rules
\eqref{eq:superpotential-selection-BB} apply in supergravity, with the
non-vanishing automorphy quantum numbers of Eq.~\eqref{eq:SUGRA-W-BB}.
In the phenomenological applications below we shall work in rigid
supersymmetry.
\section{The zero section and the eclectic limit}
\label{sec:eclectic-limit}
The relation with the conventional eclectic flavour picture can be understood in a particular limit of our construction.  The elliptic coordinate defines a point $[z]$ on the torus  $E_\tau=\mathbb C/(\mathbb Z\tau+\mathbb Z)$, 
through the equivalence
\begin{equation}
z\sim z+\lambda\tau+\mu\,.
 \label{eq:equivalence}
\end{equation}
Among these points, consider 
\begin{equation}
 [z]=[0]\qquad\hbox{in }E_\tau,
 \label{eq:zero-section}
\end{equation}
or equivalently $z=\lambda\tau+\mu$ for integers $\lambda,\mu$~\footnote{Since the equation $z=\lambda\tau+\mu$ depends on the torus $E_\tau$, the choice $ [z]=[0]$ defines a section.}.
The torus does not degenerate in this limit: its complex structure $\tau$ remains a dynamical modulus.  Only the independent elliptic variable is removed.
At generic $z$ the finite Heisenberg transformations are part of the modular action on the Jacobi fibre.  On the zero section their action on the modulus reduces to lattice equivalences, while the finite matrices acting on matter multiplets and couplings remain non-trivial.  In this sense the Heisenberg subgroup becomes indistinguishable from a traditional internal flavour symmetry.  Moreover $S^2$ acts as $z\mapsto-z$ and fixes the zero section.  Together with $\Delta(27)$ it completes the traditional $\Delta(54)$ factor, whereas the modular $T'\simeq\SL(2,3)$ continues to act on $\tau$.  The same order-$648$ core may therefore be organized as
\begin{equation}
 J_3\simeq\Delta(27)\rtimes T'
 \simeq\Omega(1)=\langle\Delta(54),T'\rangle.
 \label{eq:J3-Omega1-zero}
\end{equation}
Away from the zero section, however, the realization is genuinely Jacobi-modular because the would-be traditional transformations also act on the elliptic modulus.
The connection is equally transparent at the level of the couplings.  The weight-one Jacobi multiplets obey
\begin{equation}
E(\tau,0)=D(\tau),
 \label{eq:E-zero-repeat}
\end{equation}
and
\begin{equation}
F(\tau,0)=F_0(\tau)
 =
 \begin{pmatrix}
 -D_2(\tau)\\[1mm]
 D_1(\tau)/\sqrt2\\[1mm]
 D_1(\tau)/\sqrt2
 \end{pmatrix},
 \qquad
 G(\tau,0)=-F_0(\tau).
 \label{eq:FG-zero}
\end{equation}
Consequently the two independent index-twelve sextets collapse,
\begin{equation}
D\otimes G\longrightarrow-D\otimes F_0,
 \qquad
 E\otimes F\longrightarrow D\otimes F_0.
 \label{eq:sextets-zero}
\end{equation}
Thus the zero section reproduces the same group-theoretical structure as a modular doublet times a triplet flavon, but the triplet alignment is fixed by the modulus $\tau$ rather than by an independent flavon vacuum.  Moving away from $z=0$ restores the independent elliptic dependence and splits the two sextets.  The Jacobi coordinate therefore parametrizes a geometrical deformation of the eclectic texture.
\subsection{Other eclectic groups}
The above interpretation is not peculiar to the $\mathbb Z_3$ case.
The eclectic symmetries of the $T^2/\mathbb Z_K$ building blocks have
been systematically analysed in Refs.~\cite{Baur:2020Z2,Baur:2024qzo}.
It suggests, more generally, that the traditional and modular factors
of the other $T^2/\mathbb Z_K$ eclectic building blocks may also be
viewed as finite remnants of a larger geometric symmetry, although the
appropriate parent geometry is not identical in all cases.
The closest analogue is $T^2/\mathbb Z_4$.  Here no increase in the
number of moduli is required.  
At level two the same integral Heisenberg group introduced above
admits the finite quotient
\begin{equation}
 {\cal P}_1=
 \left\langle X,Y,Z\,\middle|\,
 X^2=Y^2=1,\quad Z^4=1,\quad
 Z\ {\rm central},\quad XY=Z^2YX
 \right\rangle ,
\end{equation}
which is the order--16 Pauli group $[16,13]$.  This is precisely the
non-$R$ traditional flavour group of the $T^2/\mathbb Z_4$ orbifold.
The modular factor reduces at level two to
\begin{equation}
   \SL(2,\mathbb F_2)\simeq S_3 .
\end{equation}
The $\mathbb Z_4$ case can therefore be regarded, at the level of its
non-$R$ core, as the level--two counterpart of the Jacobi construction
discussed here.

The $T^2/\mathbb Z_2$ case requires a slightly larger geometric
structure.  Unlike the $\mathbb Z_3$ case, the orbifold action fixes
neither the K\"ahler modulus $T$ nor the complex-structure modulus $U$.
The duality group therefore contains two modular factors,
\begin{equation}
   \SL(2,\mathbb Z)_T\times\SL(2,\mathbb Z)_U ,
\end{equation}
supplemented by the mirror transformation which exchanges $T$ and
$U$.  These transformations constitute the familiar two-modulus
Narain duality structure.  On the other hand, the traditional flavour
symmetry at generic values of $T$ and $U$ is~\cite{Baur:2020Z2}:
\begin{equation}
   \frac{D_8\times D_8}{\mathbb Z_2}\simeq[32,49]\,.
\end{equation}
The two ingredients suggest a direct generalization of the Jacobi
picture discussed above.
Indeed, the single Heisenberg pair $(X,Y)$ can be replaced by two
pairs,
\begin{equation}
   (X_T,Y_T),\qquad (X_U,Y_U),
\end{equation}
with a common central element.  They generate the rank--two integral
Heisenberg group $H_5(\mathbb Z)$, which is a central extension of
$\mathbb Z^4$.  Correspondingly, one introduces two elliptic
coordinates, $z_T$ and $z_U$: the first pair translates $z_T$ on the
elliptic curve with modulus $T$, while the second pair translates
$z_U$ on the elliptic curve with modulus $U$.  The two modular groups
act on the respective pairs $(T,z_T)$ and $(U,z_U)$, and the mirror
symmetry exchanges them.
This suggests the rank--two analogue of the Jacobi construction,
\begin{equation}
 T^2/\mathbb Z_2:\qquad
 H_5(\mathbb Z)\rtimes
 \left[
   \bigl(\SL(2,\mathbb Z)_T\times\SL(2,\mathbb Z)_U\bigr)
   \rtimes\mathbb Z_2^{\rm mir}
 \right]
 \quad\hbox{acting on}\quad
 (T,U;z_T,z_U) .
\end{equation}
On the zero section $z_T=z_U=0$ one recovers the usual two-modulus
description in terms of $T$ and $U$.  \footnote{At level two the rank--two Heisenberg parent contains a slightly larger
group than the traditional $T^2/\mathbb Z_2$ flavour symmetry.  Keeping the
primitive central phase gives a group of order $64$, while the subgroup
generated by the four lattice translations has order $32$ and is precisely
\begin{equation}
   H_5(\mathbb F_2)
   \simeq
   \frac{D_8\times D_8}{\mathbb Z_2}
   \simeq[32,49].
\end{equation}
Thus the familiar $T^2/\mathbb Z_2$ flavour group arises naturally as
a subgroup of the level--two Heisenberg quotient.}
At the same level, the modular part of the Narain duality group reduces
to
\begin{equation}
 \bigl[
   \SL(2,\mathbb F_2)_T\times\SL(2,\mathbb F_2)_U
 \bigr]\rtimes\mathbb Z_2^{\rm mir}
 \simeq
 \bigl(S_3^T\times S_3^U\bigr)\rtimes\mathbb Z_2^{\rm mir},
\end{equation}
providing the corresponding finite modular factor.
\begin{table}[h!]
\centering
\small
\renewcommand{\arraystretch}{1.4}
\begin{tabular}{c|p{0.43\textwidth}|p{0.43\textwidth}}
\hline
$K$
&
Integral parent and coordinates
&
Finite remnants
\\
\hline\hline

$3$
&
$\Heis(\mathbb Z)\rtimes\SL(2,\mathbb Z)$
acting on $(\tau,z)$
&
Traditional:
$H_3(\mathbb F_3)\simeq\Delta(27)$

Modular:
$\SL(2,\mathbb F_3)\simeq T'$
\\
\hline

$4$
&
$\Heis(\mathbb Z)\rtimes\SL(2,\mathbb Z)$
acting on $(\tau,z)$
&
Traditional:
${\cal P}_1\simeq[16,13]$

Modular:
$\SL(2,\mathbb F_2)\simeq S_3$
\\
\hline

$2$
&
$H_5(\mathbb Z)\rtimes
\bigl[
(\SL(2,\mathbb Z)_T\times\SL(2,\mathbb Z)_U)
\rtimes\mathbb Z_2^{\rm mir}
\bigr]$

acting on $(T,U;z_T,z_U)$
&
Traditional:
$H_5(\mathbb F_2)\simeq[32,49]$

Modular:
$(S_3^T\times S_3^U)
\rtimes\mathbb Z_2^{\rm mir}$
\\
\hline

\end{tabular}
\caption{Schematic geometric interpretation of the $K=2,3,4$
eclectic building blocks.  For $K=3,4$ the parent is a rank--one
Heisenberg--modular geometry, while $K=2$ requires its rank--two
generalization.  In all three cases the traditional flavour symmetry
arises as a finite remnant of the corresponding integral Heisenberg
sector.}
\label{tab:eclectic-geometric-parents}
\end{table}

The $T^2/\mathbb Z_6$ case is instructive in a different way.  Its
finite modular symmetry is
\begin{equation}
   \Gamma'_6
   \simeq
   \SL(2,\mathbb Z/6\mathbb Z)
   \simeq S_3\times T' .
\end{equation}
The non-$R$ traditional symmetry is instead reduced to the
$\mathbb Z_6^{(\mathrm{PG})}$ point-group selection rule.
A cyclic point-group factor is also present in the $K=2,3,4$
orbifolds.  In those cases, however, it is embedded in a larger
non-Abelian traditional flavour group, generated in addition by
transformations associated with the different orbifold fixed points.
The modular transformations act non-trivially on this larger group
and thereby lead to the characteristic eclectic, or
Heisenberg--modular, structure discussed above.
The $K=6$ case lacks this additional non-Abelian traditional sector:
only the point-group factor $\mathbb Z_6^{(\mathrm{PG})}$ remains.
Moreover, this factor commutes with the finite modular group
$\Gamma'_6$.  Hence the modular transformations do not induce a
non-trivial automorphism of the traditional symmetry, and there is no
analogue of the non-trivial semidirect-product structure encountered
for $K=2,3,4$.  We therefore do not include the $K=6$ case in the
Heisenberg--modular interpretation summarized in
Table~\ref{tab:eclectic-geometric-parents}.

\subsection{The Klingen cusp}
The relation between Jacobi and Siegel modular structures has several
precedents in string theory.  In heterotic compactifications with
Wilson-line moduli, perturbative couplings and threshold corrections
can be expressed in terms of Siegel and Jacobi modular forms
\cite{Mayr:1995Stieberger,Cardoso:1996JacobiWilson,
Nazaroglu:2013Jacobi}.  In more recent heterotic constructions a
Wilson-line modulus enlarges the familiar modular dualities to an
$\Sp(4,\Z)$ symmetry, which has also been interpreted as a modular
flavour symmetry
\cite{Baur:2020Siegel,Nilles:2021Sp4Orbifolds}.
Related Fourier--Jacobi structures arise in the non-separating
degeneration of genus-two string amplitudes
\cite{DHoker:2017HigherGenus}.
It is therefore natural to compare our intrinsic Jacobi theory with
the rank-one boundary geometry of the genus-two Siegel upper
half-plane.

The limit $[z]=[0]$ has also an interesting geometrical interpretation, when the Jacobi group is embedded into $\Sp(4,\Z)$.
Introduce  the matrix
\begin{equation}
 \Omega=
 \begin{pmatrix}
 \tau&z\\ z&\sigma
 \end{pmatrix}
 \in\Htwo\,,
 \label{eq:Omega-BB}
\end{equation}
which is the period matrix of a genus-two Riemann surface. 
Let $\Sp(4,\Z)$ act by
\begin{equation}
 \Omega\longmapsto
 (\mathcal A\Omega+\mathcal B)
 (\mathcal C\Omega+\mathcal D)^{-1}.
 \label{eq:Siegel-action-BB}
\end{equation}
The modular factor of the Jacobi group is embedded through
\begin{align}
 \tau&\mapsto\frac{a\tau+b}{c\tau+d}\,,
 &z&\mapsto\frac{z}{c\tau+d}\,,
 &\sigma&\mapsto\sigma-\frac{cz^2}{c\tau+d}\,,\nonumber\\
 \tau&\mapsto\tau\,,
 &z&\mapsto z+\lambda\tau+\mu\,,
 &\sigma&\mapsto\sigma+2\lambda z+\lambda^2\tau+\kappa\,.
 \label{eq:action-Siegel-BB}
\end{align}
The transformations of $(\tau,z)$ are exactly those postulated in the bottom-up Jacobi theory.  The extra coordinate $\sigma$ supplies the compensating shifts required for a symplectic realization.
\begin{figure}[h!]
\centering
\includegraphics[width=1.0\textwidth]{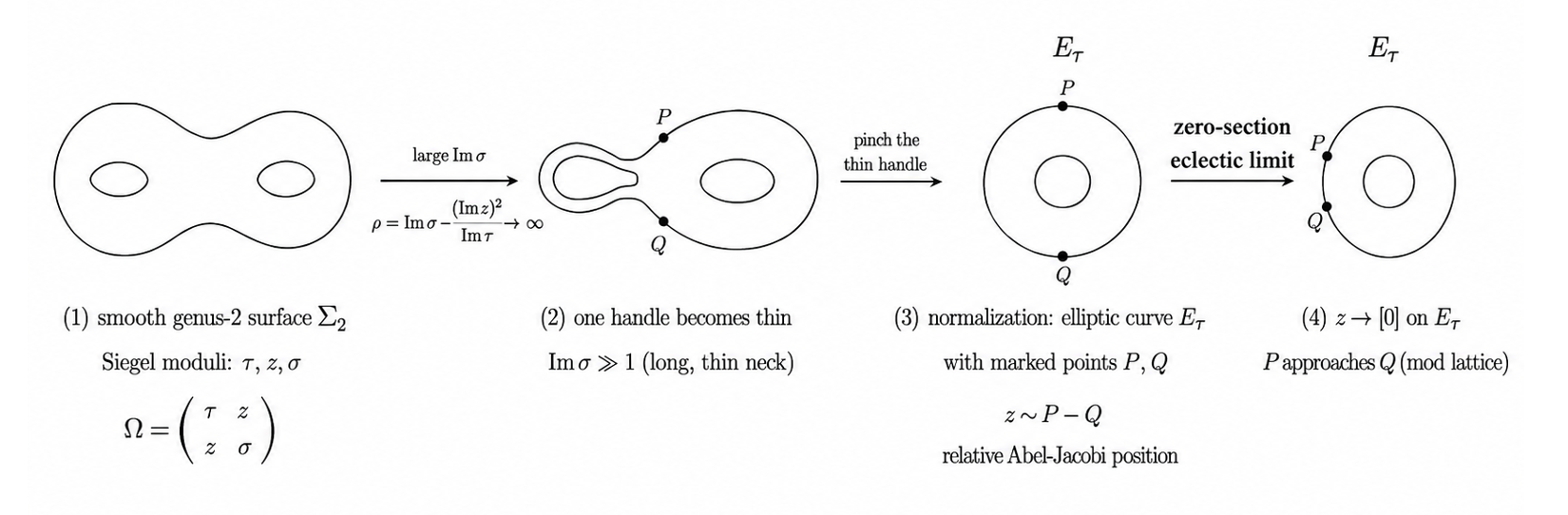}
\caption[Geometric interpretation of the Klingen-cusp and zero-section limits.]{
Geometric interpretation of the Klingen-cusp and zero-section limits.
Starting from a smooth genus-two Riemann surface with period matrix
$\Omega=\bigl(\begin{smallmatrix}\tau&z\\ z&\sigma\end{smallmatrix}\bigr)$,
the limit
$\rho=\operatorname{Im}\sigma
-(\operatorname{Im}z)^2/\operatorname{Im}\tau\to\infty$
makes one of the two handles increasingly long and thin, while the
surface still has genus two. Pinching this thin handle and normalizing
the resulting nodal curve leaves an elliptic curve $E_\tau$ with two
marked points $P$ and $Q$, corresponding to the two attachment regions
of the shrinking handle. Their relative Abel--Jacobi position is
encoded by $z\simeq P-Q\in E_\tau$. The zero-section limit
$z\to[0]$ then corresponds to $P\to Q$ modulo the elliptic lattice,
while the underlying elliptic curve $E_\tau$ remains smooth and its
complex structure $\tau$ is kept fixed.
}
\label{fig:Klingen-cusp}
\end{figure}

The Jacobi theory arises in a neighbourhood of the Klingen cusp, a rank-one boundary component of the genus-two Siegel moduli space. 
Geometrically, approaching it corresponds to degenerating only one of the two cycles of the genus-two surface, while a genus-one component with complex structure $\tau$ remains finite.
In suitable coordinates this limit is described by ${\rm Im}\,\sigma\to\infty$, with $\tau$ and $z$ kept finite. The subgroup of $Sp(4,\mathbb Z)$ relevant to this boundary component is the Jacobi group introduced above, acting on the surviving variables $(\tau,z)$. Correspondingly, Siegel modular forms admit a Fourier--Jacobi expansion near the cusp, whose coefficients are Jacobi forms.
In particular, for large $\Img\sigma$, a Siegel modular form
${\cal Y}(\Omega)$ can be expanded as
\begin{equation}
 {\cal Y}(\Omega)
 =
 \sum_r
 \exp\left(\frac{2\pi i r\sigma}{N}\right)
 Y_r(\tau,z),
 \label{eq:FJ-large-rho}
\end{equation}
where $N$ depends on the level.  The leading coefficients $Y_r(\tau,z)$ then provide a natural
effective description, higher Fourier sectors being parametrically
suppressed.
The notion of large $\Img\sigma$ can be formulated more intrinsically
in terms of the quantity
\begin{equation}
 \rho=
 \frac{\det(\operatorname{Im}\Omega)}
      {\operatorname{Im}\tau}
 =
 \Img\sigma-\frac{(\Img z)^2}{\Img\tau}\,,
 \label{eq:Klingen-height}
\end{equation}
which is invariant under the full Jacobi group.
Our construction can be interpreted as an effective description of the
Jacobi-invariant asymptotic region $\rho\gg1$, at finite values of $\Img z$ and $\Img\tau$.
In the large $\Img\sigma$ limit, one handle of the genus-two surface becomes long and thin.  After
pinching this handle, the remaining geometry can be described by an
elliptic curve $E_\tau$ with two marked points $P$ and $Q$.  The
Jacobi coordinate measures their relative position,
\begin{equation}
 z\simeq P-Q
 \qquad {\rm in}\quad E_\tau .
 \label{eq:z-PQ}
\end{equation}
This geometrical picture is summarized in
Fig.~\ref{fig:Klingen-cusp}.
Finally, the coordinate $z$ is removed by performing the limit $z\to 0$, which leaves an unmarked genus-one Riemann surface. 

\section{Finite Jacobi symmetry and low-weight forms at level three}
In this section we collect the group-theoretical and automorphic
ingredients needed for the flavour construction.  We first describe the
finite level-three Jacobi symmetry and the representations relevant for
matter multiplets and couplings.  We then classify the low-weight Jacobi
forms transforming in these representations, with particular emphasis on
the weights and indices entering the phenomenological models discussed
below.  This provides a direct bridge between the finite symmetry
structure and the holomorphic textures used in the quark and lepton
sectors.
\subsection{The finite Jacobi group}
For the finite Jacobi group
\begin{equation}
 \Jthree=\Dtw\rtimes\SL(2,3)\,,
 \label{eq:J3-finite-repeat}
\end{equation}
one has $|\Jthree|=648$. Its irreducible representations are
 distributed as
\begin{center}
\begin{tabular}{c@{\qquad}c@{\qquad}c}
\toprule
Dimension & Multiplicity & Contribution to $\sum_r d_r^2$\\
\midrule
$1$ & $3$ & $3$\\
$2$ & $3$ & $12$\\
$3$ & $7$ & $63$\\
$6$ & $6$ & $216$\\
$8$ & $3$ & $192$\\
$9$ & $2$ & $162$\\
\midrule
& & $648$\\
\bottomrule
\end{tabular}
\end{center}
We use throughout the standard modular generators
\begin{equation}
 S=\begin{pmatrix}0&-1\\ 1&0\end{pmatrix}\,,
 \qquad
 T=\begin{pmatrix}1&1\\ 0&1\end{pmatrix}\,,
 \label{eq:J3-standard-ST}
\end{equation}
whose images in $\SL(2,3)$ obey
\begin{equation*}
 S^4=T^3=\one\,,
 \qquad
 (ST)^3=S^2\,,
 \qquad
 [S^2,T]=0\,.
\end{equation*}
All representation labels introduced below refer to this convention.
The quotient $\Jthree/\Dtw\simeq \SL(2,3)$
provides seven representations which are trivial on $\Dtw$: three
singlets, three doublets and one triplet. We denote the one-dimensional
characters by
\begin{equation}
 \mathbf1^{(p)}\,,
 \qquad p=0,1,2\,,
 \label{eq:J3-singlets}
\end{equation}
with $\mathbf1^{(0)}=\mathbf1$. The three doublets are denoted by
\begin{equation}
 \mathbf2^{(p)}
 =
 \mathbf1^{(p)}\otimes\mathbf2^{(0)}\,,
 \qquad p=0,1,2\,.
 \label{eq:J3-doublets}
\end{equation}
All superscripts are understood modulo three.
The unique triplet inflated from $\SL(2,3)$ is denoted by
$\mathbf3_{\rm mod}$. It is trivial on the full Heisenberg subgroup.
The two three-dimensional irreducible representations of $\Dtw$ with
non-trivial central character give instead the Heisenberg-charged
sector. Each admits three extensions to $\Jthree$,
\begin{equation}
 \mathbf3^{(p)}
 =
 \mathbf1^{(p)}\otimes\mathbf3^{(0)}\,,
 \qquad
 \bar{\mathbf3}^{(p)}
 =
 \mathbf1^{(p)}\otimes\bar{\mathbf3}^{(0)}\,,
 \qquad p=0,1,2\,.
 \label{eq:J3-charged-triplets}
\end{equation}
The seven three-dimensional irreducible representations are therefore
\begin{equation}
 \mathbf3_{\rm mod}\,,\qquad
 \mathbf3^{(0)},\mathbf3^{(1)},\mathbf3^{(2)}\,,\qquad
 \bar{\mathbf3}^{(0)},\bar{\mathbf3}^{(1)},
 \bar{\mathbf3}^{(2)}\,.
 \label{eq:seven-triplets}
\end{equation}
The remaining Heisenberg-charged representations consist of three
sextets and one nine-dimensional representation in each of the two
non-trivial central sectors. Three octets complete the centre-neutral
sector.
For phenomenological family assignments the representations
$\mathbf3^{(p)}$ and $\bar{\mathbf3}^{(p)}$ are particularly useful,
since, contrary to $\mathbf3_{\rm mod}$, they carry a non-trivial
Heisenberg central charge. An explicit basis of the relevant irreducible representations and the Clebsch--Gordan decompositions of their tensor products is presented in Appendix \ref{sec:J3-lowdim-reps}. 
\subsection{Low-weight Jacobi forms at level three}
\label{sec:low-weight-Jacobi}
We now collect the low-weight holomorphic Jacobi forms needed in the flavour construction.  We restrict throughout to forms with trivial scalar multiplier on the level-three Jacobi kernel and denote the corresponding space by
\begin{equation}
 {\cal J}^{(3)}_{k,m}\,,
 \label{eq:Jkm-definition}
\end{equation}
where $k$ is the modular weight and $m$ the classical index in the coordinate $w=z/3$.  The dimensions quoted below follow from the theta decomposition and the associated finite Weil representation; general results and dimension formulae can be found in Refs.~\cite{Skoruppa:2008Critical,Skoruppa:2007DegreeOne,Skoruppa:2006Dimensions}.
The level-three central consistency condition derived in Eqs.~\eqref{eq:m-central-character-BB}--\eqref{eq:m-multiple-three-BB} gives
\begin{equation}
 m\in3\Z_{\geq0},
 \qquad
 \rho(Z)=\omega^{m/3}\one\,.
 \label{eq:index-central-summary-BB}
\end{equation}
This relation is intrinsic to the Jacobi theory.  
\begin{table}[h!]
\centering
\begin{tabular}{c@{\qquad}c@{\qquad}c@{\qquad}c}
\toprule
$m$ & $\rho(Z)$ & $\dim{\cal J}^{(3)}_{1,m}$ & $\dim{\cal J}^{(3)}_{2,m}$\\
\midrule
$0$  & $1$        & $2$ & $3$\\
$3$  & $\omega$   & $3$ & $6$\\
$6$  & $\omega^2$ & $0$ & $6$\\
$9$  & $1$        & $2$ & $4$\\
$12$ & $\omega$   & $3$ & $12$\\
\bottomrule
\end{tabular}
\caption{Dimensions of the first low-weight spaces of holomorphic level-three Jacobi forms with trivial scalar multiplier on the level-three kernel.}
\label{tab:low-weight-dimensions}
\end{table}

For later use we summarize their $J_3$ content in Table~\ref{tab:low-weight-combined}.  The explicit constructions and the basis conventions are given in Appendix~\ref{sec:Jacobi-forms}.
\begin{table}[h!]
\centering
\begin{tabular}{c@{\quad}c@{\quad}c@{\qquad}c@{\qquad}c}
\toprule
$k$ & $m$ & Dimension & $J_3$ content & Form / construction\\
\midrule
$1$ & $0$  & $2$  & $\mathbf2^{(0)}$ & $D=Y^{(1,0)}$\\
$1$ & $3$  & $3$  & $\mathbf3^{(0)}$ & $F=Y^{(1,3)}$\\
$1$ & $6$  & $0$  & --- & ---\\
$1$ & $9$  & $2$  & $\mathbf2^{(0)}$ & $E=Y^{(1,9)}$\\
$1$ & $12$ & $3$  & $\mathbf3^{(0)}$ & $G=Y^{(1,12)}$\\
\midrule
$2$ & $0$  & $3$  & $\mathbf3_{\rm mod}$ & $\mathrm{Sym}^2D$\\
$2$ & $3$  & $6$  & $\mathbf6^{(0)}$ & $D\otimes F$\\
$2$ & $6$  & $6$  & $\bar{\mathbf6}^{(0)}$ & $\mathrm{Sym}^2F$\\
$2$ & $9$  & $4$  & $\mathbf1^{(1)}\oplus\mathbf3_{\rm mod}$ & $D\otimes E$\\
$2$ & $12$ & $12$ & $\mathbf6^{(0)}\oplus\mathbf6^{(0)}$ & $(D\otimes G)\oplus(E\otimes F)$\\
\bottomrule
\end{tabular}
\caption{Low-weight catalogue of holomorphic level-three Jacobi forms with trivial scalar multiplier on the level-three kernel.}
\label{tab:low-weight-combined}
\end{table}

Two properties will be especially important in model building.  First, the charged triplets $F$ and $G$ provide moduli-dependent flavour directions without introducing independent triplet fields.  Second, at weight two and index twelve there are two independent sextets,
\begin{equation}
 {\cal J}^{(3)}_{2,12}
 \simeq
 \mathbf6_A^{(0)}\oplus\mathbf6_B^{(0)},
 \qquad
 \mathbf6_A^{(0)}\sim D\otimes G,
 \qquad
 \mathbf6_B^{(0)}\sim E\otimes F.
 \label{eq:two-sextets-BB}
\end{equation}
This multiplicity is absent in the ordinary modular limit and provides the basic holomorphic freedom exploited in the flavour applications.
\section{CP transformations}
\label{sec:generalized-CP}
CP transformations in the presence of discrete flavour symmetries have been extensively studied, with the consistency condition relating CP to automorphisms of the flavour group playing a central role~\cite{Feruglio:2012cw,Holthausen:2012dk,Ding:2013hpa,Fallbacher:2015pga}.  This framework was subsequently extended to modular flavour symmetries, where CP acts also on the modulus and can be combined consistently with modular transformations~\cite{Baur:2019kwi,Novichkov:2019sqv,Yao:2020qyy,Ding:2021iqp}. In a suitable CP basis the symmetry can enforce real coupling constants, so that CP violation originates entirely from the complex vacuum expectation values of the moduli~\cite{Novichkov:2019sqv,Yao:2020qyy}.
In our case, the discrete infinite symmetry acting on the moduli is the integral
Jacobi group $\GJ=\Heis(\Z)\rtimes\SL(2,\Z)$ of
Eq.~\eqref{eq:Jacobi-infinite}.  A convenient representative of 
CP is the anti-holomorphic involution
\begin{equation}
CP:\qquad
 (\tau,z)\longmapsto(-\bar\tau,-\bar z)\,.
 \label{eq:CP-moduli}
\end{equation}
It induces the automorphism $u(g)=CP\,g\,{CP}^{-1}$.  With the Heisenberg conventions
of Eq.~\eqref{eq:integral-Heis}, we find
\begin{equation}
u(S)=S^{-1}\,,\qquad
u(T)=T^{-1}\,,\qquad
 u(X)=X\,,\qquad
 u(Y)=Y^{-1}\,,\qquad
 u(Z)=Z^{-1}\,.
 \label{eq:CP-Heisenberg-automorphism}
\end{equation}
Since $Z$ is central, while $u(Z)=Z^{-1}\neq Z$, $u$ is an outer
automorphism.  It preserves the level-three kernel and therefore descends
to an automorphism of the finite group $J_3$.  Other generalized CP
transformations are obtained by composing Eq.~\eqref{eq:CP-moduli} with
elements of $\GJ$.
For a multiplet in an irreducible representation $r$, generalized CP is
consistent with $J_3$ provided
\begin{equation}
 {\cal X}_r\,\rho_r(g)^*\,{\cal X}_r^{-1}
 =\rho_r\!\left(u(g)\right)
 \qquad (g\in J_3)\,.
 \label{eq:CP-consistency-J3}
\end{equation}
In the representation bases adopted in
Appendix~\ref{sec:J3-lowdim-reps}, this condition is satisfied by the
particularly simple choice
\begin{equation}
 {\cal X}_r=\one_{d_r}
 \label{eq:CP-Xr-identity}
\end{equation}
for
\begin{equation}
 r=\mathbf1^{(p)},\ \mathbf2^{(p)},\ \mathbf3_{\rm mod},\
 \mathbf3^{(p)},\ \bar{\mathbf3}^{(p)},\
 \mathbf6^{(p)},\ \bar{\mathbf6}^{(p)},\
 \mathbf8^{(p)},\ \mathbf9,\ \bar{\mathbf9}\,,
 \qquad p=0,1,2\,.
 \label{eq:CP-all-irreps}
\end{equation}
Thus, suppressing the usual space-time parity action, CP acts simply as
complex conjugation on all the flavour multiplets used below.

The explicit bases of Appendix~\ref{sec:Jacobi-forms} are already adapted
to the same CP transformation.  With $w=z/3$, all Jacobi multiplets used
in the phenomenological construction satisfy
\begin{equation}
 {Y}^{(k,m)}_r(-\bar\tau,-\bar w)
 =
 {Y}^{(k,m)}_r(\tau,w)^* \,.
 \label{eq:CP-Jacobi-forms}
\end{equation}
This holds in particular for $D,E,F,G$, for their weight-two products,
including the two independent sextets at $(k,m)=(2,12)$, and for the
primitive $\eps^2$ sextet of
Appendix~\ref{app:primitive-eps2-sextet}.  For the theta-function
constructions this property follows from complex conjugation of the
Fourier--theta series; in the charged triplets $F$ and $G$, the phases
entering the projectors ${\cal P}_a$ are compensated by
${\cal U}\to{\cal U}^{-1}$.  No additional phase or mixing of equivalent
multiplets is required in the bases used here.

Consider an expansion of the superpotential in independent Jacobi- and
$J_3$-invariant contractions,
\begin{equation}
 W=\sum_\alpha c_\alpha\,{\cal I}_\alpha(\tau,w,\Phi)\,.
 \label{eq:CP-superpotential}
\end{equation}
Because Eqs.~\eqref{eq:CP-Xr-identity} and
\eqref{eq:CP-Jacobi-forms} imply
${\cal I}_\alpha\stackrel{\rm CP}{\longmapsto}{\cal I}_\alpha^*$,
CP invariance is equivalent, in this basis, to
\begin{equation}
 c_\alpha=c_\alpha^*\,,
 \qquad\text{i.e.}\qquad
 c_\alpha\in\R\,.
 \label{eq:CP-real-coefficients}
\end{equation}
In particular, when several independent Jacobi multiplets occur in the
same representation, as for the two sextets at $(k,m)=(2,12)$, their
coefficients are independently real.  CP violation can therefore arise
spontaneously from vacuum values of $(\tau,z)$ which are not invariant
under Eq.~\eqref{eq:CP-moduli}.
\section{Charged-fermion phenomenology}
\label{sec:charged-fermion-phenomenology}
\label{sec:quark-phenomenology}

We now illustrate the phenomenological potential of the Jacobi construction
in the full charged-fermion sector, postponing neutrino masses and lepton
mixing to the next section.  We adopt a universal matter assignment and a
common Yukawa structure spanned by two independent textures for up quarks,
down quarks and charged leptons.  The main purpose of this section is to
show that the observed charged-fermion spectrum and CKM mixing can be
reproduced with a minimal set of Jacobi-covariant kinetic deformations, and
to relate the successful solutions to the geometry associated with an
order-two point on the torus.

\subsection{Matter assignment and Yukawa sector}

We assign all charged matter multiplets to the same charged triplet of
$J_3$, while the two Higgs multiplets are trivial under the Jacobi symmetry,
\begin{equation}
 q,\;l,\;u^c,\;d^c,\;e^c
 \sim
 (k_I,m_I,\rho_I)
 =
 (1,6,\mathbf3^{(0)})\,,
 \qquad
 H_u,\;H_d\sim(0,0,\mathbf1)\,.
 \label{eq:quark-assignment}
\end{equation}
The rigid-supersymmetry selection rules therefore require the
charged-fermion Yukawa couplings to have $(k_Y,m_Y)=(2,12)$.  As discussed
in Sec.~\ref{sec:weight-two-Jacobi}, this space contains two independent
sextets,
\begin{equation}
 {\cal J}^{(3)}_{2,12}
 \simeq
 \mathbf6_A^{(0)}\oplus\mathbf6_B^{(0)},
 \qquad
 \mathbf6_A^{(0)}\sim D\otimes G,\qquad
 \mathbf6_B^{(0)}\sim E\otimes F.
 \label{eq:quark-two-sextets}
\end{equation}
For a doublet $X=(X_1,X_2)^T$ and a triplet
$Y=(Y_1,Y_2,Y_3)^T$, define
\begin{equation}
 {\cal M}[X,Y]=
 \begin{pmatrix}
 X_2Y_1 & -X_1Y_3/\sqrt2 & -X_1Y_2/\sqrt2\\[1mm]
 -X_1Y_3/\sqrt2 & X_2Y_2 & -X_1Y_1/\sqrt2\\[1mm]
 -X_1Y_2/\sqrt2 & -X_1Y_1/\sqrt2 & X_2Y_3
 \end{pmatrix}.
 \label{eq:MXY-quark}
\end{equation}
The two independent Yukawa textures are
\begin{equation}
 {\cal M}_{DG}\equiv{\cal M}[D,G],
 \qquad
 {\cal M}_{EF}\equiv{\cal M}[E,F],
\end{equation}
and all three charged sectors are described by the same two-texture
Yukawa structure,
\begin{equation}
 Y_f=y_f\left({\cal M}_{DG}+r_f{\cal M}_{EF}\right),
 \qquad f=u,d,e\,.
 \label{eq:quark-Yukawas}
\end{equation}
The overall coefficients $y_u,y_d,y_e$ fix the three absolute mass scales
and are not included in the fit.  The holomorphic structure of the
charged-fermion sector is therefore determined by the two common moduli
$(\tau,w)$ and by the three ratios $r_u,r_d,r_e$, which are initially
allowed to be complex.
For later use, the same Yukawa matrices can equivalently be obtained by
assigning
\begin{equation}
 q,\;l\sim\mathbf3^{(2)},
 \qquad
 u^c,\;d^c,\;e^c\sim\mathbf3^{(1)},
 \label{eq:charged-alternative-reps}
\end{equation}
with unchanged weights and indices.  The relevant
$\bar{\mathbf6}^{(0)}$ contraction has the same Clebsch--Gordan
coefficients in the basis used here.  This equivalent realization will be
useful when the neutrino sector is introduced.

\subsection{Input observables and statistical treatment}

We fit ten dimensionless observables: the four nearest-neighbour quark mass
ratios, the three CKM mixing angles and the CKM phase, together with the two
charged-lepton mass ratios.  All charged-fermion quantities are evaluated
at
\[
 M_{\rm GUT}=2\times10^{16}\ {\rm GeV},
\]
using the GUT-scale benchmark of Ref.~\cite{Antusch:2025fpm} with
$M_{\rm SUSY}=10\,{\rm TeV}$, vanishing SUSY threshold corrections and
$\tan\bar\beta=10$.  The numerical inputs are collected in
Table~\ref{tab:charged-inputs}.
All residuals are treated as linear Gaussians.  When an observable is
quoted with asymmetric uncertainties $\sigma_+$ and $\sigma_-$, we retain
a two-piece Gaussian if the two sides differ by more than $10\%$ of the
larger one. Otherwise we use a symmetric Gaussian with the larger
uncertainty.  Thus the four quark mass ratios and $\theta_{13}^q$ are
treated with two-piece Gaussians, whereas $\theta_{12}^q$,
$\theta_{23}^q$, $\delta_{\rm CKM}$ and the two charged-lepton ratios use
symmetric errors.  For an asymmetric observable the pull is therefore
\[
 p_i=
 \begin{cases}
 ({\cal O}_i^{\rm th}-{\cal O}_i^{\rm exp})/\sigma_{i,+},
 &{\cal O}_i^{\rm th}>{\cal O}_i^{\rm exp},\\[1mm]
 ({\cal O}_i^{\rm th}-{\cal O}_i^{\rm exp})/\sigma_{i,-},
 &{\cal O}_i^{\rm th}\leq{\cal O}_i^{\rm exp},
 \end{cases}
\]
with the difference in $\delta_{\rm CKM}$ understood modulo $2\pi$, and
$\chi^2_{\rm ch}=\sum_i p_i^2$.

\begin{table}[t]
\centering
\small
\renewcommand{\arraystretch}{1.15}
\begin{tabular}{lcc}
\hline\hline
Observable & central value & $1\sigma$ uncertainty\\
\hline
$m_u/m_c$
 & $2.014\times10^{-3}$ & ${}^{+4.36}_{-2.69}\times10^{-4}$\\
$m_c/m_t$
 & $2.836\times10^{-3}$ & ${}^{+1.11}_{-0.919}\times10^{-4}$\\
$m_d/m_s$
 & $5.00\times10^{-2}$ & ${}^{+5.48}_{-4.38}\times10^{-3}$\\
$m_s/m_b$
 & $1.831\times10^{-2}$ & ${}^{+1.45}_{-0.935}\times10^{-3}$\\
$\theta_{12}^q$
 & $0.2270$ & $8.30\times10^{-4}$\\
$\theta_{13}^q$
 & $3.44\times10^{-3}$ & ${}^{+0.080}_{-0.070}\times10^{-3}$\\
$\theta_{23}^q$
 & $3.90\times10^{-2}$ & $3.90\times10^{-4}$\\
$\delta_{\rm CKM}$
 & $1.139$ & $0.023$\\
\hline
$m_e/m_\mu$
 & $4.7468\times10^{-3}$ & $3.65\times10^{-6}$\\
$m_\mu/m_\tau$
 & $5.8609\times10^{-2}$ & $4.54\times10^{-5}$\\
\hline\hline
\end{tabular}
\caption{Charged-fermion observables used in the numerical analysis.
All quantities refer to the GUT-scale benchmark at
$M_{\rm GUT}=2\times10^{16}\,{\rm GeV}$, and the CKM angles and phase are
given in radians.  Asymmetric uncertainties are treated according to the
prescription described in the text.}
\label{tab:charged-inputs}
\end{table}

\subsection{Reference model}
\label{subsec:charged-reference}

The universal matter metric is proportional to the identity in family
space and its overall normalization can be absorbed into the coefficients
$y_f$.  A fit with all matter metrics proportional to the identity,
however, does not reproduce the charged-fermion data at their present
experimental accuracy.  The holomorphic Yukawa structure must therefore
be supplemented by non-universal kinetic terms.  The departure of the
corresponding metrics from universality provides a useful measure of how
strongly the successful fit relies on non-holomorphic effects.

Non-universal kinetic terms may be generated by Jacobi-covariant
Hermitian tensors.  The only direction needed in the reference model is
built from the triplet
$G(\tau,w)\in{\cal J}^{(3)}_{1,12}\simeq\mathbf3^{(0)}$.  A convenient
polynomial normalization of the rank-one tensors generated by the
weight-one triplets $A=F,G$ is
\begin{equation}
{\cal Q}_A=
 {\cal H}^{J}_{1,m_A}\,AA^\dagger,
 \qquad
 (m_F,m_G)=(3,12),
 \label{eq:QA-polynomial}
\end{equation}
where ${\cal H}^{J}_{k,m}$ is defined in
Eq.~\eqref{eq:H-Jacobi}.  In particular,
\begin{equation}
{\cal H}^{J}_{1,12}
 =
 (\Img\tau)
 \exp\left[-\frac{48\pi(\Img w)^2}{\Img\tau}\right].
 \label{eq:QG-quark}
\end{equation}
For the numerical scans and for quantifying the size of the kinetic
corrections it is more convenient to use instead the normalized projector
\begin{equation}
P_A=\frac{AA^\dagger}{A^\dagger A},\qquad A=F,G,
 \label{eq:projector-convention}
\end{equation}
so that, at any fixed point in moduli space,
\begin{equation}
\one+c_A{\cal Q}_A=\one+\kappa_A P_A,
 \qquad
 \kappa_A=c_A\,{\cal H}^{J}_{k_A,m_A}\,A^\dagger A .
 \label{eq:kappa-c-relation}
\end{equation}
The two parametrizations are therefore locally equivalent, although the
map between $c_A$ and $\kappa_A$ is moduli dependent.  In what follows we
use the projector convention throughout.  We take
\begin{equation}
K_q=\one,\qquad
 K_{u^c}=\one+\kappa_G^{u^c}P_G,\qquad
 K_{d^c}=\one+\kappa_G^{d^c}P_G,\qquad
 K_l=K_{e^c}=\one .
 \label{eq:reference-Kahler}
\end{equation}
No independent flavour spurion is introduced: the kinetic tensors are
fixed functions of the same moduli that determine the holomorphic Yukawa
couplings.
The physical Yukawa matrices are obtained by canonical normalization,
\begin{equation}
 \widehat Y_u
 =\left(K_{u^c}^{-1/2}\right)^*Y_uK_q^{-1/2},
 \qquad
 \widehat Y_d
 =\left(K_{d^c}^{-1/2}\right)^*Y_dK_q^{-1/2},
 \qquad
 \widehat Y_e=Y_e .
 \label{eq:canonical-Yukawa-reference}
\end{equation}

The reference model contains twelve real shape parameters for ten fitted
observables and is therefore underconstrained by two parameters.  A
representative best-fit point is
\begin{align}
 \tau&=-0.918226+1.122170\,\ii,
 &
 w&=-0.802283+0.373290\,\ii,
 \nonumber\\
 r_u&=-1.960961-0.148746\,\ii,
 &
 r_d&=-3.065699-0.714184\,\ii,
 \nonumber\\
 r_e&=-2.083849-0.253508\,\ii,
 &
\kappa_G^{u^c}&=-0.998588,
 \qquad
\kappa_G^{d^c}=-0.761812.
 \label{eq:reference-bestfit}
\end{align}
At this point
\begin{equation}
 \chi^2_{\rm ch}=0.1688 .
 \label{eq:reference-chi2}
\end{equation}
The largest pull is only about $-0.40\sigma$, in $m_d/m_s$; all other
observables are reproduced within about $0.08\sigma$.  Thus the common
holomorphic two-texture Yukawa structure is fully compatible with the
charged-fermion data once the two $G$-induced right-handed quark kinetic
directions are included.  The very small $\chi^2$ should not be interpreted
as a statistical prediction, since the fit has ${\rm dof}=-2$.

\subsection{Geometry of the reference minimum}
\label{subsec:charged-anatomy}
Special loci in moduli space where part of the modular symmetry survives are known to play a distinguished role in flavour model building.  In the one-modulus case, fixed points and their neighbourhoods have been extensively exploited to generate fermion-mass hierarchies
~\cite{Feruglio:2021Hierarchies,Novichkov:2021Residual,Novichkov:2022Stabilisation,Ishiguro:2022Residual,Chen:2025Hierarchies},
to constrain lepton mixing
~\cite{Feruglio:2023Universal,Feruglio:2023Critical,Kashav:2025Residual}
and as a guide in model building
~\cite{Ding:2019FixedPoints,King:2019vhv,Petcov:2023A4,Abe:2023S4prime,Kikuchi:2023A4cubed}.
The same general idea extends to theories with several moduli and symplectic modular symmetry, where fixed points and higher-dimensional invariant regions can control the scaling of masses and mixing observables~\cite{Ding:2020Automorphic,Ding:2021iqp,Ding:2024SiegelFixed,Carducci:2026Siegel}.

The mechanism encountered here is, however, qualitatively different.  The best-fit points lie close to a locus which is not an isolated fixed point of moduli space, but rather a complex one-dimensional section mapped into itself by a non-trivial subgroup of the Jacobi group.  The two independent Yukawa textures develop a common zero mode, which is protected by the setwise stabilizer of this section and persists all along the locus.  To the best of our knowledge, this provides a new realization of symmetry protection in modular flavour theories.  We shall refer to this mechanism as \emph{modular protection}.

Indeed, the successful fits display a recurrent geometric feature: their moduli lie
close to the two-torsion locus.  In the coordinate
$w=z/3$ the two-torsion sections are
\begin{equation}
 w_0=\frac{a\tau+b}{6},
 \qquad a,b\in\mathbb Z,
 \label{eq:general-two-torsion}
\end{equation}
modulo the Jacobi lattice.  Their relevance can be seen directly from the
closed expression
\begin{equation}
 \det{\cal M}_{DG}(\tau,w)
 =
 2\,\eta(\tau)^{10}\,
 \vartheta(\tau,6w)^2 ,
 \label{eq:det-MDG-torsion}
\end{equation}
whose zero set is precisely Eq.~\eqref{eq:general-two-torsion}.
The same rank reduction holds for the second independent texture and, more
strongly, the two textures have the same null direction on the torsion
section.  Denoting this direction by $e_+$, one has
\begin{equation}
 \left.
 \left({\cal M}_{DG}+r\,{\cal M}_{EF}\right)e_+
 \right|_{w=w_0}=0
 \qquad\text{for any }r\in\mathbb C\,,
 \label{eq:common-torsion-zero-mode}
\end{equation}
so that the zero mode is common to all three charged-fermion Yukawa
matrices and is independent of $r_f$. A symmetry-based derivation of this common zero mode is given in Appendix~\ref{app:torsion-zero-mode}.

Since $\vartheta$ is a section rather than an ordinary function,
$|\vartheta|$ by itself is not a meaningful measure of proximity to its
zero set.  We therefore use the Petersson-normalized quantity
\begin{equation}
 \Vert\vartheta\Vert
 \equiv
 (\Img\tau)^{1/4}\,|\vartheta(\tau,6w)|
 \exp\left[
 -\frac{\pi\,(\Img 6w)^2}{\Img\tau}
 \right],
 \label{eq:Petersson-theta-norm}
\end{equation}
which vanishes on the two-torsion locus and provides a Jacobi-invariant
diagnostic of proximity~\footnote{For orientation we also quote the coordinate
distance normalized to the linear size of a Jacobi cell,
\[
 d_{\rm cell}\equiv
 \frac{|w-w_0|}{\sqrt{\Img\tau}/6}.
\]
The reference minimum has coordinate distance $d_{\rm cell}\simeq2.2\%$.}
For the reference minimum the nearest representative is
\begin{equation}
 w_0=\frac{2\tau-3}{6},
 \qquad
 |w-w_0|\simeq3.9\times10^{-3},
 \qquad
 \Vert\vartheta\Vert\simeq 0.062 .
 \label{eq:reference-torsion-distance}
\end{equation}
The proximity is not imposed in the fit.

The point $w_0=(2\tau-3)/6$ corresponds to
$z_0=\tau-3/2$, and hence, modulo the Jacobi lattice, to the
point $z=-1/2$.
The region $z=-1/2$ is a non-trivial two-torsion locus,
and it is left invariant, as a whole, by an entire subgroup of
$SL(2,\mathbb Z)$ which can be identified by a direct calculation.
Under a Jacobi transformation
\begin{equation}
 (\tau,-\frac{1}{2})\longmapsto
 \left(
 \frac{a\tau+b}{c\tau+d},
 \frac{-\frac{1}{2}+\lambda\tau+\mu}{c\tau+d}
 \right)\,,
\end{equation}
and the condition for maintaining the equality $z=-1/2$ reads
\begin{equation}
 \lambda=-\frac{c}{2}\,,
 \qquad
 \mu=\frac{1-d}{2}\,.
 \label{eq:stabilizer-half}
\end{equation}
Since $\lambda,\mu\in\mathbb Z$, this requires $c$ even and $d$ odd.
Moreover, since $ad-bc=1$, an even $c$ automatically forces both
$a$ and $d$ to be odd.  Hence the whole condition reduces to
\begin{equation}
 c\equiv0\pmod2\,,
\end{equation}
which is precisely the defining condition of the group $\Gamma_0(2)$.

\subsection{Residual parity and modular protection}

The stabilizer of the torsion section should be distinguished
from the symmetry of a generic point on that section.  For generic $\tau$
no non-trivial element of $\Gamma_0(2)$ fixes $\tau$ itself.  Nevertheless,
the full Jacobi group contains the order-two transformation
\begin{equation}
 R=Y^{-1}S^2,
 \qquad
 R^2=\one\,.
 \label{eq:R-half}
\end{equation}
Since $S^2$ leaves $\tau$ unchanged and sends
$z\mapsto-z$, the combination $Y^{-1}S^2$ fixes $z=-1/2$ exactly.  It is
the residual $\mathbb Z_2$ parity of a generic point of this section.
In the $\mathbf3^{(0)}$ representation one finds
\begin{equation}
 \rho_{\mathbf3^{(0)}}(R)
 =
 \begin{pmatrix}
 -1&0&0\\
 0&0&-\omega^2\\
 0&-\omega&0
 \end{pmatrix}.
 \label{eq:R-half-triplet}
\end{equation}
A convenient orthonormal basis with definite parity is
\begin{equation}
 e_+=\frac1{\sqrt2}(0,1,-\omega)^T,
 \qquad
 e_-^{(1)}=(1,0,0)^T,
 \qquad
 e_-^{(2)}=\frac1{\sqrt2}(0,1,\omega)^T,
 \label{eq:half-parity-basis}
\end{equation}
with
\begin{equation}
 \rho_{\mathbf3^{(0)}}(R)e_+=e_+,
 \qquad
 \rho_{\mathbf3^{(0)}}(R)e_-^{(a)}=-e_-^{(a)},
 \quad a=1,2.
 \label{eq:half-parity-eigenvalues}
\end{equation}
Thus the triplet decomposes as one parity-even direction plus a
two-dimensional parity-odd subspace.  The residual $\mathbb Z_2$ by itself
forbids mixing between these two sectors, but it does not require the
even--even entry of a Yukawa matrix to vanish.  The additional rank
reduction of Eq.~\eqref{eq:common-torsion-zero-mode} is a property of the
Jacobi forms on the torsion section.  The theta-function identities there
imply directly the common-zero-mode relations in
Eq.~\eqref{eq:common-torsion-zero-mode}.

In the basis \eqref{eq:half-parity-basis}, both independent charged-fermion
textures therefore take the universal form
\begin{equation}
 {\cal M}'_{DG}=
 \begin{pmatrix}
 0&0&0\\
 0&A_D&B_D\\
 0&B_D&C_D
 \end{pmatrix},
 \qquad
 {\cal M}'_{EF}=
 \begin{pmatrix}
 0&0&0\\
 0&A_E&B_E\\
 0&B_E&C_E
 \end{pmatrix}.
 \label{eq:half-block-textures}
\end{equation}
Consequently
\begin{equation}
 Y'_f=
 y_f\left({\cal M}'_{DG}+r_f{\cal M}'_{EF}\right),
 \qquad f=u,d,e,
\end{equation}
has rank at most two on the exact two-torsion locus, with the same null
vector $e_+$ in all three sectors.  Thus
\[
 m_u=m_d=m_e=0
\]
before the transverse breaking of the residual symmetry is switched on.
In the quark sector the common $1\oplus2$ decomposition also leaves only
one non-trivial mixing inside the two-dimensional block: two CKM angles
vanish and there is no physical CP violation,
\[
 J_{\rm CKM}=0\,.
\]
We refer to the additional rank reduction imposed by the full torsion
structure, beyond what follows from an ordinary residual $\mathbb Z_2$,
as \textit{modular protection}.

Away from the exact section one may introduce a local transverse
coordinate
\[
 \delta w=w-w_0(\tau).
\]
The residual parity reverses $\delta w$.  Consequently,
the $(1,2)$ and $(1,3)$ entries connecting the protected one-dimensional
direction to the $2\times2$ block start linearly in $\delta w$, whereas the
lifting of the $(1,1)$ entry starts quadratically.
The same geometric displacement therefore controls the first-generation
masses and the CKM structures that vanish on the exact torsion locus.

\subsection{Size of the kinetic deformations}

In the projector convention the coefficient $\kappa$ directly measures
the non-trivial eigenvalue of a single rank-one deformation.  For
\begin{equation}
K=\one+\kappa P,
\end{equation}
the eigenvalues are $\{1,1,1+\kappa\}$.  A convenient
normalization-independent diagnostic of the deformation is
\begin{equation}
 {\rm cond}\,K_f
 =
 \frac{\lambda_{\rm max}(K_f)}
      {\lambda_{\rm min}(K_f)} .
 \label{eq:quark-condition-number}
\end{equation}
Positive kinetic energy requires the metric to be positive definite and
therefore
\begin{equation}
 1+\kappa>0\,.
 \label{eq:Kahler-positivity-projector}
\end{equation}
This positivity condition is imposed throughout the numerical analysis.
All the minima quoted below lie inside the allowed domain, although the
best reference fit approaches its boundary rather closely in the $u^c$
direction.  At the reference minimum the projector coefficients are given in
Eq.~\eqref{eq:reference-bestfit}.  Therefore
\begin{equation}
 {\rm cond}\,K_{u^c}\simeq7.1\times10^2,
 \qquad
 {\rm cond}\,K_{d^c}\simeq4.2 .
 \label{eq:reference-condition-numbers}
\end{equation}
The corresponding hierarchies in the canonical-normalization matrices are
$\sqrt{{\rm cond}\,K_{u^c}}\simeq26.6$ and
$\sqrt{{\rm cond}\,K_{d^c}}\simeq2.0$.  The reference fit therefore uses a
near-singular direction in the right-handed up metric.  This should be
viewed as a genuine feature of the phenomenological realization rather
than as a perturbatively small correction to a universal K\"ahler metric.

The comparison with a fit containing a single kinetic
correction is particularly instructive.  If only the $G$ direction in
$K_{d^c}$ is switched on, one finds
\begin{equation}
 \chi^2_{\rm ch}\simeq10.7,
 \qquad
 {\rm cond}\,K_{d^c}\simeq62 .
 \label{eq:one-correction-fit}
\end{equation}
The remaining tension is concentrated mainly in the light-quark mass
ratios and in $m_c/m_t$.  Thus a qualitatively successful description does
not require a condition number of order $10^3$: 
the second kinetic correction is what turns this economical
one-correction solution into the high-accuracy charged-fermion reference
fit discussed above.
The one-correction fit also exhibits a distinct large-$\Img\tau$ branch.
Its minimum lies close to $w=-\tau/6$, with $\Img\tau\simeq3.66$.  In this
regime
\begin{equation}
 t=e^{\pi i\tau/3},
 \qquad |t|\ll1,
\end{equation}
and the two non-zero singular values of the parity-odd block satisfy
\begin{equation}
 \frac{s_2^{(f)}}{s_3^{(f)}}
 \simeq
 2\,\big|(r_f-1)(r_f+3)\big|\,|t|^2,
 \qquad
 \theta_{\rm CKM}^{(2)}
 =O\!\left(|r_d-r_u|\,|t|\right).
 \label{eq:large-Imtau-hierarchy}
\end{equation}
The large-$\Img\tau$ limit therefore supplies an additional source of
hierarchy on top of modular protection.  This mechanism is not selected by
the minimum of the reference model, for which $\Img\tau\simeq1.12$.  There
the stronger quantitative reshaping instead comes from canonical
normalization.

\subsection{Imposing CP}
\label{subsec:charged-CP}

We finally ask whether the charged-fermion sector requires intrinsically
complex Wilson coefficients.  In the basis used throughout this work the
CP transformation of Sec.~\ref{sec:generalized-CP} is a CP
basis: imposing CP at the Lagrangian level requires
\begin{equation}
 r_u,\;r_d,\;r_e\in\mathbb R .
 \label{eq:real-r-charged}
\end{equation}
A generic complex expectation value of $(\tau,w)$ may nevertheless break
CP spontaneously.
Keeping exactly the two kinetic corrections of the reference model is too
restrictive once the $r_f$ are required to be real.  A good fit is recovered
by switching on the same $G$-induced direction also in the left-handed
quark metric,
\begin{equation}
K_q=\one+\kappa_G^q P_G,\qquad
 K_{u^c}=\one+\kappa_G^{u^c}P_G,\qquad
 K_{d^c}=\one+\kappa_G^{d^c}P_G,
 \qquad
 K_l=K_{e^c}=\one .
 \label{eq:gcp-charged-Kahler}
\end{equation}
The resulting fit has ten real parameters for ten observables.  A
representative minimum is
\begin{align}
 \tau&=0.393846+1.207386\,\ii,
 &
 w&=-0.105015+0.207805\,\ii,
 \nonumber\\
 r_u&=-1.320669,
 &
 r_d&=0.464127,
 &
 r_e&=1.134404 .
 \label{eq:gcp-charged-bestfit}
\end{align}
with
\begin{equation}
 \chi^2_{\rm ch}=6.14,
 \qquad {\rm dof}=0 .
 \label{eq:gcp-charged-chi2}
\end{equation}
The largest pulls are $1.69\sigma$ in $m_u/m_c$ and $1.68\sigma$ in
$m_d/m_s$; all remaining observables are reproduced within about
$0.62\sigma$, and the CKM phase itself is fitted to better than
$0.1\sigma$.  Since all Wilson ratios are real, the physical CKM phase is
generated entirely by spontaneous CP breaking through the complex moduli.
The kinetic coefficients in the convention used throughout are
\[
\kappa_G^q=3.60\times10^5,\qquad
 \kappa_G^{u^c}=-0.997385,\qquad
 \kappa_G^{d^c}=5.10 ,
\]
corresponding to
\begin{equation}
 {\rm cond}\,K_q\simeq3.6\times10^5,\qquad
 {\rm cond}\,K_{u^c}\simeq3.8\times10^2,\qquad
 {\rm cond}\,K_{d^c}\simeq6.1 .
 \label{eq:gcp-charged-condition-numbers}
\end{equation}
Thus a CP-invariant charged-fermion superpotential is compatible with the
observed masses, mixings and CKM phase, with CP violation generated
spontaneously by the moduli.  The price is a substantially more
hierarchical quark K\"ahler sector, most notably in the left-handed
doublets.
\section{Neutrino phenomenology}
\label{sec:lepton-phenomenology}

We now extend the construction to the neutrino sector.  A literal use of
the same untwisted holomorphic structures for the charged-lepton, Dirac
and Majorana matrices leads to an excessively aligned seesaw and does not
reproduce the observed large leptonic mixing.  We therefore keep the same
matter weights and Jacobi indices as in the charged-fermion sector, but
enlarge the holomorphic coupling space by allowing a non-trivial scalar
multiplier.  This provides a genuinely new Majorana structure without
introducing higher weights or indices.  The resulting type-I seesaw shares
the same two moduli $(\tau,w)$ with the quark sector.

\subsection{Matter assignment, scalar multiplier and neutrino sector}
\label{subsec:neutrino-input}
It is useful first to make explicit how the scalar
multiplier enters the construction.  Keeping the matter weight and index
fixed at $(k,m)=(1,6)$, one can enlarge the coupling space without
introducing higher weights or indices by supplementing the finite $J_3$
representation with a one-dimensional multiplier system.  We write
\begin{align}
 \Phi_I
 &\longmapsto
 \varepsilon(\gamma)^{p_I}\,
 J_{k_I,m_I}(g;\tau,w)^{-1}\,
 \rho_I(\widetilde g)\,
 \Phi_I,
 \nonumber\\
 Y(\tau,w)
 &\longmapsto
 Y(g\!\cdot\!(\tau,w))
 =
 \varepsilon(\gamma)^p\,
 J_{k_Y,m_Y}(g;\tau,w)\,
 \rho_Y(\widetilde g)\,
 Y(\tau,w),
 \label{eq:multiplier}
\end{align}
where $\widetilde g$ denotes the finite level-three image
and $J_{k,m}$ is the Jacobi automorphy factor defined above.  Here
$\varepsilon$ is the Dedekind eta multiplier,
\begin{equation}
 \eta(\gamma\tau)
 =
 \varepsilon(\gamma)(c\tau+d)^{1/2}\eta(\tau),
 \qquad
 \varepsilon(T)=e^{\pi i/12},
 \qquad
 \varepsilon(S)=e^{-\pi i/4},
 \label{eq:epsilon-definition-pheno}
\end{equation}
trivial on the Heisenberg generators,
$\varepsilon(X)=\varepsilon(Y)=\varepsilon(Z)=1$
\footnote{Strictly speaking, odd powers $\varepsilon^{\pm1}$ are multiplier
systems on the metaplectic cover $Mp(2,\mathbb Z)$ rather than ordinary
characters of $\SL(2,\mathbb Z)$.  Accordingly, whenever such odd powers
are assigned to matter fields, Eq.~\eqref{eq:multiplier} is understood on
the metaplectic Jacobi group $\Heis(\mathbb Z)\rtimes Mp(2,\mathbb Z)$.
The even multiplier $\varepsilon^2$ entering the Majorana coupling descends
to the corresponding character of the ordinary modular group.}.
Thus the multiplier
changes only the scalar modular transformation law, leaving the finite
$J_3$ representation $\rho_I$ unchanged.  We denote by
${\cal J}^{(3)}_{k,m}[\varepsilon^p]$ the corresponding space of
level-three vector-valued Jacobi forms; the trivial-multiplier spaces used
in the charged-fermion sector are ${\cal J}^{(3)}_{k,m}[1]$.
For a rigid-supersymmetric coupling
$Y\,\Phi_{I_1}\cdots\Phi_{I_n}$, invariance requires, in addition to the
usual weight, index and $J_3$ representation selection rules, a trivial
total multiplier.  In terms of the chosen exponent representatives, this means
$ p_Y+\sum_{a=1}^{n}p_{I_a}=0$, modulo the order of the multiplier.
Non-trivial matter multipliers can therefore select new
holomorphic coupling spaces while the matter weights and indices remain
unchanged.

We use the alternative triplet realization already mentioned in
Sec.~\ref{sec:charged-fermion-phenomenology}.  A particularly symmetric
choice is to assign the same finite $J_3$ representation and scalar
multiplier to the two left-handed doublets and, separately, to all
right-handed singlets,
\begin{equation}
q,l\sim(1,6,\mathbf3^{(2)})_{\varepsilon^{+1}},
\qquad
u^c,d^c,e^c,\nu^c
\sim(1,6,\mathbf3^{(1)})_{\varepsilon^{-1}} .
\label{eq:unified-multiplier-assignment}
\end{equation}
The Higgs multiplets are taken to have trivial scalar multiplier.  With
this assignment all Dirac-type bilinears carry trivial total multiplier,
so that the holomorphic coefficients of the up-quark, down-quark,
charged-lepton and Dirac-neutrino operators continue to be drawn from the
untwisted space ${\cal J}^{(3)}_{2,12}[1]$.  By contrast, the Majorana
bilinear $\nu^c\nu^c$ carries multiplier $\varepsilon^{-2}$ and therefore
requires a coupling in
\begin{equation}
{\cal J}^{(3)}_{2,12}[\varepsilon^2].
\end{equation}
Thus the non-trivial multiplier singles out the Majorana sector while
leaving the holomorphic Dirac structures unchanged~\footnote{
For the K\"ahler sector the convention in which the quark triplets are
written as $\mathbf{3}^{(0)}$ is equivalent, since the one-dimensional
characters and scalar multipliers cancel in Hermitian bilinears.
Hence the covariant tensors $FF^\dagger$ and $GG^\dagger$, and therefore
the quark kinetic corrections, are unchanged.
}.
In the numerical realizations studied below we keep the lepton metrics
canonical,
\begin{equation}
K_l=K_{e^c}=K_{\nu^c}=\one ,
\label{eq:canonical-lepton-metrics}
\end{equation}
and allow non-universal kinetic terms only in the quark sector.
The theta decomposition and the associated finite Weil representation
give $\dim {\cal J}^{(3)}_{2,12}[\varepsilon^2]=33$ and
\begin{equation}
 {\cal J}^{(3)}_{2,12}[\varepsilon^2]
 =
 \mathbf3^{(0)}
 \oplus\mathbf6^{(1)}
 \oplus\mathbf6^{(2)}
 \oplus2\,\mathbf9 .
 \label{eq:eps2-decomposition}
\end{equation}
Here and below ``primitive'' means a form which is not generated by
products of lower-weight forms in the relevant multiplier sector.  The
primitive $\mathbf6^{(1)}$ multiplet is unique up to an overall
normalization. Appendix~\ref{app:primitive-eps2-sextet} gives an explicit
theta realization.  This is the important new ingredient in the Majorana
sector.  We denote its components by
${\boldsymbol Y}^{[2]}_{\mathbf6^{(1)}}(\tau,w)$ and associate with it the
symmetric matrix
\begin{equation}
 {\cal M}^{[2]}_{\mathbf6^{(1)}}(\tau,w)=
 \begin{pmatrix}
 Y^{[2]}_{\mathbf6^{(1)},4}
 & -Y^{[2]}_{\mathbf6^{(1)},3}/\sqrt2
 & -Y^{[2]}_{\mathbf6^{(1)},2}/\sqrt2\\
 -Y^{[2]}_{\mathbf6^{(1)},3}/\sqrt2
 & Y^{[2]}_{\mathbf6^{(1)},5}
 & -Y^{[2]}_{\mathbf6^{(1)},1}/\sqrt2\\
 -Y^{[2]}_{\mathbf6^{(1)},2}/\sqrt2
 & -Y^{[2]}_{\mathbf6^{(1)},1}/\sqrt2
 & Y^{[2]}_{\mathbf6^{(1)},6}
 \end{pmatrix}.
 \label{eq:M-six1-eps2}
\end{equation}
Suppressing irrelevant overall factors, the Yukawa and neutrino mass matrices are
\begin{align}
 Y_f&={\cal M}_{DG}+r_f{\cal M}_{EF},\quad(f=u,d,e)
 &
 M_D&={\cal M}_{DG}+r_D{\cal M}_{EF},
 \nonumber\\
 M_R&=\Lambda_R\,{\cal M}^{[2]}_{\mathbf6^{(1)}},
 &
 M_\nu&=-M_D^T M_R^{-1}M_D .
 \label{eq:neutrino-matrices}
\end{align}
The same moduli $(\tau,w)$ enter all quark and lepton matrices.
The oscillation inputs used in the numerical analysis are collected in
Table~\ref{tab:neutrino-inputs}.  The two charged-lepton mass ratios and
the eight quark observables are those of Table~\ref{tab:charged-inputs}.
The leptonic Dirac phase is displayed there only as a phenomenological
reference; the model values are reported below together with the
absolute-mass predictions.

\begin{table}[t]
\centering
\small
\renewcommand{\arraystretch}{1.15}
\begin{tabular}{lcc}
\hline\hline
Observable & normal ordering & fitted \\
\hline
$\delta m^2\,[\mathrm{eV}^2]$
& $(7.48\pm0.20)\times10^{-5}$ & yes \\
$\Delta m^2\,[\mathrm{eV}^2]$
& $(2.495\pm0.020)\times10^{-3}$ & yes \\
$\sin^2\theta_{12}$
& $0.3085\pm0.0114$ & yes \\
$\sin^2\theta_{13}$
& $0.0223\pm0.00062$ & yes \\
$\sin^2\theta_{23}$
& $0.473\pm0.024$ & yes \\
$\delta^\ell_{\rm CP}/\pi$
& $1.20\pm0.21$ & no \\
\hline\hline
\end{tabular}
\caption{Neutrino oscillation inputs used in the 
normal-ordering fits. The atmospheric-sector inputs follow the Bari global analysis~\cite{Capozzi:2025wyn}, while the solar parameters $\delta m^2$ and $\sin^2\theta_{12}$ are updated with the first JUNO results~\cite{Capozzi:2025ovi}.  The Dirac phase is shown but is not
included in the fit.}
\label{tab:neutrino-inputs}
\end{table}
\subsection{Two global realizations}
\label{subsec:neutrino-global-fits}
We now compare two realizations with identical holomorphic assignments and
canonical lepton metrics.  They differ only in the non-universal quark
kinetic terms.  The specific two-correction pattern used as a reference in the
charged-fermion analysis is useful for illustrating the geometry of the
charged sector, but it is not competitive once the neutrino observables are
included.  In our scan, we covered the $2^6=64$ patterns obtained by switching
the two rank--one directions $P_F$ and $P_G$ independently in the three quark
metrics $K_q$, $K_{u^c}$ and $K_{d^c}$.  Within the subset with exactly two
active corrections, we found a unique pattern with
$\chi^2_{q+\ell}<1$; the next-best two-correction pattern lies around
$\chi^2_{q+\ell}\simeq50$.  We denote the corresponding model by $G_2$.
We also consider a four-correction realization providing a very accurate
global fit, denoted by $G_4$.

We use throughout the projector convention defined in
Eq.~\eqref{eq:projector-convention}.  The two models are
\begin{align}
 G_2:\quad&
 K_q=\one,
 \qquad
 K_{u^c}=\one+\kappa_F^{u^c} P_F+\kappa_G^{u^c} P_G,
 \qquad
 K_{d^c}=\one,
 \nonumber\\
 G_4:\quad&
 K_q=\one+\kappa_G^q P_G,
 \qquad
 K_{u^c}=\one+\kappa_G^{u^c} P_G,
 \qquad
 K_{d^c}=\one+\kappa_F^{d^c} P_F+\kappa_G^{d^c} P_G .
 \label{eq:G2-G4-Kahler}
\end{align}
In both cases Eq.~\eqref{eq:canonical-lepton-metrics} holds.  When two
non-orthogonal projectors act on the same field, positivity is imposed on
the full metric and cannot in general be reduced to the separate
conditions $1+\kappa_A>0$.
We first allow $r_u,r_d,r_e,r_D$ to be complex.  The global fit contains
fifteen observables: eight quark quantities, two charged-lepton mass
ratios, two neutrino mass splittings and three leptonic mixing angles.
Independent re-optimizations with inverted ordering give fits of
comparable quality, so the construction is not tied to a particular mass
ordering.  For simplicity of presentation, and to keep the discussion of the
geometric mechanism focused, we report normal ordering throughout the tables
and figures below; this choice is not meant to express a phenomenological
preference for NO.  The main numerical results and representative best-fit
points are collected in Table~\ref{tab:G2-G4-global-fit}.
It is convenient to factor the light-neutrino mass matrix as
\begin{equation}
 M_\nu=k_\nu\,\widehat M_\nu(\tau,w,r_D),
 \label{eq:knu-definition}
\end{equation}
where $\widehat M_\nu$ is dimensionless in the normalization used in the
numerical scan.  Thus $k_\nu$ is an overall mass scale, quoted below in
eV.  For fixed dimensionless parameters it is profiled analytically
against the two mass-squared splittings and counts as one real fitted
parameter.
\begin{table}[t]
\centering
\small
\renewcommand{\arraystretch}{1.12}
\begin{tabular}{lcc}
\hline\hline
 & $G_2$ & $G_4$ \\
\hline
kinetic corrections
& $\kappa_F^{u^c},\ \kappa_G^{u^c}$
& $\kappa_G^q,\ \kappa_G^{u^c},\ \kappa_F^{d^c},\ \kappa_G^{d^c}$ \\
\hline
$\tau$
& $-0.382389+0.942271\,\ii$
& $-0.382336+0.942272\,\ii$ \\
$w$
& $0.229419-0.156600\,\ii$
& $0.229198-0.156945\,\ii$ \\
$r_e$
& $-2.92810-0.541718\,\ii$
& $-2.70133-0.416525\,\ii$ \\
$r_D$
& $-3.00007-0.002350\,\ii$
& $-2.99921-0.002469\,\ii$ \\
$r_u$
& $-3.24774+0.583740\,\ii$
& $-1.42870+0.201122\,\ii$ \\
$r_d$
& $-3.15901-0.069170\,\ii$
& $-1.99471+1.66883\,\ii$ \\
$k_\nu$ (overall mass scale) $[{\rm eV}]$
& $0.00455056$
& $0.00453899$ \\
\hline
$\kappa_F^{u^c}$
& $-0.894219$
& $0$ \\
$\kappa_G^{u^c}$
& $5.83963\times10^5$
& $-0.997412$ \\
$\kappa_G^q$
& $0$
& $200.363$ \\
$\kappa_F^{d^c}$
& $0$
& $21.7330$ \\
$\kappa_G^{d^c}$
& $0$
& $79.7520$ \\
\hline
$\chi^2_{q+\ell}$
& $0.3238$
& $0.1151$ \\
$\chi^2_q$
& $0.0851$
& $4.35\times10^{-5}$ \\
$\chi^2_\ell$
& $0.2387$
& $0.1150$ \\
$N_{\rm par}$
& $15$
& $17$ \\
${\rm dof}$
& $0$
& $-2$ \\
largest pull
& $\sin^2\theta_{23}^\ell:\ -0.49\,\sigma$
& $\sin^2\theta_{23}^\ell:\ -0.34\,\sigma$ \\
\hline
\multicolumn{3}{c}{Predictions evaluated at the best-fit point}\\
\hline
$\delta^\ell_{CP}/\pi$
& $1.46537$
& $1.47915$ \\
$J_{\rm PMNS}$
& $-0.03342$
& $-0.03356$ \\
$\sum_i m_i\,[{\rm eV}]$
& $0.21420$
& $0.21395$ \\
$m_\beta\,[{\rm eV}]$
& $0.06611$
& $0.06602$ \\
$m_{\beta\beta}\,[{\rm eV}]$
& $0.06540$
& $0.06524$ \\
\hline\hline
\end{tabular}
\caption{
Representative global best-fit points for the two-correction model $G_2$
and the four-correction model $G_4$, with complex Wilson coefficients and
normal neutrino mass ordering.  The final block reports the leptonic
Dirac phase and absolute neutrino-mass observables as predictions
evaluated at the same best-fit points.  
}
\label{tab:G2-G4-global-fit}
\end{table}

The $G_2$ realization already provides a very good global description
with only two non-universal kinetic corrections and zero degrees of
freedom.  Its residuals are small in both sectors:
$\chi^2_q=0.0851$ and $\chi^2_\ell=0.2387$.
The more flexible $G_4$ realization further
reduces the residuals, but its very small $\chi^2$ should not be
interpreted as a statistical prediction, since $G_4$ has two more
parameters than fitted observables.  
The last block of Table~\ref{tab:G2-G4-global-fit} displays quantities
which are not included in the fit: the leptonic Dirac phase and the
absolute neutrino-mass observables.
A striking feature is that the two best-fit vacua are almost identical.
This will be useful in the geometric discussion below.
With canonical lepton metrics, the leptonic residual is essentially
saturated by the atmospheric angle.  The fit predicts
$\sin^2\theta_{23}\simeq0.461$ in $G_2$ and
$\sin^2\theta_{23}\simeq0.465$ in $G_4$, corresponding respectively to
$-0.49\,\sigma$ and $-0.34\,\sigma$ relative to the input central value.
The stability of this value across the fits suggests that the model has a
genuine preference for this region of $\theta_{23}$.
\subsection{Geometry of the global minima}
\label{subsec:neutrino-torsion}
The global minima display the same geometric behavior already encountered
in the charged-fermion sector: they approach a non-trivial two-torsion
section without this being imposed in the fit.  Both $G_2$ and $G_4$ lie
close to
\begin{equation}
 w_0=\frac{1-\tau}{6},
 \qquad
 z_0=\frac{1-\tau}{2}.
 \label{eq:neutrino-nearest-torsion}
\end{equation}
We find $|w-w_0|\simeq1.1\times10^{-3}$ ($\Vert\vartheta\Vert\simeq0.019$) for $G_2$ 
and $|w-w_0|\simeq1.2\times10^{-3}$ ($\Vert\vartheta\Vert\simeq0.022$) for $G_4$.
It is enough to analyze the representative section
$z=(1-\tau)/2$.
Its modular stabilizer is the theta subgroup
\begin{equation}
 \Gamma^\theta=
 \left\{\gamma\in SL(2,\mathbb Z):
 \gamma\equiv I\ \hbox{or}\ S\pmod2\right\},
 \label{eq:theta-stabilizer}
\end{equation}
which is conjugate to the $\Gamma_0(2)$ stabilizer discussed in the
charged-fermion sector.  For generic $\tau$ on the section, the symmetry
of the full Jacobi point reduces to a residual $\mathbb Z_2$ parity.
The corresponding parity-adapted basis may be chosen as
\begin{equation}
 e_-^{(\theta)}=\frac{1}{\sqrt2}(1,0,-\omega)^T,
 \qquad
 e_+^{(\theta)}=\frac{1}{\sqrt2}(1,0,\omega)^T,
 \qquad
 e_2=(0,1,0)^T .
 \label{eq:theta-parity-basis}
\end{equation}
The identities relating the untwisted Jacobi forms on this section imply
that every texture entering $Y_e$ and $M_D$ takes the protected form
\begin{equation}
 {\cal M}'=
 \begin{pmatrix}
 0&0&0\\
 0&2\omega a&\sqrt2\,\omega b\\
 0&\sqrt2\,\omega b&c
 \end{pmatrix},
 \label{eq:neutrino-untwisted-block}
\end{equation}
whereas the primitive $\varepsilon^2\mathbf6^{(1)}$ Majorana matrix has
the complementary structure
\begin{equation}
 M_R'\propto
 \begin{pmatrix}
 0&x&y\\
 x&0&0\\
 y&0&0
 \end{pmatrix}.
 \label{eq:neutrino-MR-torsion}
\end{equation}
Thus, on the exact section,
\begin{equation}
 {\rm rank}\,M_D\leq2,
 \qquad
 {\rm rank}\,M_R=2
\end{equation}
generically, and the conventional seesaw is singular there.  The physical
minima lie only at a distance of order $10^{-3}$ in the $w$ coordinate,
so the light-neutrino spectrum probes a correlated deformation of both
the Dirac and Majorana matrices away from a symmetry-controlled singular
limit. 
A more detailed local picture emerges by expanding around the two-torsion
locus.  Let
\begin{equation}
 \delta w=w-w_0(\tau),
 \qquad
 \Delta r_D=r_D+3,
 \qquad
 \rho=\frac{\Delta r_D}{\delta w}\,.
 \label{eq:neutrino-local-rho}
\end{equation}
At $r_D=-3$, the Dirac matrix develops an
additional zero mode which is aligned with the null direction of the
Majorana matrix.  Consequently, the Dirac zero mode and the vanishing
Majorana eigenvalue compensate each other in the seesaw.
Keeping $\rho$ fixed as $\delta w\to0$, one finds at leading order
\begin{equation}
\widehat M_\nu(\tau,w,r_D)
=
\delta w\,{\cal A}_\nu(\tau,\rho)
+O(\delta w^2)\,.
 \label{eq:neutrino-local-expansion}
\end{equation}
The displacement from the locus therefore controls primarily the overall
neutrino-mass scale, whereas the eigenvectors of $M_\nu$, and hence the
mixing angles, approach finite limiting values.

The atmospheric angle is particularly stable in this regime.  On the
two-torsion locus the residual parity separates the protected direction
from the two-dimensional subspace containing the dominant $\mu$--$\tau$
mixing.  Once $w=w_0(\tau)$ is imposed, the matrix elements within this
subspace are determined mainly by the single modulus $\tau$, and one finds
schematically
\begin{equation}
 \theta_{23}
 =
 \theta_{23}^{(0)}(\tau,\rho)
 +O(\delta w),
 \qquad
 \theta_{23}^{(0)}(\tau,\rho)
 \simeq
 \theta_{23}^{(0)}(\tau).
 \label{eq:theta23-local-stability}
\end{equation}
The near-independence of $\theta_{23}^{(0)}$ from $\rho$ is a property of
the explicit model rather than an exact consequence of the residual
symmetry.  Thus the observed stability of $\theta_{23}$ has a simple
geometric interpretation: the modulus fixes to a large extent the
orientation of the $\mu$--$\tau$ sector, while the finite displacement from
the two-torsion locus produces only a subleading correction.

\subsection{Size of the kinetic deformations}

The economical character of a model should be distinguished from the
size of its kinetic deformation.  In the $G_2$ realization only two
K\"ahler coefficients are present, but both act on the same right-handed
up-quark metric.  As shown in Table~\ref{tab:G2-G4-global-fit}, their
best-fit values are
\begin{equation}
 \kappa_F^{u^c}=-0.894219,
 \qquad
\kappa_G^{u^c}=5.83963\times10^5 .
 \label{eq:G2-kappa}
\end{equation}
Because $P_F$ and $P_G$ are in general non-orthogonal, the condition
number of the full metric cannot be inferred from either coefficient
separately.  Nevertheless, the very large value of $\kappa_G^{u^c}$ makes it
clear that the successful two-correction fit is not a perturbatively small
deformation of a universal K\"ahler metric.
For $G_4$, the single-projector metrics already show sizeable hierarchies.
At the free-coupling minimum,
\begin{equation}
 \kappa_G^q\simeq200.36,
 \qquad
 \kappa_G^{u^c}\simeq-0.997412 .
\end{equation}
For a single projector, $K=\one+\kappa P$, so that
${\rm cond}\,K=\max(1,1+\kappa)/\min(1,1+\kappa)$.  Hence
\begin{equation}
 {\rm cond}\,K_q\simeq2.01\times10^2,
 \qquad
 {\rm cond}\,K_{u^c}\simeq3.86\times10^2 .
 \label{eq:G4-free-condition}
\end{equation}
The $d^c$ metric contains two non-orthogonal directions and its condition
number must instead be evaluated from the full matrix.  The excellent
global agreement obtained with $G_4$ should therefore be viewed as relying
on a genuinely non-universal quark kinetic sector.

\subsection{Imposing CP}
\label{subsec:neutrino-CP}

\begin{figure}[t]
\centering
\includegraphics[width=0.82\textwidth]{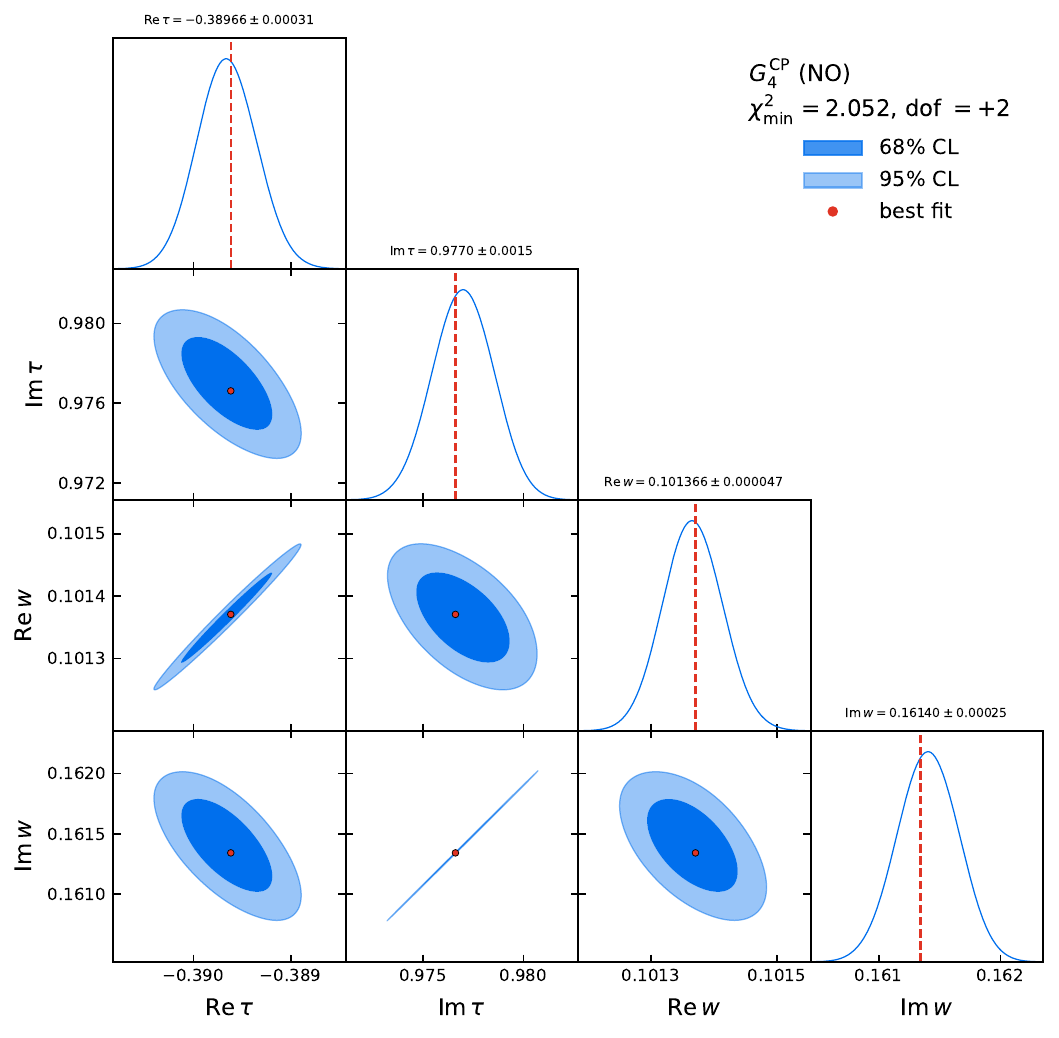}
\caption{Allowed regions for the two moduli in the
$G_4^{\rm CP}$ model (Normal Ordering).  The narrow bands follow the
two-torsion section $w=(1+\tau)/6$; the red point marks the best-fit point.}
\label{fig:gcp-moduli-triangle}
\end{figure}

\begin{figure}[t]
\centering
\includegraphics[width=0.90\textwidth]{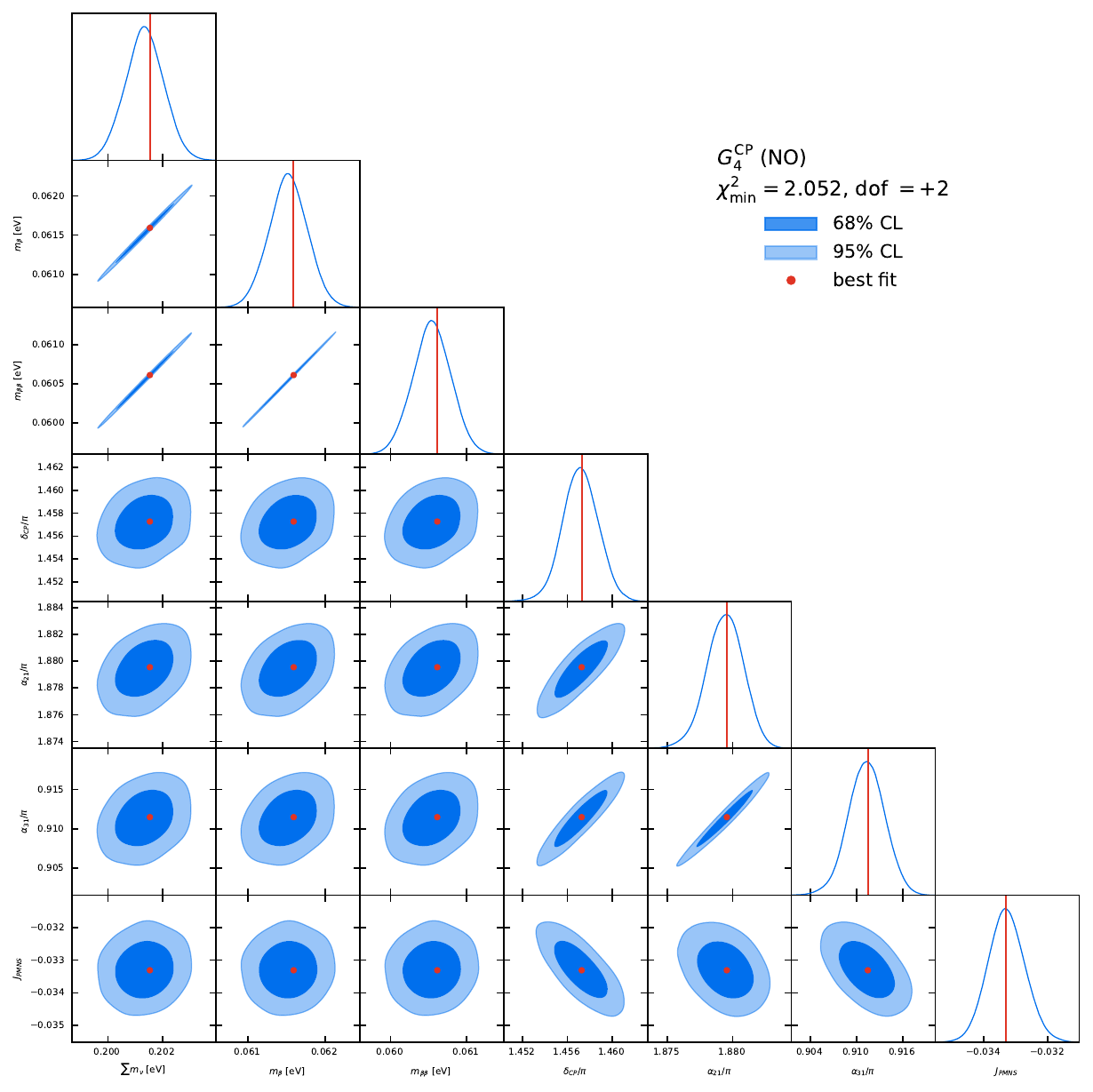}
\caption{Allowed regions for the predicted quantities in the
$G_4^{\rm CP}$ model (Normal Ordering).  The plotted quantities correspond
to the prediction block reported in Table~\ref{tab:neutrino-bestfit-summary}.}
\label{fig:gcp-predictions-triangle}
\end{figure}

We finally impose the CP symmetry discussed in
Sec.~\ref{sec:generalized-CP}.  In the CP basis used here this amounts to
requiring
\begin{equation}
 r_u,\;r_d,\;r_e,\;r_D\in\mathbb R ,
 \label{eq:real-r-neutrino}
\end{equation}
while a generic complex vacuum of $(\tau,w)$ can still break CP
spontaneously.
The two-correction $G_2$ pattern is not the preferred
realization once CP is imposed.  We therefore use $G_4$ as the
representative CP-invariant global model. We denote this CP-restricted realization by $G_4^{\rm CP}$.  With four real Wilson ratios,
four real moduli, four K\"ahler coefficients and one overall neutrino mass
scale, it contains thirteen real parameters for fifteen fitted
observables.
The normal-ordering fit gives
\begin{align}
 \tau&=-0.389617+0.976616\,\ii,
 &
 w&=0.101371+0.161342\,\ii,
 \nonumber\\
 r_e&=-2.52120,
 &
 r_D&=-2.99655,
 \nonumber\\
 r_u&=2.57839,
 &
 r_d&=43.9479,
 \nonumber\\
 \kappa_G^q&=2.32347\times10^5,
 &
 \kappa_G^{u^c}&=-0.999722,
 \nonumber\\
 \kappa_F^{d^c}&=5.13287\times10^4,
 &
 \kappa_G^{d^c}&=1.44257\times10^3,
 \nonumber\\
 k_\nu&=0.00439046\ {\rm eV}.
 \label{eq:G4CP-bestfit}
\end{align}
At this point
\begin{equation}
 \chi^2_{q+\ell}=2.0520,
 \qquad
 N_{\rm par}=13,
 \qquad
 {\rm dof}=+2 .
 \label{eq:G4CP-chi2}
\end{equation}
The largest pull is $-1.32\,\sigma$ in $m_u/m_c$.
In the lepton sector the atmospheric angle is again the irreducible
residual of the canonical-metric setup: the model predicts
$\sin^2\theta_{23}\simeq0.460$, about $0.54\,\sigma$ below the input central
value.
All Wilson ratios are real, so the CKM phase and any leptonic CP violation
arise entirely from spontaneous CP breaking by the complex moduli.
The CP minimum remains close to a non-trivial two-torsion section,
\begin{equation}
 w_0=\frac{1+\tau}{6},
 \qquad
 |w-w_0|\simeq1.5\times10^{-3},
 \qquad
 \Vert\vartheta\Vert\simeq0.026 .
 \label{eq:G4CP-torsion}
\end{equation}
The allowed regions and correlations in
Figs.~\ref{fig:gcp-moduli-triangle} and~\ref{fig:gcp-predictions-triangle}
are obtained with a Markov chain Monte Carlo (MCMC) analysis, an approach
previously applied to a combined quark and lepton fit in a modular flavour
model in Ref.~\cite{Ding:2025QuarkLeptonCorrelations}.
The scan samples show that this is not merely a property of the
best-fit point.  As illustrated in Fig.~\ref{fig:gcp-moduli-triangle}, the
allowed region follows a narrow ridge in the $(\tau,w)$ plane.  Writing
\begin{equation}
 w=\frac{1+\tau}{6}+\eta ,
 \label{eq:G4CP-eta}
\end{equation}
the fit keeps the small displacement $\eta$ approximately fixed, so that
$dw\simeq d\tau/6$.  Quantitatively, using $\tau=x+\ii y$ and $w=u+\ii v$, we define the
linear correlation coefficient along the normal-ordering chain as
\begin{equation}
 {\rm Corr}(a,b)
 =
 \frac{{\rm Cov}(a,b)}{\sigma_a\sigma_b}.
 \label{eq:correlation-coefficient}
\end{equation}
We find
\begin{equation}
 {\rm Corr}(x,u)=0.987,
 \qquad
 {\rm Corr}(y,v)=0.99997,
\end{equation}
with least-squares slopes
${\rm Cov}(x,u)/{\rm Var}(x)=0.153$ and
${\rm Cov}(y,v)/{\rm Var}(y)=0.165$, respectively, close to $1/6$.
CP does not impose this correlation by itself; rather,
by making the Wilson ratios real, it removes alternative phase directions
and exposes the geometric two-torsion valley in the moduli.

Similarly, correlations are induced among the genuine predictions,
shown in Fig.~\ref{fig:gcp-predictions-triangle}.  The absolute-mass
observables $\sum_i m_i$, $m_\beta$ and $m_{\beta\beta}$ are tightly
correlated because they are determined by the same quasi-degenerate
neutrino spectrum and by the same overall mass scale.  There is also a
correlation, not displayed in this prediction triangle, between the two
fitted mass splittings $\delta m^2$ and $\Delta m^2$.  The overall scale
$k_\nu$ is fixed by matching these two splittings simultaneously; therefore
residual motion in the two-dimensional mass-splitting plane can be
reflected in the widths and correlations of the sampled absolute-mass
predictions.  These correlations should therefore be read as properties of
the allowed fit region, not as independent additional constraints.

Thus the geometric interpretation found in the free-coupling fits
survives the imposition of CP.
The price is again a strongly hierarchical quark kinetic sector.  In
particular, the single-projector metrics satisfy approximately
\begin{equation}
 {\rm cond}\,K_q\simeq2.3\times10^5,
 \qquad
 {\rm cond}\,K_{u^c}\simeq3.6\times10^3,
\end{equation}
while the $d^c$ metric contains the two non-orthogonal $F$ and $G$
directions and must be characterized from its full spectrum.

\subsection{Summary}

The neutrino sector illustrates why scalar multipliers are useful in the
Jacobi construction.  Keeping the lowest matter weight and the common
index $(k,m)=(1,6)$, the $\varepsilon^2$ Majorana channel supplies a
primitive holomorphic sextet that is absent from the untwisted
charged-fermion construction.  The charged-lepton and Dirac-neutrino
matrices can nevertheless retain exactly the same two-texture structure
used for the charged fermions.

The global fits show that a quantitative description does not
require a proliferation of kinetic terms.  The two-correction model $G_2$
already gives $\chi^2_{q+\ell}\simeq0.324$ with zero degrees of freedom,
whereas the four-correction model $G_4$ further improves the residuals for
unconstrained complex Wilson coefficients.  Most importantly,
$G_4^{\rm CP}$ remains successful after CP is imposed,
with $\chi^2_{q+\ell}\simeq2.05$ and two positive degrees of freedom.  In
all these successful solutions the moduli lie very close to a non-trivial
two-torsion section, linking the numerical fits to the same modular
protection mechanism identified in the charged-fermion sector.
For completeness, Table~\ref{tab:neutrino-bestfit-summary} collects 
the neutrino observables and absolute-mass predictions at the three
representative best-fit points.

\begin{table}[t]
\centering
\small
\renewcommand{\arraystretch}{1.15}
\setlength{\tabcolsep}{3.5pt}
\begin{tabular}{lcccc}
\hline\hline
Observable & $G_2$ & $G_4$ & $G_4^{\rm CP}$ & Experimental input/reference \\
\hline
$\delta m^2\,[{\rm eV}^2]$
& $7.48\times10^{-5}$
& $7.48\times10^{-5}$
& $7.48\times10^{-5}$
& $(7.48\pm0.20)\times10^{-5}$ \\
$\Delta m^2\,[{\rm eV}^2]$
& $2.495\times10^{-3}$
& $2.495\times10^{-3}$
& $2.495\times10^{-3}$
& $(2.495\pm0.020)\times10^{-3}$ \\
$\sin^2\theta_{12}$
& $0.3085$
& $0.3085$
& $0.3086$
& $0.3085\pm0.0114$ \\
$\sin^2\theta_{13}$
& $0.02230$
& $0.02230$
& $0.02230$
& $0.0223\pm0.00062$ \\
$\sin^2\theta_{23}$
& $0.461$
& $0.465$
& $0.460$
& $0.473\pm0.024$ \\
\hline
$\delta^\ell_{CP}/\pi$
& $1.47$
& $1.48$
& $1.46$
& \begin{tabular}{@{}c@{}}$[1.20\pm0.21]$\\ not used in the fit\end{tabular} \\
$J_{\rm PMNS}$
& $-0.0334$
& $-0.0336$
& $-0.0333$
& -- \\
$\sum_i m_i\,[{\rm eV}]$
& $0.214$
& $0.214$
& $0.202$
& -- \\
$m_\beta\,[{\rm eV}]$
& $0.0661$
& $0.0660$
& $0.0616$
& -- \\
$m_{\beta\beta}\,[{\rm eV}]$
& $0.0654$
& $0.0652$
& $0.0606$
& -- \\
\hline\hline
\end{tabular}
\caption{
Neutrino observables at the representative normal-ordering best-fit
points of $G_2$, $G_4$ and $G_4^{\rm CP}$.  The leptonic Dirac phase and
absolute neutrino-mass observables are reported in the final rows as
predictions evaluated at the same best-fit points.  The experimental
column shows the inputs used in the numerical fit from the global
analyses in Refs.~\cite{Capozzi:2025wyn,Capozzi:2025ovi}.  The bracketed
Dirac phase is shown as an experimental reference and is not used in the
fit.
}
\label{tab:neutrino-bestfit-summary}
\end{table}

The three benchmarks correspond to a quasi-degenerate absolute spectrum.
Their predicted values, $m_\beta\simeq0.06\,{\rm eV}$ and
$m_{\beta\beta}\simeq0.06\,{\rm eV}$, are safely below the present
direct $\beta$-decay and neutrinoless-double-beta bounds summarized in
Ref.~\cite{Capozzi:2025wyn}, but they lie in a phenomenologically
interesting region.  The $m_\beta$ values are below the final reach of
KATRIN alone, while they are of the order targeted by future endpoint
programmes aiming at the few-$10\,{\rm meV}$ scale.  If neutrinos are
Majorana particles, the predicted $m_{\beta\beta}$ is especially
testable: next-generation searches such as LEGEND-1000, nEXO and CUPID
are designed to explore the inverted-ordering and quasi-degenerate
regions, modulo the usual nuclear-matrix-element uncertainties.
Cosmology gives the sharpest external pressure.  The sums
$\sum_i m_i\simeq0.20$--$0.21\,{\rm eV}$ sit close to the representative
current scale $\Sigma\simeq0.2\,{\rm eV}$ discussed in
Ref.~\cite{Capozzi:2025wyn}, whose precise interpretation depends
strongly on the cosmological model and data combination.  A robust future
upper limit significantly below this scale would therefore directly
stress these quasi-degenerate benchmarks, whereas a positive signal in
the $m_\beta$ or $m_{\beta\beta}$ ranges above would be very naturally
compatible with them.

\section{Conclusions}

Eclectic flavour symmetries provide an elegant and general framework that may ultimately contribute to a solution of the flavour puzzle. They encompass modular and traditional flavour symmetries, as well as $R$ and $CP$ symmetries. In the conventional formulation, based on a single modulus $\tau$, traditional symmetries are distinguished by their trivial action on $\tau$, while acting non-trivially on matter and modular multiplets. Their breaking generally requires a suitable set of flavons, whose magnitudes and orientations in flavour space become an essential part of the phenomenological construction and lead to the familiar vacuum-alignment problem.

In this work we have generalized the notion of a traditional flavour symmetry by preserving its defining feature --- a trivial action on $\tau$ --- while allowing it to act through a $\tau$-dependent translation on a second modulus $z$. This generalization is continuously connected to the conventional picture. On the zero section, $[z]=[0]$, the generalized traditional transformations reduce effectively to transformations of matter and modular multiplets only. The group generated jointly by modular and generalized traditional transformations can therefore be regarded as a geometric generalization of an eclectic flavour symmetry.

Focusing on $\Omega(1)$, we have shown that the relevant infinite
symmetry acting on the moduli $(\tau,z)$ is the Jacobi group
\begin{equation}
\Heis(\Z)\rtimes\SL(2,\Z).
\end{equation}
The modulus $\tau$ is left unchanged by the Heisenberg subgroup
$\Heis(\Z)$, which plays the role of the parent traditional flavour
symmetry. Reduction modulo three gives the finite group
\begin{equation}
\Delta(27)\rtimes T'\simeq\Omega(1),
\end{equation}
which determines the finite transformations of matter and Jacobi-form
multiplets. On the zero section the elliptic coordinate becomes
redundant, and the Heisenberg transformations act on matter and form
multiplets through their finite image $\Delta(27)$, thereby
reproducing the traditional flavour interpretation. The main
conceptual difference with the usual eclectic construction is that
the modular and traditional components now have a common geometric
origin: they arise as different components of a single infinite
symmetry acting on moduli.

This geometric interpretation is not restricted to the
$T^2/\mathbb Z_3$ building block.  We have argued that the
$T^2/\mathbb Z_4$ case can be obtained from the same integral
Heisenberg--modular parent by taking its finite image at level two,
while the $T^2/\mathbb Z_2$ case admits a natural rank--two
generalization involving $H_5(\mathbb Z)$ and two elliptic
coordinates.  In all three cases $K=2,3,4$, the non-$R$ traditional
flavour symmetry can therefore be interpreted as a finite remnant of
an integral Heisenberg sector, whereas the modular transformations act
non-trivially on it by conjugation.  This observation strengthens the
main motivation of the present framework: the breaking of the
traditional flavour symmetry can be controlled by geometric moduli,
rather than by introducing independent flavon fields.
The $T^2/\mathbb Z_6$ case is structurally different.  Its surviving
non-$R$ traditional symmetry is only the point-group factor
$\mathbb Z_6^{(\mathrm{PG})}$, which commutes with the finite modular
group.  Hence no non-trivial Heisenberg--modular semidirect-product
structure analogous to the $K=2,3,4$ cases is present.

We have constructed supersymmetric Jacobi-invariant effective theories, both in global supersymmetry and in supergravity. We identified the relevant automorphy factors and formulated the general Jacobi transformation laws for chiral supermultiplets and level-three Jacobi forms. Minimal and non-minimal K\"ahler potentials were constructed, together with the corresponding invariance conditions for the superpotential. We have formulated a consistent CP symmetry, which can be imposed at the
Lagrangian level and broken spontaneously by the moduli. We also provided a catalogue of low-weight, low-index holomorphic level-three Jacobi forms suitable for phenomenological applications.

Finally, we developed a model of fermion masses and mixing. Rather than scanning over a large number of possible assignments of fermion quantum numbers, we deliberately adopted a simple and economical choice. In all charged sectors the matter fields transform in irreducible triplets of $\Delta(27)\rtimes T'$, with the lowest non-trivial modular weight $k=1$ and a common Jacobi index $m=6$. As a consequence, only a limited number of structurally distinct Yukawa matrices occur. Although the number of input parameters is comparable to the number of fitted observables, and the resulting predictivity is therefore limited, the model reveals several structural features that appear to be more significant than the numerical quality of the fits alone.

Already in the charged-fermion sector, the common two-texture
holomorphic structure can reproduce the observed masses and CKM
parameters once a small number of Jacobi-covariant kinetic directions is
included.  The reference fit, with two such corrections, reaches an
essentially exact description of the ten charged-fermion observables,
although it is underconstrained by two real parameters.  More
significantly, a fit with a single kinetic correction is already
qualitatively successful.  This indicates that the agreement is not
simply a consequence of introducing a large number of independent
flavour structures.  At the same time, some of the preferred kinetic
metrics are strongly hierarchical, and in particular the reference fit
approaches a nearly singular direction in the right-handed up-quark
metric.  The non-holomorphic sector should therefore not in general be
regarded as a small perturbation of the universal K\"ahler metric.
Accordingly, the projector metrics used here should be viewed as finite
phenomenological ans\"atze along symmetry-selected flavour directions,
rather than as a controlled truncation in small K\"ahler corrections.

A particularly robust feature of the numerical analysis is geometric.
The successful charged-fermion and global quark--lepton fits repeatedly
place the moduli close to non-trivial two-torsion sections, even though
proximity to these loci is not used as a fit criterion.  On the exact
two-torsion locus the two independent untwisted Yukawa structures acquire
a common zero mode.  First-generation masses, two CKM mixing angles and
CP violation then vanish, and are generated by the displacement away
from the torsion section.  This ``modular protection'' provides a
symmetry-based origin for part of the observed flavour hierarchy and, in
our view, is more informative than the very small $\chi^2$ values of the
most flexible fits.

The neutrino sector can be incorporated without increasing the matter
weights or Jacobi indices by allowing a non-trivial Dedekind multiplier.
The $\varepsilon^2$ sector supplies a primitive Majorana structure while
the charged-lepton and Dirac-neutrino matrices retain the same
two-texture form as the charged fermions.  In the global analysis, a
model with only two quark kinetic corrections, $G_2$, already gives an
acceptable fit with zero degrees of freedom,
$\chi^2_{q+\ell}\simeq0.324$, whereas the four-correction realization
$G_4$ gives essentially exact compatibility with all fitted observables,
at the price of having two more parameters than data.  Most notably,
after imposing CP so that all Wilson coefficients are real,
the $G_4^{\rm CP}$ realization still provides a good global fit,
$\chi^2_{q+\ell}\simeq2.05$ for two positive degrees of freedom.  In
this case the observed CKM phase, and potentially leptonic CP violation,
originate entirely from the complex vacuum expectation values of the
moduli.  The corresponding minimum remains close to a two-torsion
section, showing that the same geometric mechanism persists after CP is
imposed.  Within the canonical-lepton setup the atmospheric angle is
also a genuine output of the construction, with
$\sin^2\theta_{23}\simeq0.46$ across the representative global fits.
Independent inverted-ordering fits are comparably successful. We have
restricted the numerical presentation to normal ordering only for
simplicity.

These results should nevertheless be interpreted with some caution.
The present construction is primarily a proof of principle rather than a
fully predictive flavour model: several fits contain as many parameters
as observables, or more, and the kinetic deformations required by the
best solutions can be sizeable.  A more constrained ultraviolet
realization could substantially sharpen the framework by selecting the
allowed matter assignments, multiplier sectors and K\"ahler structures.
The repeated emergence of the two-torsion geometry across fits of rather
different complexity suggests, however, that the underlying mechanism is
not merely an artefact of parameter counting.

The approach pursued in this work is deliberately bottom-up. Nevertheless, several of its geometric ingredients have concrete string-theoretic precedents. In heterotic compactifications with Wilson-line moduli, effective couplings and threshold corrections have long been known to be organised by Jacobi and Siegel modular forms
\cite{Mayr:1995Stieberger,Cardoso:1996JacobiWilson,Nazaroglu:2013Jacobi}.
More recently, Wilson-line moduli have been shown to enlarge ordinary modular dualities to symplectic modular groups such as $\Sp(4,\Z)$, which can themselves play the role of modular flavour symmetries
\cite{Baur:2020Siegel,Nilles:2021Sp4Orbifolds,
Ishiguro:2020nuf,Ishiguro:2021Symplectic}.
Related Fourier--Jacobi structures also arise naturally in rank-one degenerations of genus-two string amplitudes
\cite{DHoker:2017HigherGenus}.

These results make it plausible that the Jacobi geometry used here may admit a more fundamental string-theoretic origin. What remains to be established is whether the specific flavour construction developed in this work---in which the Jacobi group itself acts as the parent flavour symmetry, its finite Heisenberg image provides the traditional flavour component, and the elliptic modulus replaces the flavon degrees of freedom---can arise from a consistent compactification. Such a top-down realization could determine, or at least strongly constrain, the Jacobi weights, indices, finite representations and multiplier systems that have been treated here as phenomenological input. The natural embedding of the Jacobi group into the Klingen parabolic subgroup of $\Sp(4,\Z)$ provides one possible route towards such an ultraviolet interpretation.

\section*{Acknowledgements}
Several aspects of this work benefited substantially from the use of artificial-intelligence tools, in particular ChatGPT (OpenAI, GPT-5.6 Sol) and Claude (Anthropic, Claude Opus 5). These tools were used as interactive research assistants in exploring mathematical and phenomenological aspects of the project, performing and checking symbolic and numerical calculations, surveying the literature, and assisting with the preparation and revision of the manuscript. Their contribution was particularly valuable during the exploratory stages of the work. All scientific judgments, validation of the results, and responsibility for the contents of the paper remain with the authors.


\appendix
\section{Explicit low-dimensional representations of $J_3$}
\label{sec:J3-lowdim-reps}

We keep the notation of Sec.~\ref{TJG} for the Heisenberg generators
$X,Y,Z$, with $Z$ the central generator and
\begin{equation*}
 XZ=ZX\,,
 \qquad
 YZ=ZY\,,
 \qquad
 XY=Z^2YX\,.
\end{equation*}
We set $\omega=e^{2\pi i/3}$.
For the singlets one has
\begin{equation}
 \rho_{\mathbf1^{(p)}}(X)
 =\rho_{\mathbf1^{(p)}}(Y)
 =\rho_{\mathbf1^{(p)}}(Z)
 =\rho_{\mathbf1^{(p)}}(S)=1\,,
 \qquad
 \rho_{\mathbf1^{(p)}}(T)=\omega^p\,.
 \label{eq:singlet-generators}
\end{equation}
A convenient basis for $\mathbf2^{(0)}$ is
\begin{equation}
 S_{\mathbf2}
 =
 \frac{i}{\sqrt3}
 \begin{pmatrix}
 1&\sqrt2\\
 \sqrt2&-1
 \end{pmatrix}\,,
 \qquad
 T_{\mathbf2}
 =
 \begin{pmatrix}
 \omega&0\\
 0&1
 \end{pmatrix}\,.
 \label{eq:doublet-generators}
\end{equation}
The three doublets are then represented by
\begin{equation}
 \begin{aligned}
 \rho_{\mathbf2^{(p)}}(X)
 &=\rho_{\mathbf2^{(p)}}(Y)
 =\rho_{\mathbf2^{(p)}}(Z)=\one_2\,,\\
 \rho_{\mathbf2^{(p)}}(S)&=S_{\mathbf2}\,,
 &
 \rho_{\mathbf2^{(p)}}(T)&=\omega^pT_{\mathbf2}\,.
 \end{aligned}
 \label{eq:three-doublets}
\end{equation}
In particular,
\begin{equation*}
 S_{\mathbf2}^2=-\one_2\,,
 \qquad
 S_{\mathbf2}^4=T_{\mathbf2}^3=\one_2\,,
 \qquad
 (S_{\mathbf2}T_{\mathbf2})^3=S_{\mathbf2}^2\,.
\end{equation*}
The centre-neutral triplet is represented by
\begin{equation}
 \rho_{\mathbf3_{\rm mod}}(X)
 =\rho_{\mathbf3_{\rm mod}}(Y)
 =\rho_{\mathbf3_{\rm mod}}(Z)=\one_3\,,
 \qquad
 S_{\rm mod}
 =
 \frac13
 \begin{pmatrix}
 -1&2&2\\
 2&-1&2\\
 2&2&-1
 \end{pmatrix}\,,
 \qquad
 T_{\rm mod}
 =
 \begin{pmatrix}
 1&0&0\\
 0&\omega&0\\
 0&0&\omega^2
 \end{pmatrix}\,.
 \label{eq:neutral-triplet-generators}
\end{equation}
Here $S_{\rm mod}^2=\one_3$, so that $\mathbf3_{\rm mod}$ factors
through $\SL(2,3)/\mathbb Z_2\simeq A_4$.
For the charged Schr\"odinger--Weil triplet we choose
\begin{equation}
 X_{\mathbf3}
 =
 \begin{pmatrix}
 0&0&1\\
 1&0&0\\
 0&1&0
 \end{pmatrix}\,,
 \qquad
 Y_{\mathbf3}
 =
 \begin{pmatrix}
 1&0&0\\
 0&\omega&0\\
 0&0&\omega^2
 \end{pmatrix}\,,
 \qquad
 Z_{\mathbf3}=\omega\one_3\,,
 \label{eq:charged-triplet-XYZ}
\end{equation}
and
\begin{equation}
 S_{\mathbf3}
 =
 -\frac{i}{\sqrt3}
 \begin{pmatrix}
 1&1&1\\
 1&\omega&\omega^2\\
 1&\omega^2&\omega
 \end{pmatrix}\,,
 \qquad
 T_{\mathbf3}
 =
 \begin{pmatrix}
 1&0&0\\
 0&\omega&0\\
 0&0&\omega
 \end{pmatrix}\,.
 \label{eq:charged-triplet-ST}
\end{equation}
These matrices satisfy
\begin{equation*}
 X_{\mathbf3}Y_{\mathbf3}
 =\omega^2Y_{\mathbf3}X_{\mathbf3}\,,
 \qquad
 S_{\mathbf3}^4=T_{\mathbf3}^3=\one_3\,,
 \qquad
 (S_{\mathbf3}T_{\mathbf3})^3=S_{\mathbf3}^2\,.
\end{equation*}
They also implement precisely the automorphisms induced by the
standard generators in Eq.~\eqref{eq:J3-standard-ST},
\begin{align}
 S_{\mathbf3}X_{\mathbf3}S_{\mathbf3}^{-1}
 &=Y_{\mathbf3}\,,
 &
 S_{\mathbf3}Y_{\mathbf3}S_{\mathbf3}^{-1}
 &=X_{\mathbf3}^{-1}\,,
 \nonumber\\
 T_{\mathbf3}Y_{\mathbf3}T_{\mathbf3}^{-1}
 &=Y_{\mathbf3}\,,
 &
 T_{\mathbf3}X_{\mathbf3}T_{\mathbf3}^{-1}
 &=Z_{\mathbf3}X_{\mathbf3}Y_{\mathbf3}^{-1}\,.
 \label{eq:modular-Heisenberg-action}
\end{align}
The three charged triplets are therefore
\begin{equation}
 \begin{aligned}
 \rho_{\mathbf3^{(p)}}(X)&=X_{\mathbf3}\,,&
 \rho_{\mathbf3^{(p)}}(Y)&=Y_{\mathbf3}\,,&
 \rho_{\mathbf3^{(p)}}(Z)&=\omega\one_3\,,\\
 \rho_{\mathbf3^{(p)}}(S)&=S_{\mathbf3}\,,&
 \rho_{\mathbf3^{(p)}}(T)&=\omega^pT_{\mathbf3}\,.
 \end{aligned}
 \label{eq:three-triplets}
\end{equation}
The conjugate triplets are obtained from
\begin{equation}
 \begin{aligned}
 \rho_{\bar{\mathbf3}^{(p)}}(X)&=X_{\mathbf3}\,,&
 \rho_{\bar{\mathbf3}^{(p)}}(Y)&=Y_{\mathbf3}^{-1}\,,&
 \rho_{\bar{\mathbf3}^{(p)}}(Z)&=\omega^2\one_3\,,\\
 \rho_{\bar{\mathbf3}^{(p)}}(S)&=S_{\mathbf3}^{*}\,,&
 \rho_{\bar{\mathbf3}^{(p)}}(T)&=\omega^pT_{\mathbf3}^{*}\,.
 \end{aligned}
 \label{eq:three-antitriplets}
\end{equation}
With these conventions
\begin{equation*}
 \bigl[\mathbf3^{(p)}\bigr]^*
 \simeq
 \bar{\mathbf3}^{(-p)}\,.
\end{equation*}
The matrices above generate a faithful order-$648$ realization of
$\Jthree$.

\begin{table}[t]
\centering
\begin{tabular}{c@{\qquad}c@{\qquad}c@{\qquad}c}
\toprule
Irrep & $\rho(Z)$ & $\Delta(27)$ restriction & Origin\\
\midrule
$\mathbf1^{(p)}$ & $1$ & singlet & $\SL(2,3)$ character\\
$\mathbf2^{(p)}$ & $1$ & $\mathbf1\oplus\mathbf1$ & $\SL(2,3)$ doublet\\
$\mathbf3_{\rm mod}$ & $1$ & $\mathbf1\oplus\mathbf1\oplus\mathbf1$ & $\SL(2,3)$ triplet\\
$\mathbf3^{(p)}$ & $\omega$ & $\mathbf3$ & Schr\"odinger--Weil $\otimes\,\mathbf1^{(p)}$\\
$\bar{\mathbf3}^{(p)}$ & $\omega^2$ & $\bar{\mathbf3}$ & conjugate $\otimes\,\mathbf1^{(p)}$\\
\bottomrule
\end{tabular}
\caption{Low-dimensional irreducible representations of $J_3$.
For the families labelled by $p$, one has $p=0,1,2$.}
\label{tab:J3-lowdim}
\end{table}

\subsection{Tensor products and Clebsch--Gordan coefficients}
\label{sec:J3-CG}

We collect the tensor products involving singlets, doublets and
triplets. All labels $p,q$ and their sums are understood modulo three.
For the sextets we define
\begin{equation}
 \mathbf6^{(0)}
 \equiv
 \mathbf2^{(0)}\otimes\mathbf3^{(0)}\,,
 \qquad
 \mathbf6^{(p)}
 =
 \mathbf1^{(p)}\otimes\mathbf6^{(0)}\,,
 \qquad
 \bar{\mathbf6}^{(0)}
 \equiv
 \bigl(\mathbf6^{(0)}\bigr)^*\,,
 \qquad
 \bar{\mathbf6}^{(p)}
 =
 \mathbf1^{(p)}\otimes\bar{\mathbf6}^{(0)}\,.
 \label{eq:J3-sextet-definition}
\end{equation}
Thus $[\mathbf6^{(p)}]^*\simeq\bar{\mathbf6}^{(-p)}$.
The octets are defined by
\begin{equation*}
 \mathbf3^{(0)}\otimes\bar{\mathbf3}^{(0)}
 =
 \mathbf1^{(0)}\oplus\mathbf8^{(0)}\,,
 \qquad
 \mathbf8^{(p)}
 =
 \mathbf1^{(p)}\otimes\mathbf8^{(0)}\,.
\end{equation*}
The two nine-dimensional irreducible representations, with central
characters $\omega$ and $\omega^2$, are denoted by $\mathbf9$ and
$\bar{\mathbf9}$, respectively.
The tensor-product rules needed below are
\begin{align}
 \mathbf1^{(p)}\otimes\mathbf1^{(q)}
 &=\mathbf1^{(p+q)}\,,&
 \mathbf1^{(p)}\otimes\mathbf2^{(q)}
 &=\mathbf2^{(p+q)}\,,
 \nonumber\\
 \mathbf1^{(p)}\otimes\mathbf3_{\rm mod}
 &=\mathbf3_{\rm mod}\,,
 \nonumber\\
 \mathbf1^{(p)}\otimes\mathbf3^{(q)}
 &=\mathbf3^{(p+q)}\,,&
 \mathbf1^{(p)}\otimes\bar{\mathbf3}^{(q)}
 &=\bar{\mathbf3}^{(p+q)}\,,
 \label{eq:prod-1-lowdim}
\end{align}
and
\begin{align}
 \mathbf2^{(p)}\otimes\mathbf2^{(q)}
 &=\mathbf1^{(p+q+1)}\oplus\mathbf3_{\rm mod}\,,
 \label{eq:prod-22}\\
 \mathbf2^{(p)}\otimes\mathbf3_{\rm mod}
 &=\mathbf2^{(0)}\oplus\mathbf2^{(1)}\oplus\mathbf2^{(2)}\,,
 \label{eq:prod-23mod}\\
 \mathbf2^{(p)}\otimes\mathbf3^{(q)}
 &=\mathbf6^{(p+q)}\,,
 \label{eq:prod-23}\\
 \mathbf2^{(p)}\otimes\bar{\mathbf3}^{(q)}
 &=\bar{\mathbf6}^{(p+q+1)}\,.
 \label{eq:prod-2bar3}
\end{align}
For products of triplets one finds
\begin{align}
 \mathbf3_{\rm mod}\otimes\mathbf3_{\rm mod}
 &=
 \mathbf1^{(0)}\oplus\mathbf1^{(1)}\oplus\mathbf1^{(2)}
 \oplus2\,\mathbf3_{\rm mod}\,,
 \label{eq:prod-3mod3mod}\\
 \mathbf3_{\rm mod}\otimes\mathbf3^{(p)}
 &=\mathbf9\,,&
 \mathbf3_{\rm mod}\otimes\bar{\mathbf3}^{(p)}
 &=\bar{\mathbf9}\,,
 \label{eq:prod-3modcharged}\\
 \mathbf3^{(p)}\otimes\mathbf3^{(q)}
 &=
 \bar{\mathbf3}^{(p+q+2)}
 \oplus
 \bar{\mathbf6}^{(p+q)}\,,
 \label{eq:prod-33}\\
 \bar{\mathbf3}^{(p)}\otimes\bar{\mathbf3}^{(q)}
 &=
 \mathbf3^{(p+q+1)}
 \oplus
 \mathbf6^{(p+q)}\,,
 \label{eq:prod-bar3bar3}\\
 \mathbf3^{(p)}\otimes\bar{\mathbf3}^{(q)}
 &=
 \mathbf1^{(p+q)}
 \oplus
 \mathbf8^{(p+q)}\,.
 \label{eq:prod-3bar3}
\end{align}
The central character provides an immediate check of these rules.
In particular,
\begin{align*}
 \rho(Z)&=1
 &&\text{for }\mathbf1^{(p)},\mathbf2^{(p)},\mathbf3_{\rm mod},\mathbf8^{(p)}\,,\\
 \rho(Z)&=\omega
 &&\text{for }\mathbf3^{(p)},\mathbf6^{(p)},\mathbf9\,,\\
 \rho(Z)&=\omega^2
 &&\text{for }\bar{\mathbf3}^{(p)},\bar{\mathbf6}^{(p)},\bar{\mathbf9}\,.
\end{align*}
For later numerical applications it is useful to record Clebsch--Gordan
coefficients in the bases of Sec.~\ref{sec:J3-lowdim-reps}. It is
sufficient to give the untwisted products; non-zero $p,q$ are obtained
by tensoring with the corresponding one-dimensional characters.
Whenever a twist of $\mathbf3_{\rm mod}$ occurs, it is brought back to
our standard basis by the cyclic matrix
\begin{equation*}
 P=
 \begin{pmatrix}0&0&1\\1&0&0\\0&1&0\end{pmatrix}\,,
 \qquad
 P^r\bigl(\omega^rT_{\rm mod}\bigr)P^{-r}=T_{\rm mod}\,,
 \qquad
 P^rS_{\rm mod}P^{-r}=S_{\rm mod}\,.
\end{equation*}
Let $a=(a_1,a_2)^T$ and $b=(b_1,b_2)^T$ transform as
$\mathbf2^{(0)}$. Then
\begin{equation}
 (ab)_{\mathbf1^{(1)}}
 =
 \frac{a_1b_2-a_2b_1}{\sqrt2}\,,
 \qquad
 (ab)_{\mathbf3_{\rm mod}}
 =
 \begin{pmatrix}
 a_2b_2\\[1mm]
 (a_1b_2+a_2b_1)/\sqrt2\\[1mm]
 -a_1b_1
 \end{pmatrix}\,.
 \label{eq:CG-22}
\end{equation}
For $a\sim\mathbf2^{(0)}$ and
$b=(b_1,b_2,b_3)^T\sim\mathbf3_{\rm mod}$,
\begin{align}
 (ab)_{\mathbf2^{(0)}}
 &=
 \frac1{\sqrt3}
 \begin{pmatrix}
 a_1b_1+\sqrt2a_2b_2\\
 \sqrt2a_1b_3-a_2b_1
 \end{pmatrix}\,,
 \nonumber\\[1mm]
 (ab)_{\mathbf2^{(1)}}
 &=
 \frac1{\sqrt3}
 \begin{pmatrix}
 a_1b_2+\sqrt2a_2b_3\\
 \sqrt2a_1b_1-a_2b_2
 \end{pmatrix}\,,
 \nonumber\\[1mm]
 (ab)_{\mathbf2^{(2)}}
 &=
 \frac1{\sqrt3}
 \begin{pmatrix}
 a_1b_3+\sqrt2a_2b_1\\
 \sqrt2a_1b_2-a_2b_3
 \end{pmatrix}\,.
 \label{eq:CG-2-3mod}
\end{align}
For $a,b\sim\mathbf3_{\rm mod}$ the three singlets are
\begin{align*}
 (ab)_{\mathbf1^{(0)}}
 &=\frac{a_1b_1+a_2b_3+a_3b_2}{\sqrt3}\,,\\
 (ab)_{\mathbf1^{(1)}}
 &=\frac{a_1b_2+a_2b_1+a_3b_3}{\sqrt3}\,,\\
 (ab)_{\mathbf1^{(2)}}
 &=\frac{a_1b_3+a_3b_1+a_2b_2}{\sqrt3}\,,
\end{align*}
and the two triplets can be chosen as
\begin{align}
 (ab)_{\mathbf3_{\rm mod},s}
 &=
 \frac1{\sqrt6}
 \begin{pmatrix}
 2a_1b_1-a_2b_3-a_3b_2\\
 2a_3b_3-a_1b_2-a_2b_1\\
 2a_2b_2-a_1b_3-a_3b_1
 \end{pmatrix}\,,
 \nonumber\\[1mm]
 (ab)_{\mathbf3_{\rm mod},a}
 &=
 \frac1{\sqrt2}
 \begin{pmatrix}
 a_2b_3-a_3b_2\\
 a_1b_2-a_2b_1\\
 a_3b_1-a_1b_3
 \end{pmatrix}\,.
 \label{eq:CG-3mod3mod}
\end{align}
For $a,b\sim\mathbf3^{(0)}$, the antisymmetric triplet and symmetric
sextet are
\begin{equation}
 (ab)_{\bar{\mathbf3}^{(2)}}
 =
 \frac1{\sqrt2}
 \begin{pmatrix}
 a_2b_3-a_3b_2\\
 a_3b_1-a_1b_3\\
 a_1b_2-a_2b_1
 \end{pmatrix}\,,
 \label{eq:CG-33-triplet}
\end{equation}
\begin{equation}
 (ab)_{\bar{\mathbf6}^{(0)}}
 =
 \begin{pmatrix}
 a_1b_1\\
 a_2b_2\\
 a_3b_3\\
 (a_2b_3+a_3b_2)/\sqrt2\\
 (a_3b_1+a_1b_3)/\sqrt2\\
 (a_1b_2+a_2b_1)/\sqrt2
 \end{pmatrix}\,.
 \label{eq:CG-33-sextet}
\end{equation}
The same real coefficients apply to
$\bar{\mathbf3}^{(0)}\otimes\bar{\mathbf3}^{(0)}$, with outputs
$\mathbf3^{(1)}$ and $\mathbf6^{(0)}$, respectively. Hence
\begin{equation*}
 {\rm Sym}^2\mathbf3^{(0)}=\bar{\mathbf6}^{(0)}\,,
 \qquad
 \wedge^2\mathbf3^{(0)}=\bar{\mathbf3}^{(2)}\,.
\end{equation*}
For $a\sim\mathbf3^{(0)}$ and
$b\sim\bar{\mathbf3}^{(0)}$ the invariant singlet is
\begin{equation}
 (ab)_{\mathbf1^{(0)}}
 =
 \frac{a_1b_1+a_2b_2+a_3b_3}{\sqrt3}\,.
 \label{eq:CG-3bar3-singlet}
\end{equation}
A convenient orthonormal basis for the complementary octet is
\begin{equation*}
 (ab)_{\mathbf8^{(0)}}
 =
 \begin{pmatrix}
 (a_1b_1-a_2b_2)/\sqrt2\\
 (a_1b_1+a_2b_2-2a_3b_3)/\sqrt6\\
 a_1b_2\\a_1b_3\\a_2b_1\\a_2b_3\\a_3b_1\\a_3b_2
 \end{pmatrix}\,.
\end{equation*}
Finally, products which are already irreducible require no further
projection. In particular, in the direct-product basis,
\begin{equation*}
 \mathbf2^{(0)}\otimes\mathbf3^{(0)}:\quad
 (ab)_{\mathbf6^{(0)}}
 =
 (a_1b_1,a_1b_2,a_1b_3,a_2b_1,a_2b_2,a_2b_3)^T\,,
\end{equation*}
and similarly the nine products $a_ib_j$ furnish the irreducible
$\mathbf9$ in
$\mathbf3_{\rm mod}\otimes\mathbf3^{(0)}$.

\section{Low-weight Jacobi forms}
\label{sec:Jacobi-forms}

\subsection{Weight-one forms}
\label{sec:weight-one-Jacobi}

\begin{table}[t]
\centering
\begin{tabular}{c@{\qquad}c@{\qquad}c@{\qquad}c}
\toprule
$m$ & Dimension & $J_3$ representation & Explicit realization\\
\midrule
$0$  & $2$ & $\mathbf2^{(0)}$ & modular doublet\\
$3$  & $3$ & $\mathbf3^{(0)}$ & torsion orbit of $Q_{1,1}$\\
$6$  & $0$ & --- & ---\\
$9$  & $2$ & $\mathbf2^{(0)}$ & theta-decomposition doublet\\
$12$ & $3$ & $\mathbf3^{(0)}$ & torsion orbit of $Q_{2,2}$\\
\bottomrule
\end{tabular}
\caption{Weight-one holomorphic Jacobi forms for the first few
classical indices. Representation labels refer to the bases fixed in
Sec.~\ref{sec:J3-lowdim-reps}.}
\label{tab:weight-one-Jacobi}
\end{table}

\paragraph{Index $m=0$.}

At zero index the elliptic dependence disappears. A convenient basis
for the weight-one doublet is
\begin{equation}
\displaystyle 
 D(\tau)
 \equiv
 {\boldsymbol Y}^{(1,0)}(\tau)
 =
 \begin{pmatrix}
 D_1(\tau)\\[1mm]D_2(\tau)
 \end{pmatrix}
 =
 \begin{pmatrix}
 -3\sqrt2\,\dfrac{\eta^3(3\tau)}{\eta(\tau)}\\[4mm]
 3\,\dfrac{\eta^3(3\tau)}{\eta(\tau)}
 +\dfrac{\eta^3(\tau/3)}{\eta(\tau)}
 \end{pmatrix}\,.
 \label{eq:k1m0-doublet}
\end{equation}
It transforms as $\mathbf2^{(0)}$ with the matrices in
Eq.~\eqref{eq:doublet-generators}. In the notation commonly used for
$T'$ in Ref.~\cite{Baur:2022hma}, the same irrep is denoted
$\mathbf2^{\prime\prime}$. We use only the uniform notation
$\mathbf2^{(p)}$ in the present work; the two descriptions are related
by translating the convention for the modular generators.

\paragraph{Indices $m=3$ and $m=12$.}

We use the odd Jacobi theta function in the convention
\begin{equation}
 \vartheta(\tau,w)
 =
 \sum_{n\in\mathbb Z}(-1)^n
 q^{(n+1/2)^2/2}\,\xi^{\,n+1/2}\,,
 \qquad
 q=e^{2\pi i\tau}\,,
 \qquad
 \xi=e^{2\pi iw}\,.
 \label{eq:odd-theta-convention}
\end{equation}
The theta blocks of Ref.~\cite{Gritsenko:2026ThetaBlocks} are
\begin{equation}
 Q_{a,b}(\tau,w)
 =
 \frac{
 \vartheta(\tau,aw)\,
 \vartheta(\tau,bw)\,
 \vartheta(\tau,(a+b)w)
 }{\eta(\tau)}\,,
 \label{eq:theta-quark}
\end{equation}
with weight one and classical index $m=a^2+ab+b^2$. The two seeds
needed here are
\begin{align}
 q_3(\tau,w)
 &\equiv
 Q_{1,1}(\tau,w)
 =
 \frac{\vartheta(\tau,w)^2\vartheta(\tau,2w)}{\eta(\tau)}\,,
 \label{eq:q3}\\
 q_{12}(\tau,w)
 &\equiv
 Q_{2,2}(\tau,w)
 =
 \frac{\vartheta(\tau,2w)^2\vartheta(\tau,4w)}{\eta(\tau)}
 =q_3(\tau,2w)\,.
 \label{eq:q12}
\end{align}
Introduce the three-torsion operators
\begin{align}
 ({\cal U}f)(\tau,w)
 &=f\!\left(\tau,w+\frac13\right)\,,
 \nonumber\\
 ({\cal V}_mf)(\tau,w)
 &=
 \exp\!\left[
 2\pi im\left(\frac{\tau}{9}+\frac{2w}{3}\right)
 \right]
 f\!\left(\tau,w+\frac{\tau}{3}\right)\,,
 \label{eq:UV-torsion}
\end{align}
and
\begin{equation*}
 {\cal P}_a
 =
 \frac13\sum_{j=0}^2\omega^{-aj}{\cal U}^{\,j}\,,
 \qquad a=0,1,2\,.
\end{equation*}
A basis adapted simultaneously to the Heisenberg and modular matrices
of Eqs.~\eqref{eq:charged-triplet-XYZ} and
\eqref{eq:charged-triplet-ST} is
\begin{equation}
\displaystyle 
 {\boldsymbol Y}^{(1,m)}(\tau,w)
 =
 \begin{pmatrix}
 {\cal P}_0{\cal V}_m q_m\\[1mm]
 {\cal P}_1q_m\\[1mm]
 -{\cal P}_2q_m
 \end{pmatrix}\,,
 \qquad m=3,12\,.
 \label{eq:charged-triplet-Jacobi}
\end{equation}
We denote
\begin{equation*}
\displaystyle 
 F(\tau,w)\equiv{\boldsymbol Y}^{(1,3)}(\tau,w)\,,
 \qquad
 G(\tau,w)\equiv{\boldsymbol Y}^{(1,12)}(\tau,w)\,.
\end{equation*}
With the basis ordering in Eq.~\eqref{eq:charged-triplet-Jacobi},
the finite transformations are exactly those of $\mathbf3^{(0)}$,
including $X_{\mathbf3},Y_{\mathbf3},S_{\mathbf3},T_{\mathbf3}$.
Thus
\begin{equation}
 {\cal J}^{(3)}_{1,3}\simeq\mathbf3^{(0)}\,,
 \qquad
 {\cal J}^{(3)}_{1,12}\simeq\mathbf3^{(0)}\,.
 \label{eq:k1-charged-triplets}
\end{equation}

\paragraph{Index $m=9$.}

Define the index-$m$ theta functions
\begin{equation}
 \vartheta_{m,r}(\tau,w)
 =
 \sum_{n\in\mathbb Z}
 q^{(2mn+r)^2/(4m)}\xi^{\,2mn+r}\,,
 \qquad r\pmod{2m}\,.
 \label{eq:theta-mr}
\end{equation}
For $m=9$ introduce
\begin{align}
 A_9(\tau,w)
 &=
 \vartheta_{9,0}(\tau,w)
 +\vartheta_{9,6}(\tau,w)
 +\vartheta_{9,12}(\tau,w)\,,
 \nonumber\\
 B_9(\tau,w)
 &=
 \vartheta_{9,3}(\tau,w)
 +\vartheta_{9,9}(\tau,w)
 +\vartheta_{9,15}(\tau,w)\,,
 \label{eq:A9B9}
\end{align}
and the index-three theta constants
\begin{equation*}
 C_r(\tau)=\vartheta_{3,r}(\tau,0)\,,
 \qquad r=0,1,2,3\,,
\end{equation*}
where $\vartheta_{3,4}(\tau,0)=C_2(\tau)$ and
$\vartheta_{3,5}(\tau,0)=C_1(\tau)$.
An explicit basis of ${\cal J}^{(3)}_{1,9}$ is
\begin{equation}
\displaystyle 
 E(\tau,w)
 \equiv
 {\boldsymbol Y}^{(1,9)}(\tau,w)
 =
 \begin{pmatrix}
 E_1(\tau,w)\\[1mm]E_2(\tau,w)
 \end{pmatrix}
 =
 \begin{pmatrix}
 -\sqrt2\,[C_1(\tau)B_9(\tau,w)+C_2(\tau)A_9(\tau,w)]\\[2mm]
 C_0(\tau)A_9(\tau,w)+C_3(\tau)B_9(\tau,w)
 \end{pmatrix}\,.
 \label{eq:k1m9-explicit}
\end{equation}
It is Heisenberg-neutral after extraction of the standard Jacobi
automorphy factor. Explicitly,
\begin{align*}
 E\!\left(\tau,w+\frac13\right)
 &=E(\tau,w)\,,\\
 \exp\!\left[
 2\pi i9\left(\frac{\tau}{9}+\frac{2w}{3}\right)
 \right]
 E\!\left(\tau,w+\frac{\tau}{3}\right)
 &=E(\tau,w)\,.
\end{align*}
Its modular transformations are
\begin{align}
 E(\tau+1,w)
 &=T_{\mathbf2}E(\tau,w)\,,
 \nonumber\\
 E\!\left(-\frac1\tau,\frac{w}{\tau}\right)
 &=
 \tau\,
 \exp\!\left(\frac{2\pi i9w^2}{\tau}\right)
 S_{\mathbf2}E(\tau,w)\,.
 \label{eq:k1m9-ST}
\end{align}
Hence
\begin{equation}
 {\cal J}^{(3)}_{1,9}\simeq\mathbf2^{(0)}\,.
 \label{eq:k1m9-doublet}
\end{equation}
A useful check is the exact restriction
\begin{equation}
 E(\tau,0)=D(\tau)\,.
 \label{eq:k1m9-z0}
\end{equation}
Thus the positive-index doublet is a genuine Jacobi deformation of
the ordinary modular doublet.
The original elliptic coordinate is recovered throughout from
$z=3w$.

\subsection{Weight-two forms}
\label{sec:weight-two-Jacobi}

Multiplication respects the bigrading by weight and index,
\begin{equation}
 {\cal J}^{(3)}_{k_1,m_1}
 {\cal J}^{(3)}_{k_2,m_2}
 \subset
 {\cal J}^{(3)}_{k_1+k_2,m_1+m_2}\,.
 \label{eq:Jacobi-product-bigrading}
\end{equation}
For the low indices displayed in Table~\ref{tab:low-weight-dimensions},
the product spaces generated by $D,E,F,G$ have the maximal ranks
allowed by the independently determined dimensions. No additional
primitive weight-two form is needed up to $m=12$.

\begin{table}[t]
\centering
\begin{tabular}{c@{\qquad}c@{\qquad}c@{\qquad}c}
\toprule
$m$ & Dimension & Construction & $J_3$ decomposition\\
\midrule
$0$ & $3$ & ${\rm Sym}^2D$ & $\mathbf3_{\rm mod}$\\
$3$ & $6$ & $D\otimes F$ & $\mathbf6^{(0)}$\\
$6$ & $6$ & ${\rm Sym}^2F$ & $\bar{\mathbf6}^{(0)}$\\
$9$ & $4$ & $D\otimes E$ & $\mathbf1^{(1)}\oplus\mathbf3_{\rm mod}$\\
$12$ & $12$ & $(D\otimes G)\oplus(E\otimes F)$ & $\mathbf6^{(0)}\oplus\mathbf6^{(0)}$\\
\bottomrule
\end{tabular}
\caption{Weight-two Jacobi forms for the first few classical indices.
All spaces shown are saturated by products of the explicit weight-one
forms.}
\label{tab:weight-two-Jacobi}
\end{table}

\paragraph{Index $m=0$.}

Using the Clebsch--Gordan coefficients of Eq.~\eqref{eq:CG-22}, a
basis in exactly the same $\mathbf3_{\rm mod}$ convention as
Eq.~\eqref{eq:neutral-triplet-generators} is
\begin{equation}
 {\boldsymbol Y}^{(2,0)}_{\mathbf3_{\rm mod}}(\tau)
 =
 \begin{pmatrix}
 D_2^2\\[1mm]
 \sqrt2D_1D_2\\[1mm]
 -D_1^2
 \end{pmatrix}\,.
 \label{eq:k2m0-triplet}
\end{equation}
Therefore
\begin{equation*}
 {\cal J}^{(3)}_{2,0}\simeq\mathbf3_{\rm mod}\,.
\end{equation*}

\paragraph{Index $m=3$.}

The unique sextet is the direct product
\begin{equation}
 {\boldsymbol Y}^{(2,3)}_{\mathbf6^{(0)}}(\tau,w)
 =
 \begin{pmatrix}
 D_1F_1\\D_1F_2\\D_1F_3\\
 D_2F_1\\D_2F_2\\D_2F_3
 \end{pmatrix}\,.
 \label{eq:k2m3-sextet}
\end{equation}
Thus ${\cal J}^{(3)}_{2,3}\simeq\mathbf6^{(0)}$. This is the direct
Jacobi analogue of a modular doublet times a flavour triplet.

\paragraph{Index $m=6$.}

Since ${\cal J}^{(3)}_{1,6}=0$, this space is generated by
${\rm Sym}^2F$. In the sextet basis of Eq.~\eqref{eq:CG-33-sextet},
\begin{equation}
 {\boldsymbol Y}^{(2,6)}_{\bar{\mathbf6}^{(0)}}(\tau,w)
 =
 \begin{pmatrix}
 F_1^2\\
 F_2^2\\
 F_3^2\\
 \sqrt2F_2F_3\\
 \sqrt2F_3F_1\\
 \sqrt2F_1F_2
 \end{pmatrix}\,.
 \label{eq:k2m6-sextet}
\end{equation}
Hence
\begin{equation*}
 {\cal J}^{(3)}_{2,6}\simeq\bar{\mathbf6}^{(0)}\,,
 \qquad
 \rho(Z)=\omega^2\one_6\,.
\end{equation*}
\begin{table}[h!]
\centering
\begin{tabular}{c@{\quad}c@{\quad}c@{\qquad}c}
\toprule
$k$ & $m$ & Dimension & $J_3$ content\\
\midrule
$1$ & $0$  & $2$  & $\mathbf2^{(0)}$\\
$1$ & $3$  & $3$  & $\mathbf3^{(0)}$\\
$1$ & $6$  & $0$  & ---\\
$1$ & $9$  & $2$  & $\mathbf2^{(0)}$\\
$1$ & $12$ & $3$  & $\mathbf3^{(0)}$\\
\midrule
$2$ & $0$  & $3$  & $\mathbf3_{\rm mod}$\\
$2$ & $3$  & $6$  & $\mathbf6^{(0)}$\\
$2$ & $6$  & $6$  & $\bar{\mathbf6}^{(0)}$\\
$2$ & $9$  & $4$  & $\mathbf1^{(1)}\oplus\mathbf3_{\rm mod}$\\
$2$ & $12$ & $12$ & $\mathbf6^{(0)}\oplus\mathbf6^{(0)}$\\
\bottomrule
\end{tabular}
\caption{Low-weight catalogue of holomorphic level-three Jacobi forms
with trivial scalar multiplier on the level-three kernel.}
\label{tab:low-weight-combined2}
\end{table}
\paragraph{Index $m=9$.}

The product $D\otimes E$ decomposes according to
$\mathbf2^{(0)}\otimes\mathbf2^{(0)}
=\mathbf1^{(1)}\oplus\mathbf3_{\rm mod}$. The singlet is
\begin{equation}
 Y_{\mathbf1^{(1)}}^{(2,9)}(\tau,w)
 =
 \frac{D_1E_2-D_2E_1}{\sqrt2}\,,
 \label{eq:k2m9-singlet}
\end{equation}
and the triplet in the standard basis is
\begin{equation}
 {\boldsymbol Y}_{\mathbf3_{\rm mod}}^{(2,9)}(\tau,w)
 =
 \begin{pmatrix}
 D_2E_2\\[1mm]
 (D_1E_2+D_2E_1)/\sqrt2\\[1mm]
 -D_1E_1
 \end{pmatrix}\,.
 \label{eq:k2m9-triplet}
\end{equation}
Therefore
\begin{equation*}
 {\cal J}^{(3)}_{2,9}
 \simeq
 \mathbf1^{(1)}\oplus\mathbf3_{\rm mod}\,.
\end{equation*}
Since $E(\tau,0)=D(\tau)$, the singlet vanishes on the ordinary
modular locus,
\begin{equation*}
 Y_{\mathbf1^{(1)}}^{(2,9)}(\tau,0)=0\,.
\end{equation*}

\paragraph{Index $m=12$.}

Two independent products have the same weight, index and finite
representation,
\begin{equation*}
 D\otimes G\sim\mathbf6^{(0)}\,,
 \qquad
 E\otimes F\sim\mathbf6^{(0)}\,.
\end{equation*}
We denote them by
\begin{equation}
 {\boldsymbol Y}^{(2,12)}_{\mathbf6,A}(\tau,w)
 =
 \begin{pmatrix}
 D_1G_1\\D_1G_2\\D_1G_3\\
 D_2G_1\\D_2G_2\\D_2G_3
 \end{pmatrix}\,,
 \label{eq:k2m12-sextet-A}
\end{equation}
and
\begin{equation}
 {\boldsymbol Y}^{(2,12)}_{\mathbf6,B}(\tau,w)
 =
 \begin{pmatrix}
 E_1F_1\\E_1F_2\\E_1F_3\\
 E_2F_1\\E_2F_2\\E_2F_3
 \end{pmatrix}\,.
 \label{eq:k2m12-sextet-B}
\end{equation}
Their twelve components have rank twelve as functions of $(\tau,w)$,
so that
\begin{equation}
 {\cal J}^{(3)}_{2,12}
 \simeq
 \mathbf6_A^{(0)}\oplus\mathbf6_B^{(0)}\,.
 \label{eq:k2m12-rep}
\end{equation}
The labels $A,B$ distinguish two copies of the same irrep, not new
group-theoretical quantum numbers. A general index-twelve sextet is
therefore
\begin{equation}
 {\boldsymbol Y}^{(2,12)}_{\mathbf6}
 =
 c_A{\boldsymbol Y}^{(2,12)}_{\mathbf6,A}
 +c_B{\boldsymbol Y}^{(2,12)}_{\mathbf6,B}\,.
 \label{eq:k2m12-general-sextet}
\end{equation}
This multiplicity will be important in the Yukawa sector.

\paragraph{Completeness at weight two.}

The numerical ranks of the explicit product bases are
\begin{center}
\begin{tabular}{c@{\qquad}c@{\qquad}c}
\toprule
$m$ & Explicit product basis & Rank\\
\midrule
$0$ & ${\rm Sym}^2D$ & $3$\\
$3$ & $D\otimes F$ & $6$\\
$6$ & ${\rm Sym}^2F$ & $6$\\
$9$ & $D\otimes E$ & $4$\\
$12$ & $(D\otimes G)\oplus(E\otimes F)$ & $12$\\
\bottomrule
\end{tabular}
\end{center}
They coincide with the dimensions in
Table~\ref{tab:low-weight-dimensions}. Consequently,
\begin{align}
 {\cal J}^{(3)}_{2,0}
 &= {\rm Sym}^2{\cal J}^{(3)}_{1,0}\,,
 \nonumber\\
 {\cal J}^{(3)}_{2,3}
 &= {\cal J}^{(3)}_{1,0}{\cal J}^{(3)}_{1,3}\,,
 \nonumber\\
 {\cal J}^{(3)}_{2,6}
 &= {\rm Sym}^2{\cal J}^{(3)}_{1,3}\,,
 \nonumber\\
 {\cal J}^{(3)}_{2,9}
 &= {\cal J}^{(3)}_{1,0}{\cal J}^{(3)}_{1,9}\,,
 \nonumber\\
 {\cal J}^{(3)}_{2,12}
 &=
 \bigl({\cal J}^{(3)}_{1,0}{\cal J}^{(3)}_{1,12}\bigr)
 \oplus
 \bigl({\cal J}^{(3)}_{1,9}{\cal J}^{(3)}_{1,3}\bigr)\,.
 \label{eq:weight-two-completeness}
\end{align}

A concise summary is provided in table \ref{tab:low-weight-combined2}.
\section{Explicit theta realization of the primitive $\eps^2$ sextet}
\label{app:primitive-eps2-sextet}

In this appendix we give an explicit theta realization of the unique
primitive multiplet
\begin{equation}
 \boldsymbol Y^{[2]}_{\six{1}}(\tau,w)
 \in {\cal J}^{(3)}_{2,12}[\eps^2]\big|_{\six{1}} .
 \label{eq:app-primitive-space}
\end{equation}
The construction is equivalent to the twisted-Weil intertwiner used in
our numerical analysis.  We choose here an overall normalization for
which all coefficients belong to $\mathbb Q(\sqrt2)$.  This differs by
an irrelevant common complex factor from the orthonormal
normalization used in the numerical implementation.

We use the index-$m$ theta functions
\begin{equation}
 \vartheta_{m,r}(\tau,w)
 =\sum_{n\in\mathbb Z}
 q^{(2mn+r)^2/(4m)}\,\xi^{2mn+r},
 \qquad
 q=e^{2\pi i\tau},\qquad \xi=e^{2\pi i w},
 \label{eq:app-theta-mr}
\end{equation}
with $r$ understood modulo $2m$.  At $w=0$ it is convenient to use the
symmetric unary-theta coordinates
\begin{equation}
 u^{(\ell)}_s(\tau)=
 \begin{cases}
 \vartheta_{\ell,s}(\tau,0),&s=0,\ell,\\[1mm]
 \sqrt2\,\vartheta_{\ell,s}(\tau,0),&1\leq s\leq\ell-1,
 \end{cases}
 \qquad \ell=2,4,6,
 \label{eq:app-unary-coordinates}
\end{equation}
and the weight-$3/2$ monomials
\begin{equation}
 P_{\alpha\beta\gamma}(\tau)
 =u^{(2)}_{\alpha}(\tau)\,
  u^{(4)}_{\beta}(\tau)\,
  u^{(6)}_{\gamma}(\tau),
 \qquad
 0\leq\alpha\leq2,\quad
 0\leq\beta\leq4,\quad
 0\leq\gamma\leq6.
 \label{eq:app-Pabc}
\end{equation}

Only six linear combinations of these $105$ monomials are required.
We define
\begin{align}
 \mathfrak a={}&
 P_{016}+P_{030}+2P_{113}+P_{210}+P_{236},
 \nonumber\\
 \mathfrak b={}&
 \sqrt2\left(P_{000}-P_{040}+P_{206}-P_{246}\right),
 \nonumber\\
 \mathfrak c={}&
 P_{010}+P_{036}+2P_{133}+P_{216}+P_{230},
 \label{eq:app-abc}\\[1mm]
 \mathfrak a'={}&
 P_{012}+P_{034}+\sqrt2\left(P_{111}+P_{115}\right)
 +P_{214}+P_{232},
 \nonumber\\
 \mathfrak b'={}&
 \sqrt2\left(P_{004}-P_{044}+P_{202}-P_{242}\right),
 \nonumber\\
 \mathfrak c'={}&
 P_{014}+P_{032}+\sqrt2\left(P_{131}+P_{135}\right)
 +P_{212}+P_{234}.
 \label{eq:app-abc-prime}
\end{align}
All six functions in Eqs.~\eqref{eq:app-abc}--\eqref{eq:app-abc-prime}
have modular weight $3/2$.

In terms of them, the six components of the primitive sextet take the
compact form
\begin{align}
 Y^{[2]}_{\six{1},1}={}&
 -\mathfrak a\,\vartheta_{12,3}
 -\mathfrak b\,\vartheta_{12,6}
 +\mathfrak c\,\vartheta_{12,9}
 -\mathfrak c\,\vartheta_{12,15}
 +\mathfrak b\,\vartheta_{12,18}
 +\mathfrak a\,\vartheta_{12,21},
 \nonumber\\
 Y^{[2]}_{\six{1},2}={}&
 \mathfrak c\,\vartheta_{12,1}
 -\mathfrak c\,\vartheta_{12,7}
 +\mathfrak b\,\vartheta_{12,10}
 +\mathfrak a\,\vartheta_{12,13}
 -\mathfrak a\,\vartheta_{12,19}
 -\mathfrak b\,\vartheta_{12,22},
 \nonumber\\
 Y^{[2]}_{\six{1},3}={}&
 \mathfrak b\,\vartheta_{12,2}
 +\mathfrak a\,\vartheta_{12,5}
 -\mathfrak a\,\vartheta_{12,11}
 -\mathfrak b\,\vartheta_{12,14}
 +\mathfrak c\,\vartheta_{12,17}
 -\mathfrak c\,\vartheta_{12,23},
 \label{eq:app-primitive-first-three}\\
 Y^{[2]}_{\six{1},4}={}&
 -\mathfrak a'\,\vartheta_{12,3}
 -\mathfrak b'\,\vartheta_{12,6}
 +\mathfrak c'\,\vartheta_{12,9}
 -\mathfrak c'\,\vartheta_{12,15}
 +\mathfrak b'\,\vartheta_{12,18}
 +\mathfrak a'\,\vartheta_{12,21},
 \nonumber\\
 Y^{[2]}_{\six{1},5}={}&
 \mathfrak c'\,\vartheta_{12,1}
 -\mathfrak c'\,\vartheta_{12,7}
 +\mathfrak b'\,\vartheta_{12,10}
 +\mathfrak a'\,\vartheta_{12,13}
 -\mathfrak a'\,\vartheta_{12,19}
 -\mathfrak b'\,\vartheta_{12,22},
 \nonumber\\
 Y^{[2]}_{\six{1},6}={}&
 \mathfrak b'\,\vartheta_{12,2}
 +\mathfrak a'\,\vartheta_{12,5}
 -\mathfrak a'\,\vartheta_{12,11}
 -\mathfrak b'\,\vartheta_{12,14}
 +\mathfrak c'\,\vartheta_{12,17}
 -\mathfrak c'\,\vartheta_{12,23}.
 \label{eq:app-primitive-last-three}
\end{align}
Here and in Eqs.~\eqref{eq:app-primitive-first-three}--\eqref{eq:app-primitive-last-three},
$\vartheta_{12,r}$ without displayed arguments denotes
$\vartheta_{12,r}(\tau,w)$.

Equations~\eqref{eq:app-primitive-first-three}--\eqref{eq:app-primitive-last-three}
are equivalent to the theta decomposition
\begin{equation}
 Y^{[2]}_{\six{1},a}(\tau,w)
 =\sum_{r\ ({\rm mod}\ 24)}
 h^{(1)}_{ar}(\tau)\,\vartheta_{12,r}(\tau,w).
 \label{eq:app-h-definition}
\end{equation}
For completeness, the non-vanishing coefficients $h^{(1)}_{ar}$ are
\begin{align}
 a=1:\quad&
 h_{1,3}=-\mathfrak a,\quad h_{1,6}=-\mathfrak b,\quad
 h_{1,9}=\mathfrak c,\quad h_{1,15}=-\mathfrak c,\quad
 h_{1,18}=\mathfrak b,\quad h_{1,21}=\mathfrak a,
 \nonumber\\
 a=2:\quad&
 h_{2,1}=\mathfrak c,\quad h_{2,7}=-\mathfrak c,\quad
 h_{2,10}=\mathfrak b,\quad h_{2,13}=\mathfrak a,\quad
 h_{2,19}=-\mathfrak a,\quad h_{2,22}=-\mathfrak b,
 \nonumber\\
 a=3:\quad&
 h_{3,2}=\mathfrak b,\quad h_{3,5}=\mathfrak a,\quad
 h_{3,11}=-\mathfrak a,\quad h_{3,14}=-\mathfrak b,\quad
 h_{3,17}=\mathfrak c,\quad h_{3,23}=-\mathfrak c,
 \label{eq:app-h-first}\\
 a=4:\quad&
 h_{4,3}=-\mathfrak a',\quad h_{4,6}=-\mathfrak b',\quad
 h_{4,9}=\mathfrak c',\quad h_{4,15}=-\mathfrak c',\quad
 h_{4,18}=\mathfrak b',\quad h_{4,21}=\mathfrak a',
 \nonumber\\
 a=5:\quad&
 h_{5,1}=\mathfrak c',\quad h_{5,7}=-\mathfrak c',\quad
 h_{5,10}=\mathfrak b',\quad h_{5,13}=\mathfrak a',\quad
 h_{5,19}=-\mathfrak a',\quad h_{5,22}=-\mathfrak b',
 \nonumber\\
 a=6:\quad&
 h_{6,2}=\mathfrak b',\quad h_{6,5}=\mathfrak a',\quad
 h_{6,11}=-\mathfrak a',\quad h_{6,14}=-\mathfrak b',\quad
 h_{6,17}=\mathfrak c',\quad h_{6,23}=-\mathfrak c'.
 \label{eq:app-h-last}
\end{align}
All coefficients not listed in Eqs.~\eqref{eq:app-h-first}--\eqref{eq:app-h-last}
vanish.

We briefly summarize how the above expressions are obtained.  Let
$\sigma_{\six{1}}$ denote the metaplectic representation induced on the
index-twelve Heisenberg-intertwiner space after projection to the
weight-$3/2$ central eigenspace.  The primitive form is selected by the
unique intertwiner
\begin{equation}
 X_{\rm prim}\in
 \operatorname{Hom}_{Mp_2}
 \!\left(
 \sigma_{\six{1}},
 W(2)^\eps\otimes W(4)^\eps\otimes W(6)^\eps
 \right),
 \qquad
 \dim\operatorname{Hom}_{Mp_2}=1,
 \label{eq:app-Xprim-hom}
\end{equation}
which obeys
\begin{equation}
 \rho_{246}(S)X_{\rm prim}=X_{\rm prim}\sigma_{\six{1}}(S),
 \qquad
 \rho_{246}(T)X_{\rm prim}=X_{\rm prim}\sigma_{\six{1}}(T).
 \label{eq:app-Xprim-equations}
\end{equation}
Writing $H$ for the Heisenberg basis and $P_c$ for the projection onto the
weight-$3/2$ central eigenspace, the coefficient matrix is generated by
\begin{equation}
 \operatorname{vec} h(\tau)
 =H\,P_c\,X_{\rm prim}^{\dagger}
 \left[
 \boldsymbol u^{(2)}(\tau)\otimes
 \boldsymbol u^{(4)}(\tau)\otimes
 \boldsymbol u^{(6)}(\tau)
 \right],
 \label{eq:app-intertwiner-master}
\end{equation}
where
$\boldsymbol u^{(\ell)}=(u^{(\ell)}_0,\ldots,u^{(\ell)}_\ell)^T$.
Choosing the common normalization and phase of $X_{\rm prim}$ so that
Eq.~\eqref{eq:app-intertwiner-master} has real coefficients gives
precisely Eqs.~\eqref{eq:app-abc}--\eqref{eq:app-h-last}.

The theta-component basis used above should not be identified directly with the
phenomenological coordinates $(A_X,A_Y,A_Z,B_X,B_Y,B_Z)$.  The invariant
contraction in
\begin{equation}
 \six{1}\otimes\three{1}\otimes\three{1}\supset\one
\end{equation}
fixes instead, up to one common normalization,
\begin{equation}
 \begin{pmatrix}
 A_X\\ A_Y\\ A_Z\\ B_X\\ B_Y\\ B_Z
 \end{pmatrix}
 =
 \begin{pmatrix}
 Y^{[2]}_{\six{1},4}\\
 Y^{[2]}_{\six{1},5}\\
 Y^{[2]}_{\six{1},6}\\
 -Y^{[2]}_{\six{1},1}/\sqrt2\\
 -Y^{[2]}_{\six{1},2}/\sqrt2\\
 -Y^{[2]}_{\six{1},3}/\sqrt2
 \end{pmatrix}.
 \label{eq:app-theta-to-pheno-six1}
\end{equation}
Consequently the symmetric Majorana texture used in the phenomenological
analysis is
\begin{equation}
 {\cal M}^{[2]}_{\six{1}}(\tau,w)=
 \begin{pmatrix}
 Y^{[2]}_{\six{1},4} & -Y^{[2]}_{\six{1},3}/\sqrt2 & -Y^{[2]}_{\six{1},2}/\sqrt2\\
 -Y^{[2]}_{\six{1},3}/\sqrt2 & Y^{[2]}_{\six{1},5} & -Y^{[2]}_{\six{1},1}/\sqrt2\\
 -Y^{[2]}_{\six{1},2}/\sqrt2 & -Y^{[2]}_{\six{1},1}/\sqrt2 & Y^{[2]}_{\six{1},6}
 \end{pmatrix} .
 \label{eq:app-primitive-majorana-matrix}
\end{equation}
This map is the unique Clebsch--Gordan contraction compatible with the
$J_3$ basis used for the matter triplets
and is the contraction used in the phenomenological analysis of
Sec.~\ref{sec:lepton-phenomenology}.
 The theta expressions
\eqref{eq:app-primitive-first-three}--\eqref{eq:app-h-last} themselves are
unchanged.  The overall normalization of
${\cal M}^{[2]}_{\six{1}}$ is absorbed into the Majorana scale
$\Lambda_R$.

\section{The zero mode on the two-torsion locus}
\label{app:torsion-zero-mode}
In this appendix we give a simple symmetry argument for the exact zero
mode on the locus
\begin{equation}
 {\cal L}_2:\qquad
 w=-\frac{\tau}{6}\,,
 \qquad\Longleftrightarrow\qquad
 z=-\frac{\tau}{2}\,.
 \label{eq:app-torsion-locus}
\end{equation}
By Jacobi conjugation, the result applies equally to the other non-trivial two-torsion sections discussed in Secs.~\ref{subsec:charged-anatomy} and \ref{subsec:neutrino-torsion}.
The point $z=-\tau/2$ represents a non-trivial two-torsion point of the
elliptic fibre.  Its stabilizer in the Jacobi group projects onto
\begin{equation}
 \Gamma^0(2)=
 \left\{
 \gamma=
 \begin{pmatrix}
 a&b\\ c&d
 \end{pmatrix}
 \in {\rm SL}(2,\mathbb Z)
 \,;\;
 b=0\ {\rm mod}\ 2
 \right\}\,.
 \label{eq:Gamma02-definition}
\end{equation}
Indeed, a transformation belonging to $\Gamma^0(2)$ maps the
two-torsion point $\tau/2$ into the corresponding two-torsion point of
the transformed elliptic curve, up to an allowed elliptic translation.
Thus, after restriction to ${\cal L}_2$, a Jacobi form can be regarded
as a modular object for $\Gamma^0(2)$, with an induced multiplier.

Consider a generic sextet
\begin{equation}
 {\cal Y}
 =
 (y_1,y_2,y_3,y_4,y_5,y_6)^T
 \sim{\bf 6}^{(0)}
\end{equation}
of weight $k=2$ and index $m=12$.  The element
\begin{equation}
 R=X^{-1}S^2
\end{equation}
belongs to the stabilizer of ${\cal L}_2$.  On the sextet it implies
\begin{equation}
 y_1=y_3\,,
 \qquad
 y_4=y_6\,,
 \label{eq:R-even-sextet}
\end{equation}
so that the restriction of ${\cal Y}$ to the locus belongs to a
four-dimensional $R$-even subspace.  In a parity basis adapted to $R$, the $(1,1)$ entry of the corresponding Yukawa
texture is proportional to the combination
\begin{equation}
 f(\tau)
 =
 y_4+\frac{y_2}{\sqrt2}\,.
 \label{eq:f11-definition}
\end{equation}
For the two sextets relevant to the charged-fermion sector this gives
\begin{equation}
 f_{DG}
 =
 D_2G_1+\frac{D_1G_2}{\sqrt2}\,,
 \qquad
 f_{EF}
 =
 E_2F_1+\frac{E_1F_2}{\sqrt2}\,.
 \label{eq:fDG-fEF}
\end{equation}
The residual $Z_2$ generated by $R$ alone does not force these
combinations to vanish.  The stronger constraint follows from the full
stabilizer~\eqref{eq:Gamma02-definition}.

It is convenient first to remove the universal factor induced by the
Jacobi index.  For $m=12$ we define
\begin{equation}
 \widehat f(\tau)
 \equiv
 q^{1/3}f(\tau)\,,
 \qquad
 q=e^{2\pi i\tau}\,.
 \label{eq:fhat-definition}
\end{equation}
Using the representation matrices of the sextet and the appropriate
Jacobi lifts of the generators of $\Gamma^0(2)$, one finds that
$\widehat f$ transforms as a scalar modular form of weight two with a
cubic character $\chi$.  A convenient choice of generators is
\begin{equation}
 A=T^2=
 \begin{pmatrix}
 1&2\\0&1
 \end{pmatrix},
 \qquad
 B=ST^{-1}S^{-1}=
 \begin{pmatrix}
 1&0\\1&1
 \end{pmatrix},
\end{equation}
for which
\begin{equation}
 \chi(A)=\chi(B)=\omega^2\,,
 \qquad
 \omega=e^{2\pi i/3}\,.
 \label{eq:cubic-character}
\end{equation}
Hence
\begin{equation}
 \widehat f(\gamma\tau)
 =
 \chi(\gamma)(c\tau+d)^2\widehat f(\tau)\,,
 \qquad
 \gamma\in\Gamma^0(2)\,.
 \label{eq:fhat-modular}
\end{equation}

This transformation law is already sufficiently restrictive to imply
$\widehat f=0$.  The cusp at $i\infty$ has width two.  Since
\begin{equation}
 \widehat f(\tau+2)=\omega^2\widehat f(\tau)\,,
\end{equation}
its expansion in the local variable
\begin{equation}
 Q_\infty=e^{\pi i\tau}
\end{equation}
starts at least as $Q_\infty^{2/3}$.  At the second cusp, $\tau=0$, the
character in~\eqref{eq:cubic-character} similarly implies an order of
at least $1/3$.  Cubing the form removes the character,
\begin{equation}
 g(\tau)\equiv\widehat f(\tau)^3\,,
\end{equation}
so that $g$ would be an ordinary modular form of weight six for
$\Gamma^0(2)$.  Its orders at the two cusps would satisfy
\begin{equation}
 {\rm ord}_{i\infty}(g)\geq2\,,
 \qquad
 {\rm ord}_{0}(g)\geq1\,.
 \label{eq:g-cusp-orders}
\end{equation}
On the other hand,
\begin{equation}
 \big[{\rm SL}(2,\mathbb Z):\Gamma^0(2)\big]=3\,,
\end{equation}
and the valence formula allows a total weighted number of zeros equal
to
\begin{equation}
 \frac{6}{12}\,
 \big[{\rm SL}(2,\mathbb Z):\Gamma^0(2)\big]
 =
 \frac32\,.
\end{equation}
The cusp contribution in~\eqref{eq:g-cusp-orders} would already be at
least three, which is impossible for a non-vanishing form.  Therefore
\begin{equation}
 \widehat f(\tau)=0\,,
 \qquad
 f(\tau)=0\,,
 \qquad
 \tau\in\mathbb H\,.
 \label{eq:f-vanishing}
\end{equation}

Applied to the two independent sextets of the charged-fermion sector,
Eq.~\eqref{eq:f-vanishing} gives
\begin{equation}
 D_2G_1+\frac{D_1G_2}{\sqrt2}=0\,,
 \qquad
 E_2F_1+\frac{E_1F_2}{\sqrt2}=0\,,
 \qquad
 w=-\frac{\tau}{6}\,.
 \label{eq:DG-EF-identities}
\end{equation}
Together with
\begin{equation}
 G_1=G_3\,,
 \qquad
 F_1=F_3\,,
\end{equation}
these relations imply
\begin{equation}
 {\cal M}'_{DG}
 =
 0\oplus B^{(2)}_{DG}\,,
 \qquad
 {\cal M}'_{EF}
 =
 0\oplus B^{(2)}_{EF}\,.
\end{equation}
Consequently,
\begin{equation}
 Y'_f
 =
 {\cal M}'_{DG}
 +r_f{\cal M}'_{EF}
 =
 0\oplus B_f^{(2)}\,,
 \qquad
 f=u,d,e\,,
\end{equation}
for arbitrary $r_f$.  The massless mode on the exact two-torsion locus
is therefore not an accidental property of the explicit theta-function
realization: it follows from the enlarged symmetry of the restriction
to the two-torsion section.  The explicit theta identities provide an
independent check of this result.

\bibliographystyle{utphys}
\bibliography{jacobi_draft}
\end{document}